\documentclass[a4paper,fleqn,usenatbib]{mnras}
\pdfoutput=1

\usepackage{newtxtext,newtxmath}

\usepackage[T1]{fontenc}
\usepackage{ae,aecompl}

\usepackage{graphicx, subfig}	
\usepackage{amsmath}	
\usepackage{amssymb}	

\usepackage{xspace}
\hypersetup{pdfauthor={Trebitsch, Volonteri, Dubois and Madau},pdftitle={Escape fraction in high-z dwarf galaxies with AGN}}

\newcommand{\hi}{\ifmmode {\mathrm{H\,\textsc{i}}} \else H\,\textsc{i} \fi}
\newcommand{\hii}{\ifmmode {\mathrm{H\,\textsc{ii}}} \else H\,\textsc{ii} \fi}
\newcommand{\hei}{\ifmmode {\mathrm{He\,\textsc{i}}} \else He\,\textsc{i} \fi}
\newcommand{\heii}{\ifmmode {\mathrm{He\,\textsc{ii}}} \else He\,\textsc{ii} \fi}
\newcommand{\heiii}{\ifmmode {\mathrm{He\,\textsc{iii}}} \else He\,\textsc{iii} \fi}
\newcommand{\Msun}{\ifmmode {\rm M}_{\odot} \else ${\rm M}_\odot$\xspace\fi}
\newcommand{\Mvir}{\ifmmode {M_{\rm vir}} \else $M_{\rm vir}$\xspace\fi}
\newcommand{\Mstar}{\ifmmode {M_\star} \else $M_{\star}$\xspace\fi}
\newcommand{\Mbh}{\ifmmode {M_\bullet} \else $M_{\bullet}$\xspace\fi}
\newcommand{\fesc}{\ifmmode {f_{\rm esc}} \else $f_{\rm esc}$\xspace\fi}
\newcommand{\fescAGN}{\ifmmode {f_{\rm esc}^{\rm AGN}} \else $f_{\rm esc}^{\rm AGN}$\xspace\fi}
\newcommand{\muv}{\ifmmode {M_{1500}} \else $M_{1500}$\xspace\fi}
\newcommand{\ramsesrt}{\textsc{Ramses-RT}\xspace}

\newcommand{\agnt}{\texttt{AGN\_SN}\xspace}
\newcommand{\agno}{\texttt{SN\_only}\xspace}
\newcommand{\agntnoSN}{\texttt{AGN\_only}\xspace}
\newcommand{\agnobh}{\texttt{BH\_SN}\xspace}
\newcommand{\agntEdd}{\texttt{AGN\_SN\_Eddington}\xspace}

\usepackage[usenames,dvipsnames,svgnames,table]{xcolor}

\title[Escape fraction in high-$z$ dwarf galaxies with AGN]{Escape of ionizing radiation from high redshift dwarf galaxies: role of AGN feedback}

\author[M. Trebitsch et al.]{
Maxime Trebitsch,$^{1}$\thanks{E-mail: maxime.trebitsch@iap.fr}
Marta Volonteri,$^{1}$
Yohan Dubois$^{1}$
and Piero Madau$^{1,2}$
\\
$^{1}$Sorbonne Universit\'{e}s, UPMC Univ Paris 6 et CNRS, UMR 7095, \\
\phantom{$^{1}$}Institut d'Astrophysique de Paris, 98 bis bd Arago, 75014 Paris, France\\
$^{2}$Department of Astronomy and Astrophysics, University of California, 1156 High Street, Santa Cruz, CA 95064, USA\\
}

\date{Accepted XXX. Received YYY; in original form ZZZ}

\pubyear{2017}

\begin{document}
\label{firstpage}
\pagerange{\pageref{firstpage}--\pageref{lastpage}}
\maketitle

\begin{abstract}
While low mass, star forming galaxies are often considered as the primary driver of reionization, their actual contribution to the cosmic ultraviolet background is still uncertain, mostly because the escape fraction of ionizing photons is only poorly constrained. Theoretical studies have shown that efficient supernova feedback is a necessary condition to create paths through which ionizing radiation can escape into the intergalactic medium.
We investigate the possibility that accreting supermassive black holes in early dwarf galaxies may provide additional feedback and enhance the leakage of ionizing radiation. We use a series of high resolution cosmological radiation hydrodynamics simulations where we isolate the different sources of feedback.
We find that supernova feedback prevents the growth of the black hole, thus quenching its associated feedback. Even in cases where the black hole can grow, the structure of the interstellar medium is strongly dominated by supernova feedback. We conclude that, in the dwarf galaxy regime, supermassive black holes do not appear to play a significant role in enhancing the escape fraction and in contributing to the early UV background.

\end{abstract}




\section{Introduction}
\label{sec:intro}

Cosmic reionization is the epoch during which the universe transitioned from the so-called Dark Ages, where most of the hydrogen in the intergalactic medium (IGM) was neutral, to the present epoch, where the IGM is fully ionized. It started as the first luminous objects in the universe formed (around $z \sim 15$), and observations of high redshift quasars suggest that it proceeded until $z \lesssim 6$ \citep[e.g.][]{Becker2001, Fan2001} with some large-scale variations at the very end \citep{Becker2015}. The latest results from the \emph{Planck} mission suggest that the reionization mid-point (when half of the volume of the universe is ionized) occurs around $z \sim 8.5$, only $600\ \mbox{Myr}$ after the Big Bang \citep{PlanckCollaboration2016}.
While the study of the imprints on the IGM of the early stages of the Epoch of Reionization (EoR) is one of the major endeavours of upcoming 21-cm experiments such as HERA and SKA, there is a strong need to understand the nature of the sources that reionized the universe, especially as the \emph{James Webb Space Telescope} (JWST) will be able to observe and characterize galaxies deep into the reionization era.

Two kinds of sources have predominantly been invoked as major contributors to the ionizing budget of the EoR since the first evidence for an ionized IGM was found by \citet{Gunn1965}: hot, massive stars in star forming galaxies, and active galactic nuclei (AGN) powered by supermassive black-holes (SMBH), with the exact contribution of each still highly uncertain.
Even though they are extremely bright, the standard picture emerging from high redshift observations is that quasars are too rare and cannot dominate the ionizing ultraviolet (UV) background during the EoR \citep[e.g.][]{Willott2010, Fontanot2012, Grissom2014}. In this paradigm, the main driver of reionization must be stellar light from star forming galaxies \citep[see e.g.][]{Robertson2010, Haardt2012, Kuhlen2012}. One of the main feature of this family of models is that they all require a large number of faint galaxies, with a UV luminosity function (LF) that must be extrapolated down to magnitudes $\muv \lesssim -13$ in order to match the observed integrated Thomson optical depth derived from CMB anisotropies \citep[e.g.][]{Robertson2013}. Such extrapolation has been validated by extremely deep surveys that have shown no sign of a turnover in the steep faint end slope of the UV LF \citep{Atek2015, Ishigaki2017, Livermore2017}, consistent with recent simulations of the EoR \citep[e.g.][]{OShea2015, Gnedin2016, Ocvirk2016}.
This galaxy-dominated picture has recently been challenged by recent measurements of the quasar UV LF by \citet{Giallongo2015}, who find many more faint AGN at high redshift than previously thought \citep[but see][for a different point of view]{Weigel2015, Vito2016, Parsa2017}. These recent observations have revitalized some models where AGN are contributing a fraction \citep{Volonteri2009} or most of the ionizing photons \citep{Madau2004, Madau2015}. Other models and sime numerical works appear to exclude an AGN-dominated scenario \citep{Mao2016, Hassan2017, Onoue2017, Qin2017, Ricci2017, Yoshiura2017}.

Besides the intrinsic luminosity or abundance of either kind of sources, the escape fraction \fesc of ionizing photons (defined as the fraction of photons that manage to escape the host halo and ionize the IGM) is a key parameter in understanding reionization. Despite more than twenty years of effort \citep{Leitherer1995, Deharveng2001}, direct observations of the Lyman continuum (LyC) below $912\ \mbox{\AA}$ are still very difficult, and there is only a few galaxies for which \fesc has been directly measured  (see e.g. \citealt{Bergvall2013} for a review). At high redshift, the increasing opacity of the IGM makes this even harder, and foreground contamination becomes a major issue \citep{Siana2015}. Moreover, observations yield a large scatter on the value of \fesc between galaxies of different types and at different redshift, with no clear pattern: \citet{Izotov2016} recently reported LyC leakage in all of the four galaxies they observed at $z \sim 0.3$, with $\fesc \simeq 8-13\%$, while at intermediate redshift, \citet{Bridge2010} present stringent upper limit of $1\%$ on a stacked sample. Higher redshift observations of \citet{Vanzella2016} report a very high $\fesc \sim 50\%$ for a compact star forming galaxy at $z \sim 3$ while \citet{Grazian2017} finds only $\fesc < 1.7\%$ for a stacked sample of bright galaxies at similar redshift, and \citet{Vasei2016} find a much lower value of $\fesc < 8\%$ at $z \sim 2.38$. This large variation from galaxy to galaxy might be explained by variations in the luminosity \citep{Japelj2017} or in the physical properties of the galaxy \citep{Izotov2016}. Numerical simulations suggest a similar large scatter in the value of \fesc at fixed halo mass \citep{Kimm2014, Paardekooper2015}, and this large variability is correlated with the intrinsic burstiness of star formation in low mass galaxies and the resulting supernova (SN) feedback \citep{Wise2014, Ma2015, Trebitsch2017}.
AGN-dominated reionization models usually assume that $\fescAGN \sim 100\%$, meaning that all the emitted ionizing photons contribute to the ionizing background \citep[e.g.][]{Giallongo2015}, consistent with observations of bright quasars \citep{Cristiani2016}. However, for fainter AGN, the observations of \citet{Micheva2017} suggest \fesc well below unity.

At faint luminosities, it becomes increasingly difficult to separate the AGN contribution from the starlight of its host, and conversely, ionizing radiation coming from a source identified as an AGN might actually be partially of stellar origin \citep[see e.g.][]{Volonteri2017}. For instance, the LyC emission coming from two AGN out of the four LyC candidates targeted by \citet{Micheva2017} could be of stellar origin. Furthermore, even when they are undetected, AGN can still have an impact on the interstellar medium (ISM) of high redshift galaxies, providing an additional source of feedback in addition to supernovae. This AGN feedback could in turn have a strong effect on the escape of ionizing radiation from galaxies. Using \emph{Chandra} data, \citet{Kaaret2017} find some evidence for the presence of a compact X-ray source in the LyC leaking galaxy Tol 1247-232 that might be powered by a low-luminosity AGN. They further suggest that the AGN could be responsible for outflows helping the escape of ionizing radiation.

While signs of nuclear activity have been detected for three decades in nearby dwarf galaxies \citep{Kunth1987, Filippenko1989}, systematic searches for BHs in the nuclei of dwarf galaxies are still fairly recent. \citet{Reines2013} (see also \citealt{Moran2014}) present a large spectroscopic sample of 136 dwarf galaxies $10^{8.5}\ \Msun \lesssim \Mstar \lesssim 10^{9.5}\ \Msun$ with BH signature, among which 35 have their spectra dominated by the AGN. This suggest that BHs, with masses sometimes as low as $5 \times 10^4 \Msun$ \citep{Baldassare2015}, can be present even in the lowest mass galaxies, such as NGC~4395, a galaxy with stellar mass below $10^8 \Msun$  \citep{Filippenko1989}. Searches for BHs in low mass galaxies are, however, extremely difficult due to the contamination by stellar light at all wavelengths \citep[see][for a review of AGN searches in low mass galaxies]{Reines2016}. At $z \lesssim 0.3$, \citet{Chen2017} find a sample of 10 more low mass galaxies harbouring AGN activity. \citet{Mezcua2016} recently uncovered a population of BHs in dwarfs at $z \lesssim 1.5$ in the \emph{COSMOS} survey. As dwarf galaxies have a simpler evolution than massive galaxies, their massive BH may be relics of seeds formed at very high redshift \citep[e.g.][]{Volonteri2008, Volonteri2010}, and the fraction of dwarf galaxies hosting massive BH could provide valuable clues to help understanding the formation of the progenitors of massive BH.

In this paper, we use high resolution radiation hydrodynamics (RHD) simulations of a galaxy in a cosmological context to investigate the growth of BHs in dwarf galaxies during the EoR. We take an agnostic view with respect to the seeding mechanism, and we only focus on the coevolution of high redshift dwarfs and their putative central BH. Our main goal is to understand the interplay between the BH growth and the various feedback processes involving massive stars and AGN. We then discuss the importance of the different sources of feedback on the escape of ionizing radiation from the galaxy.

The paper is structured as follows: we present in Sect.~\ref{sec:sims} the numerical methods employed in this work, we then discuss the connection between galaxy and BH growth in Sect.~\ref{sec:coevolution}. We investigate how this affects the production and escape of ionizing radiation in Sect.~\ref{sec:fesc}, and we discuss our results in Sect.~\ref{sec:discussion} before concluding in Sect.~\ref{sec:ccl}.

\section{Description of the simulations}
\label{sec:sims}

The simulations presented here are run with \ramsesrt \citep{Rosdahl2013, Rosdahl2015}, a public multi-group radiative transfer (RT) extension of the adaptive mesh refinement (AMR) code \textsc{Ramses}\footnote{\label{fn:ramses}\url{https://bitbucket.org/rteyssie/ramses/}} \citep{Teyssier2002}. The collisionless dark matter (DM) and stellar particles are evolved with a particle-mesh solver with cloud-in-cell interpolation. We follow the gas evolution by solving the Euler equations with a second-order Godunov scheme using the HLLC Riemann solver \citep{Toro1994}, with a MinMod total variation diminishing scheme to reconstruct the inter-cell fluxes. We impose a Courant factor of 0.8 for all simulations. We use a standard quasi-Lagrangian refinement strategy, in which a cell is refined if $\rho_{\rm DM} \Delta x^3 + \frac{\Omega_{\rm DM}}{\Omega_b}\rho_{\rm gas} \Delta x^3+ \frac{\Omega_{\rm DM}}{\Omega_b} \rho_* \Delta x^3 > 8\ m_{\rm DM}^{\rm HR}$, where $\rho_{\rm DM}$, $\rho_{\rm gas}$ and $\rho_*$ are respectively the DM, gas and stellar densities in the cell, $\Delta x$ is the cell size, and $m_{\rm DM}^{\rm HR}$ is the mass of the highest resolution DM particle; this would allow refinement as soon as there are at least 8 high-resolution DM particles in a cell in a DM-only run.

The RT module propagates radiation emitted by stars in three energy intervals between $13.6 - 24.59\ \mbox{eV}$, $24.59 - 54.4\ \mbox{eV}$, and over $54.4 \mbox{eV}$, respectively describing the \hi, \hei and \heii photon field. The equation of radiative transfer is solved on the AMR grid using a first-order Godunov method with the M1 closure \citep{Levermore1984, Dubroca1999} for the Eddington tensor. The radiation is coupled to the hydrodynamical evolution of the gas through the non-equilibrium thermochemistry for hydrogen and helium. Since we are mainly interested in the absorption of radiation within the galaxy (rather than the expansion of cosmological \hii regions in the IGM), we can use the reduced speed of light approximation, $\tilde{c} = 0.01 c$. Because \ramsesrt can handle an arbitrarily large number of sources, we emit radiation from all star particles at each timestep using the \textsc{Galaxev} model of \citet{Bruzual2003}, assuming a \citet{Chabrier2003} initial mass function (IMF). Finally, we use the on-the-spot approximation: any ionizing photon emitted by recombination is assumed to be absorbed locally, and we thus ignore direct recombinations to the ground level and the associated emission of ionizing radiation from the gas. We also ignore the radiation pressure from the UV field to the gas, which has been shown to have only a subdominant effect \citep{Rosdahl2015a}.

\subsection{Initial conditions}
\label{sec:sims:ICs}

To select the halo of interest, we first run for 1 Gyr a DM-only simulation of a region of $10\, h^{-1}$ comoving Mpc on a side, sampled with $512^3$ particles ($m_{\mathrm{DM}} \simeq 6.4\times 10^5\ h^{-1} \Msun \simeq 9.4\times 10^5\ \Msun$) initialized at $z = 150$. We use the \textsc{Music}\footnote{\label{fn:music}\url{https://bitbucket.org/ohahn/music/}} code \citep{Hahn2011} to generate the initial conditions, and assume a flat $\Lambda$CDM cosmology consistent with the \emph{Planck} results \citep[dark energy density $\Omega_\Lambda = 0.692$, total matter density $\Omega_m = 0.308$, Hubble parameter $h = 0.6781$ and baryon matter density $\Omega_b = 0.048$,][]{Planck2015}

The target halo is selected  to have a total mass of $5 \times 10^9\ \Msun$ in the final output at $z \sim 5.7$ with \textsc{HaloMaker} \citep{Tweed2009} using the \textsc{AdaptaHOP} algorithm \citep{Aubert2004}. We generate high resolution initial conditions for that halo by refining the mass distribution such that the DM particle mass is $m_{\mathrm{DM}}^{\rm HR} \sim 1.5 \times 10^4\ \Msun$ in the zoomed-in region, equivalent to a $2048^3$ particles simulation. This corresponds to an AMR coarse grid level of $\ell = 11$ in the zoomed-in region, and we allow for 10 more levels of refinement, resulting in a most refined cell size of $\Delta x = 10h^{-1}\textrm{Mpc} / 2^{21} = 7$ pc. The gas is assumed to be initially neutral, and with an initial gas phase metallicity $Z = 10^{-3} Z_\odot = 2 \times 10^{-5}$ everywhere.

\subsection{Galaxy formation physics}
\label{sec:sims:subgrid}

The subgrid recipes used in this work to model star formation, the subsequent feedback from massive stars and the various gas cooling and heating processes are very similar to those of \citet{Trebitsch2017}. In this section, we will summarize them briefly and highlight the few differences in their implementation.

\subsubsection{Star formation}
\label{sec:sims:subgrid:SF}

We model  star formation using the model presented in \citet{Trebitsch2017} and \citet{Kimm2017} (see also Devriendt et al. in preparation). The approach is similar to that of \citet{Rasera2006}: in star forming cells, $M_{\rm sf} = \epsilon \rho_{\rm gas} \Delta x^3 \Delta t/t_{\rm ff}$ of gas is converted into star particles during one timestep $\Delta t$, where $G$ is the gravitational constant, $\epsilon$ is the local star formation efficiency and $t_{\rm ff} = \sqrt{3\pi / 32 G \rho_{\rm gas}}$ is the gas free fall time. The actual number $N$ of star particles formed in one timestep $\Delta t$ is drawn from a Poisson distribution $P(N) = (\lambda^N/N!) \exp(-\lambda)$ of parameter $\lambda = M_{\rm sf} / m_\star$, where $m_\star \simeq 2000\ \Msun$ is the minimum mass of a star particle in our simulations.

This model describes star forming regions as clouds characterized by their average gas density $\rho_0$, sound speed $c_{\rm s}$ and local velocity dispersion $\sigma_{\text{gas}}$ (which accounts for the turbulence in the cloud). The local star formation efficiency $\epsilon$ is computed following the `multi-ff PN' model of \citet{Federrath2012}, based on \citet{Padoan2011}.
\begin{equation}
  \label{eq:sfr_ff}
  \epsilon \propto \text{e}^{\frac{3}{8}\sigma_s^2}\left[1 + \mathrm{erf}\left(\frac{\sigma_s^2 - s_{\rm crit}}{\sqrt{2\sigma_s^2}}\right)\right],
\end{equation}
where $\sigma_s  = \sigma_s(\sigma_{\rm gas}, c_{\rm s})$ characterizes the turbulent density fluctuations, $s_{\rm crit} \equiv \mathrm{ln}\left(\frac{\rho_{\rm gas, crit}}{\rho_0}\right)$ is the critical density above which the gas will be accreted onto stars, and $\rho_{\rm gas, crit} \propto (\sigma_{\rm gas}^2 + c_{\rm s}^2) \frac{\sigma_{\rm gas}^2}{c_{\rm s}^2}$.

The primary difference with \citet{Trebitsch2017} is in the way we identify star forming regions: instead of selecting cells that are Jeans-unstable (even with the additional turbulent support), we use a more relaxed criterion by simply enforcing the gas to be supersonic. The motivation for this modified approach is twofold. First, the models of \citet{Federrath2012} break down when the gas becomes trans- or subsonic, so we require a Mach number $\mathcal{M} \geq 2$ in star forming regions. This is well motivated by observations of molecular clouds, in which the Mach number is usually far greater than unity. Note that we only allow star formation in cells where $\rho > \rho_{\rm th} = 5\ \mbox{cm}^{-3}$ in order to avoid forming stars outside of the high resolution region.
Second, we drop the turbulent Jeans criterion to select star forming regions. Indeed, enforcing the `turbulent Jeans length' to be smaller than four cells directly translates into a constraint on the virial parameter of the cloud to be $\alpha_{\rm vir} \lesssim 5 - 6$, while observations show a large scatter with values sometimes above this threshold \citep[see for instance][]{Kauffmann2013}. In our model, the dependence of the star formation efficiency on the virial parameter is already encoded in equation~\eqref{eq:sfr_ff}, such that a cloud with high virial parameter can form stars, but at a very low efficiency, consistent with the cloud-scale, high resolution simulations of e.g. \citet{Padoan2017}.

\subsubsection{Feedback from massive stars}
\label{sec:sims:subgrid:stellar-fb}

We implement two channels of stellar feedback: radiative feedback from stars resulting from photo-ionization heating and type II supernovae (SN) explosions.
Photo-ionization heating is the main channel of radiative feedback on galactic scale \citep{Rosdahl2015a}, and is naturally included in the simulations as a consequence of the ionizing radiative transfer.
For  SN feedback, we use a mechanical feedback model similar to that first described by \citet{Kimm2014} and \citet{Kimm2015}, and later refined by \citet{Rosdahl2017}. We release momentum in the neighbouring cells (and energy in the SN host cell) in a single event 5 Myr after the birth of the star particle. This scheme was developed to capture correctly the momentum transfer at all stages of the Sedov blast wave, and the amount of momentum deposited in the simulation depends on the local properties (density and metallicity) of the gas.
We use a modified version of this scheme, implemented by \citet{Kimm2017}, which takes into account the argument of \citet{Geen2015} that the photo-ionization pre-processing of the ISM prior to the SN can augment the final radial momentum from a SN. While this should be taken into account by the radiative transfer in our simulation, \citet{Trebitsch2017} argued that a significant fraction of the radiation is actually emitted in regions where the Str{\"o}mgren radius is not resolved, leading to a likely underestimation of this momentum increase. We thus follow the subgrid model of \citet{Kimm2017} which adds this missing momentum when the Str{\"o}mgren radius is locally not resolved.

\subsubsection{Gas cooling and heating}
\label{sec:sims:subgrid:cooling}

\ramsesrt features non-equilibrium primordial cooling tracking the abundances of 5 species: H, H$^+$, He, He$^+$, He$^{++}$. We include an extra cooling term for metals, using the standard tabulated rates included in \textsc{Ramses} and computed with \textsc{Cloudy}\footnote{\label{fn:cloudy}\url{http://www.nublado.org/}} {\citep[last described in][]{Ferland2017}} above $10^4$~K. We also account for energy losses via metal line cooling below $10^4$~K following the prescription of \citet{Rosen1995}, and approximate the effect of the metallicity by scaling linearly the metal cooling enhancement. We do not take into account the impact of the local ionizing flux on metal cooling. Instead, metal cooling is computed assuming photo-ionization equilibrium with a redshift dependent \citet{Haardt1996} UV background. We stress that we do not use this UV background for the hydrogen and helium non equilibrium photo-chemistry, but rather take into account the local photo-heating rate self-consistently as we follow the radiation transport on the fly, which would be the physically reasonable assumption for an isolated galaxy before the end of the EoR.
We adopt an initial homogeneous metallicity floor of $Z = 10^{-3} Z_\odot$ in the whole box to account for the lack of molecular cooling and to allow the gas to cool down below $10^4$~K before the first stars have formed.

\subsection{BH model and AGN feedback}
\label{sec:sims:AGN}

We use the fiducial implementation of \citet{Dubois2012} for the black hole (BH) seeding and growth and the associated feedback from the AGN. In the following, we recall the main features of the model.

BHs are represented by sink particles of initial mass $M_{\bullet,0} = 10^4\ \Msun$ that are created in cells which meet the following criteria: both the gas and stellar densities must exceed a density threshold $\rho_{\rm sink}$, the cell must be Jeans-unstable, and there must be enough gas in the cell to form the sink. In order to avoid the creation of multiple BH in the same galaxies, we also enforce that there should be no other sink particle closer than an exclusion radius $r_{\rm excl}$. In our simulation, we use a density threshold $\rho_{\rm sink} = 400\ \mbox{cm}^{-3}$ (which corresponds to $10^4\ \Msun$ in a cell of $10\ \mbox{pc}$), and an exclusion radius $r_{\rm excl} = 50\ \mbox{kpc}$ (comoving).
Once the sink particle has been created, we spawn a swarm of `cloud' particles equally spaced by $\Delta x/2$ on a regular grid lattice within a sphere of radius $4 \Delta x$. These clouds particle provide a convenient way of averaging the properties of the gas around the BH.

BHs are allowed to grow both via gas accretion and via mergers. The accretion rate is computed using the Bondi-Hoyle-Lyttleton prescription \citep{Bondi1952}:
\begin{equation}
  \label{eq:accrate-bhl}
  \dot{M}_{\rm BHL} = 4\pi G^2 \Mbh^2 \frac{\bar{\rho}}{\left(\bar{c_s}^2 + \bar{u}^2\right)^{3/2}}
\end{equation}
where $\Mbh$ is the BH mass, $\bar{\rho}$ and $\bar{c_s}$ are respectively the average gas density and sound speed, and $\bar{u}$ the relative velocity between the black hole and the surrounding gas. The bar notation denotes an averaging over the cloud particles (see~\citealp{Dubois2012} for details). Note that we do not boost accretion as is usually done in kpc-resolution cosmological simulations \citep[see e.g.][]{Booth2009} because with our set of simulations we are able to resolve dense gas clumps. Even though we do not resolve the clumpy accretion onto the BH on the smallest scales, our $\sim 10\ \mbox{pc}$ resolution allows us to start to account for the multiphase structure of the ISM, and we are close to the regime where the non-boosted Bondi formalism captures correctly the accretion rate \citep{Negri2017}.
The BH accretion rate is capped at the Eddington rate, $\dot{M}_{\rm Edd} = 4\pi G \Mbh m_p / (\epsilon_r \sigma_{\rm T} c)$, where $m_p$ is the mass of a proton, $\sigma_{\rm T}$ is the Thompson cross section, $c$ is the speed of light and $\epsilon_r$ is the radiative efficiency of the accretion flow onto the black hole, taken to be the canonical value of $\epsilon_r = 0.1$.
Finally, following galaxy mergers, BH binaries are allowed to coalesce when they are closer than $4 \Delta x$ from each other and when their relative velocity is smaller than the escape velocity of the binary.

In order to avoid spurious spatial oscillations of the BH, we include a drag force to model the unresolved dynamical friction from the gas on the black hole as introduced in~\cite{Dubois2013}. The frictional force is proportional to $F_{\rm DF} = \alpha f_{\rm gas} 4\pi \rho (G \Mbh / \bar{c_s}^2)$, where $\alpha$ is an artificial boost introduced to account for unresolved small scale structures, with $\alpha  = (\rho/\rho_{\rm th})^2$ if $\rho > \rho_{\rm th}$ and 1 otherwise, and $f_{\rm gas}$ is a fudge factor varying between 0 and 2 which depends on the BH Mach number $\mathcal{M}_\bullet = \bar{u}/\bar{c_s}$ \citep{Ostriker1999, Chapon2013}.

We include the same dual mode of AGN feedback as in \citet{Dubois2012}, with ``radio mode'' feedback when the Eddington ratio $\lambda_{\rm Edd} = \dot{M}_{\rm BHL} / \dot{M}_{\rm Edd} < 0.01$ and ``quasar mode'' when $\lambda_{\rm Edd} \geq 0.01$.
For the ``quasar mode", we release thermal energy at the coarse timestep in a sphere of radius $\Delta x$ around the black hole at a rate $\dot{E}_{\rm AGN} = \epsilon_f \epsilon r \dot{M}_\bullet c^2$, where $\epsilon_f = 0.15$ is the fraction of the radiated energy that is transferred to the gas, and $\dot{M}_\bullet$ is the BH accretion rate.
In the radio mode, the energy is deposited as a bipolar outflow with a jet velocity $10^4\ \mbox{km\,s}^{-1}$, modelled as a cylinder of radius $\Delta x$ and height $2\Delta x$. The radio mode efficiency is assumed to be larger than in the ``quasar mode", with $\epsilon_f = 1$.
The feedback efficiencies in both the radio and quasar modes are empirically determined in order to reproduce the BH-to-bulge mass relations at $z=0$ \citep{Dubois2012}.

\begin{table}
  \centering
  \caption{Summary of all the simulations presented in this work and the variations on the physical processes used.}
  \begin{tabular}{lccc}
    Simulation          & SN           & AGN          & Accretion   \\
    \hline \hline
    \agnt       & $\checkmark$ & $\checkmark$ & Bondi       \\
    \hline
    \agno       & $\checkmark$ & \textendash  & \textendash \\
    \hline
    \agntnoSN   & \textendash  & $\checkmark$ & Bondi       \\
    \hline
    \agnobh     & $\checkmark$ & \textendash  & Bondi       \\
    \hline
    \agntEdd    & $\checkmark$ & $\checkmark$ & Eddington   \\
    \hline
  \end{tabular}
  \label{tab:feedback-variations}
\end{table}

\subsection{Summary of the simulations}
\label{sec:sims:summary}

In order to assess the exact role of BH growth and the associated feedback on the early evolution of low mass galaxies, we rerun the simulation while varying the origin and amount of feedback. We present on Table~\ref{tab:feedback-variations} the list of the simulations used in this work.

\begin{figure*}
  \centering
  \includegraphics[width=0.85\linewidth]{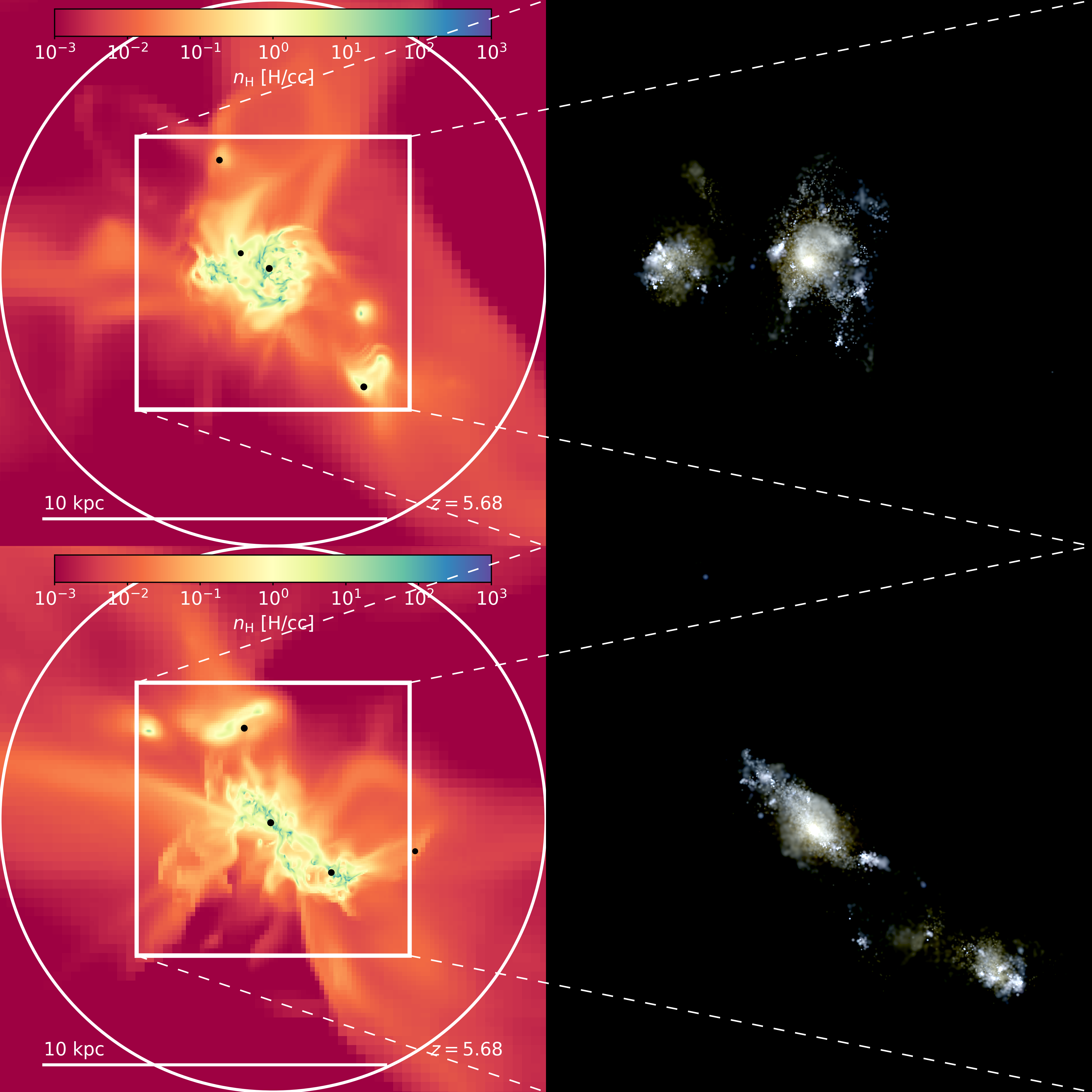}

  \caption{Views of the main galaxy in our simulation along the $x$ and $z$ lines of sight. \emph{Left}: mass-weighted projection of the gas density, with the black holes shown as black dots and the virial radius of the halo as a white circle. \emph{Right}: Mock $ugr$ image, zoomed on the galaxy at $z \sim 5.7$.}
  \label{fig:AGNT-maps}
\end{figure*}

For our fiducial model, which we will note \agnt, we use all the subgrid models described in the previous sections: mechanical feedback for supernovae, dual mode AGN feedback and Bondi accretion. For comparison with \citet{Trebitsch2017}, our \agno simulation only includes SN feedback, and no BH physics at all. We then alternatively removed SN feedback (\agntnoSN) and AGN feedback (\agnobh) to isolate their respective effects.
Finally, we perform a simulation where we enforce the BH to accrete at the Eddington limit (provided there is enough gas in its vicinity) rather than using the Bondi prescription, which we will denote by \agntEdd. Since we do not boost the accretion explicitly, this provides an extreme case to assess the robustness of our results, were we to underestimate the real accretion rate of our BH using the standard Bondi approach.

\section{Galaxy -- black hole coevolution}
\label{sec:coevolution}

In this section, we discuss how the main galaxy in our simulations assembles its stellar mass, and how this correlates with the growth of the central BH. We then discuss the role of the various feedback processes in controlling the accretion onto the BH.

\subsection{Stellar mass assembly}
\label{sec:coevolution:galaxy}

We illustrate in Fig.~\ref{fig:AGNT-maps} (left column) the gas density around the main galaxy in our \agnt simulation projected along the $x$ and $z$ lines of sight of the simulation, with the virial radius shown as a large circle. The right column presents mock $ugr$ images of the galaxy at $z \sim 5.7$, zooming on the inner region within $50\%$ of the virial radius. The galaxy resides in a large scale filament and is well isolated from its nearest neighbours. The gas distribution is very clumpy and disturbed, and the galaxy has clearly not settled in a fully formed, well defined disc, but rather shows a disturbed morphology. Black holes are shown as the black dots, and we can see that while one of the BHs resides in the centre of the galaxy, the others have not yet fallen at the bottom of the potential well and are still attached to smaller clumps.

In order to compare our simulation to high redshift observations, we derive the physical properties of the galaxy (stellar mass and SFR) using an observational approach. We compare these estimates with the quantities directly measured in the simulation in Appendix~\ref{sec:app-obs}.
We estimate the stellar mass from the UV continuum at $1500\ \mbox{\AA}$ based the empirical power-law relation obtained by \citet{Song2016}. We assume that each star particle in the simulation has the spectral energy distribution (SED) of a single stellar population (SSP) with the same age, mass and metallicity as the particle.
Using \citet{Bruzual2003} models, we compute the UV luminosity at $\lambda = 1500\ \mbox{\AA}$ of each star particle, and measure the UV magnitude of the galaxy by summing over all the luminosities.
This is likely to yield an overestimate of what would actually be measured in observations, since we assume no flux losses due to low surface brightness. However, given the very small size of the galaxy ($\lesssim 1\ \mbox{kpc}$ or $\lesssim 0.2\arcsec$ at $z\sim 6$), it would be at most barely resolved even with the high resolution imaging of \citet{Song2016}. We also assume that there is no dust attenuation, which is a fairly reasonable approximation at first order in the high redshift, low mass regime. We assume that the stellar mass can be related to the UV magnitude through $\log\Mstar = a(z) + b \times\muv$, and we linearly interpolate the best-fit normalization of \citet[Table~1]{Song2016} between $z = 5$ and $z = 8$ with fixed slope of $b = -0.5$.
We note that \citet{Song2016} assume a \citet{Salpeter1955} IMF, while we use a \citet{Chabrier2003} IMF to estimate the luminosity of our stellar particles. At fixed age, mass and metallicity, an SSP populated with a \citet{Salpeter1955} IMF will contain more low-mass stars, so its UV luminosity will be lower. Conversely, an SSP needs to be more massive to yield the same UV luminosity. We thus reduce the $\log\Mstar - \muv$ scaling of \citet{Song2016} by $0.2\ \mbox{dex}$ in \Mstar to take this into account.

\begin{figure}
  \centering
  \includegraphics[width=\linewidth]{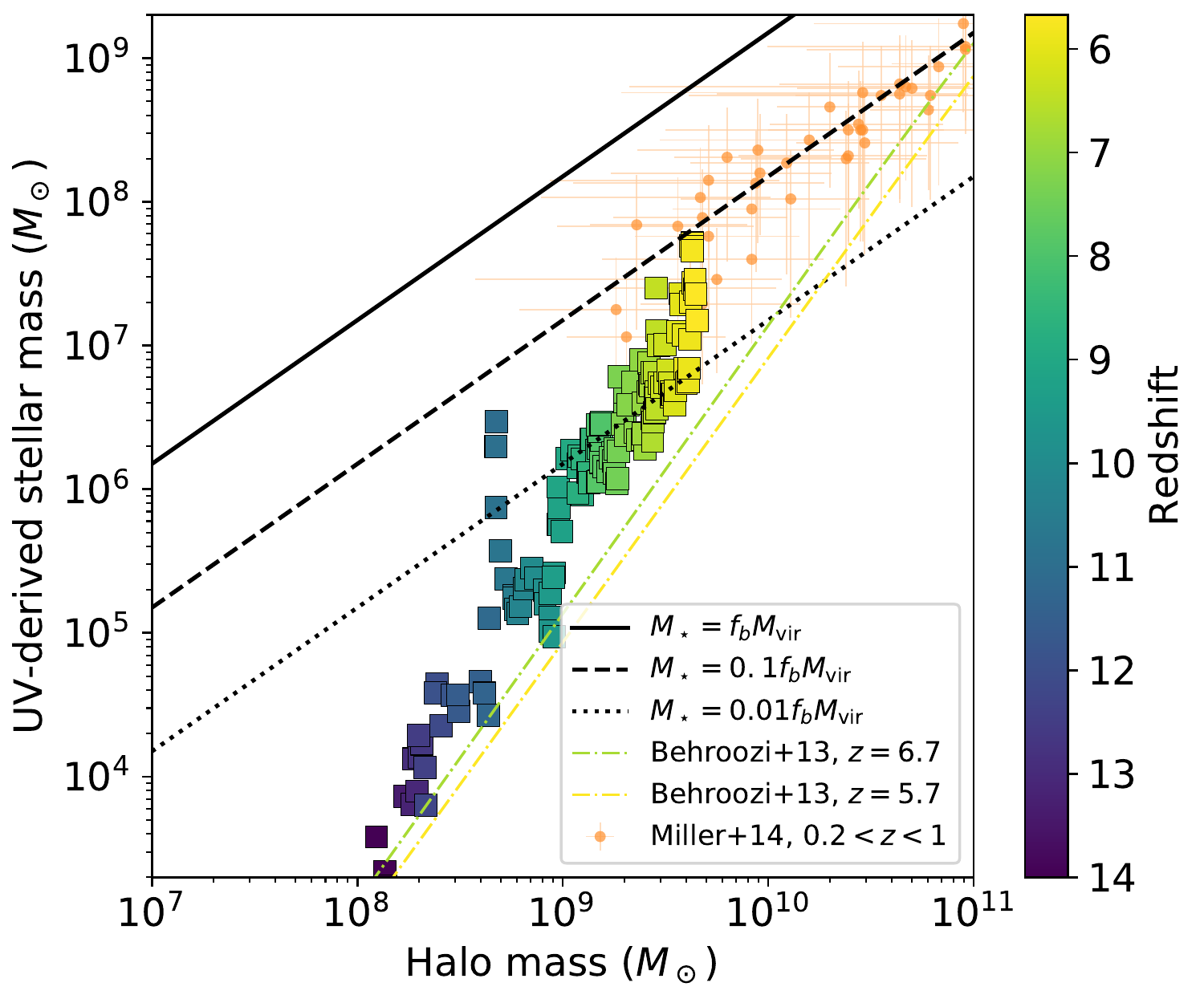}
  \caption{Stellar to halo mass relation for our simulated galaxy, using the UV estimate for the stellar mass. The solid, dashed and dotted line correspond respectively to 100\%, 10\% and 1\% of the baryons converted into stars.}
  \label{fig:AGNT-SMHM}
\end{figure}
We present in Fig.~\ref{fig:AGNT-SMHM} the resulting stellar mass to halo mass (SMHM) relation for the \agnt simulation. The solid, dashed and dotted lines respectively indicate 100\%, 10\% or 1\% of the universal baryon fraction. Using this UV estimate for the stellar mass, it appears that the galaxy converts around $1\%$ of its baryons into stars, with some increase at higher halo mass.
While this is above the predictions from abundance matching techniques \citep[e.g.][shown here as dash-dotted lines]{Behroozi2013}, we note that such models are explicitly not designed for the mass regime we are studying in this work \citep[see the discussion in][]{Moster2017}.
The large scatter at fixed halo mass mostly comes from the estimate of the stellar mass. Indeed, at low mass, galaxies assemble their mass by burst. Since the UV luminosity is dominated by massive, short-lived stars, a galaxy right after a star forming episode will appear brighter than a few tens of Myr later. At that stage, even though the real stellar mass does not decrease, our UV-based estimation will yield a lower stellar mass, and should therefore be considered as a lower limit of the true stellar mass. We still prefer to use this quantity rather than the stellar mass measured in the simulation, since it suffers from the same uncertainties as high redshift observations (e.g. degeneracy with the star formation history).
By comparison, we also show in Fig.~\ref{fig:AGNT-SMHM} the observations of \citet{Miller2014}, who present a sample of 41 galaxies at much lower redshift ($0.2 \lesssim z \lesssim 1$) and for which they compute halo masses from dynamical modelling and stellar masses from photometry. They find a SMHM systematically above the models of \citet{Behroozi2013}, with roughly $10\%$ of the universal baryon fraction converted into stars, very similar to our results in the mass range where observations and simulations overlap.

Still with the focus of comparing our simulation to observable quantities, we estimate star formation rates (SFR) using the far-UV calibration from \citet{Kennicutt2012}. While they assume a \citet{Kroupa2003}, models with a \citet{Chabrier2003} IMF yield identical results, so we include no correction for the IMF conversion.
\begin{figure}
  \centering
  \includegraphics[width=\linewidth]{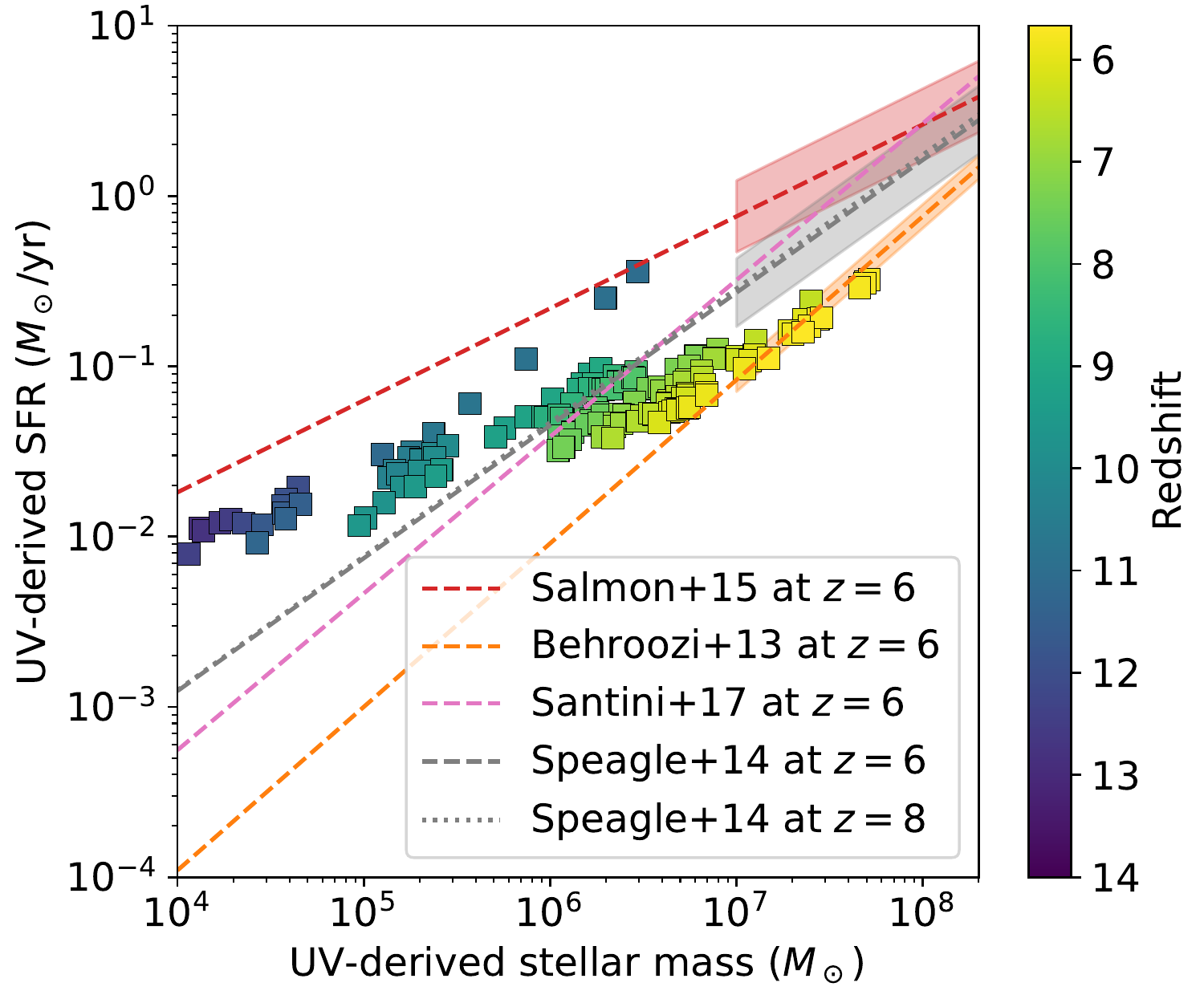}
  \caption{Star formation main sequence based on UV properties for the \agnt simulation, colour-coded by redshift. The lines correspond to SF main sequences extrapolated from the models of \citet[][orange]{Behroozi2013} and from the observations of \citet[][red]{Salmon2015}, \citet[][pink]{Santini2017} and \citet[][grey]{Speagle2014}, with their standard deviation shown as a shaded area when available.}
  \label{fig:AGNT-MS}
\end{figure}
\begin{figure*}
  \centering
  \includegraphics[width=.9\linewidth]{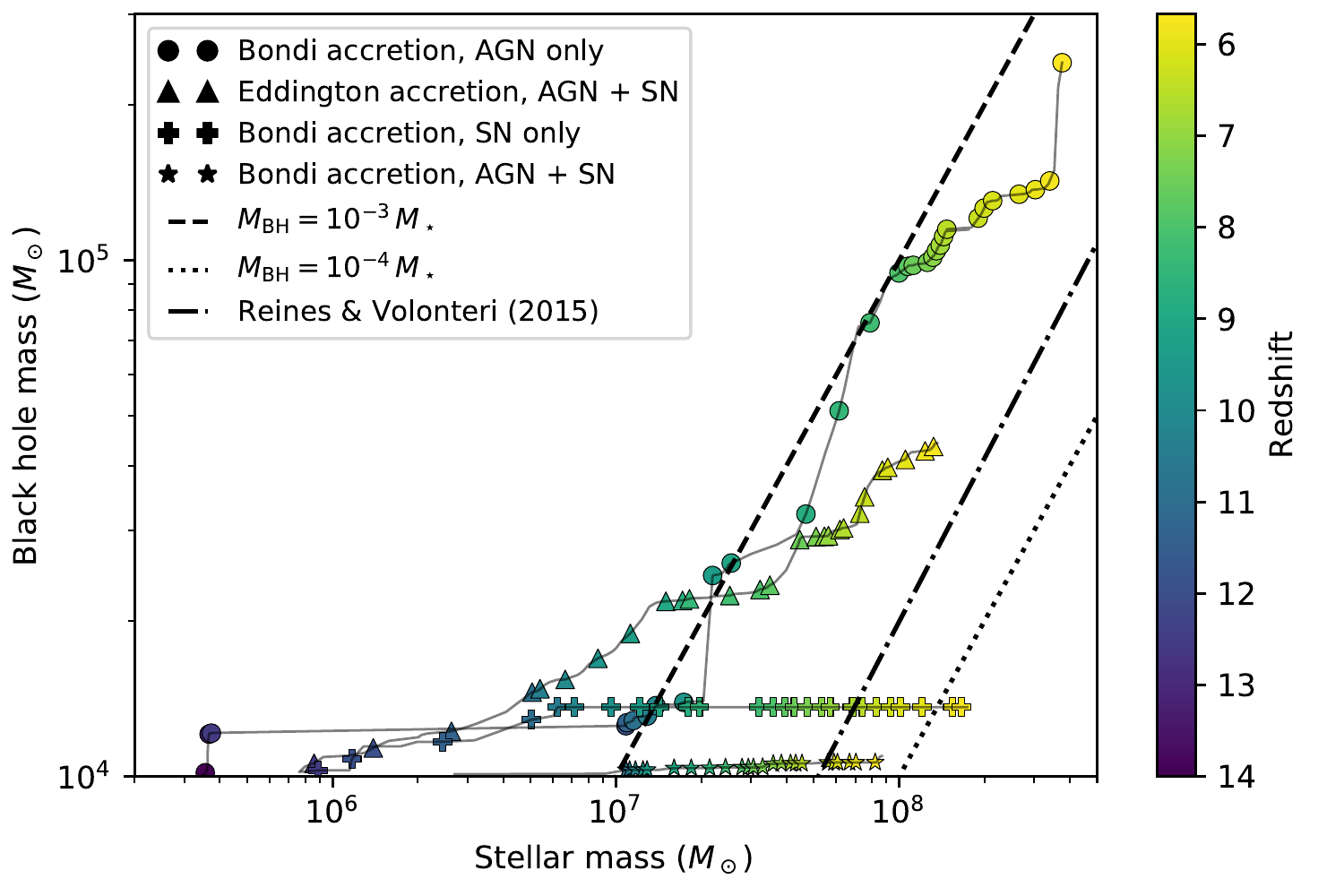}
  \caption{Black hole mass versus stellar mass colour-coded by redshift for all our simulations: stars for \agnt, circles for \agntnoSN, crosses for \agnobh and triangles for \agntEdd. For better legibility, we only show one point every five snapshots and use thin lines for the missing snapshots. SN feedback is by far the strongest mechanism to suppress BH growth in our low mass galaxy.}
  \label{fig:BH-growth}
\end{figure*}
We show in Fig.~\ref{fig:AGNT-MS} where our galaxy falls on the star forming main sequence (MS), with the UV-derived star formation rate as a function of the UV-derived stellar mass colour-coded with time since the beginning of the simulation. The various dashed lines indicate observational fits of the MS at $z \sim 6$ from \citet[][in red]{Salmon2015}, \citet[][in pink]{Santini2017} and \citet[][in grey]{Speagle2014}, and the shaded areas represent the standard deviation on the MS, shown when available. Both for \citet{Salmon2015} and \citet{Speagle2014}, the standard deviation is of order $0.2\ \mbox{dex}$. The grey dotted line shows the MS extrapolated to $z \sim 8$ from \citet{Speagle2014}, and is almost indistinguishable from the MS at $z \sim 6$. We over-plot in orange an extrapolation of the MS of \citet{Behroozi2013} as fitted by \citet{Salmon2015}.
We highlight that the apparent tight correlation between the stellar mass and the SFR comes from the fact that both quantities are derived from the UV luminosity, and we are effectively plotting against each others two functions of the same quantity. If we were to use the quantities directly measured in the simulation, we would find a much wider scatter (see Fig.~\ref{fig:AGNT-MS-direct}). Nevertheless, as stated previously, we prefer to use the UV-derived properties in order to get a fairer comparison to the observations shown on the figure.
Overall, the galaxy in our simulation falls within the various lines, with a somewhat shallower slope than most observations. We note however that the uncertainty on the exact shape of the MS at high redshift exceeds by far the standard deviation of $\sim 0.2\ \mbox{dex}$. As a result,  extrapolations to low masses such as those we present in Fig.~\ref{fig:AGNT-MS} are only trends around which some scatter is naturally expected due to intrinsic burstiness of star formation in low mass galaxies. While we should not consider quantitatively the position and evolution of our galaxy on the MS, the broad consistency between our simulation and observations suggests that our simulated galaxy assembles its mass at a reasonable pace.

The other simulations including SN feedback (\agno, \agnobh and \agntEdd) present a very similar stellar mass assembly and star formation history, while the simulation without SN feedback (\agntnoSN) forms much more stars, as expected.

\subsection{Black hole growth}
\label{sec:coevolution:BH-growth}

We present in Fig.~\ref{fig:BH-growth} the mass growth of the most massive BH of our simulations as a function of the stellar mass of the host galaxy; the colour coding indicates the redshift. Note that contrary to the previous section, we use here the stellar mass directly measured in the simulation. The various symbols represent various simulations: stars for the \agnt case, circles for \agntnoSN, crosses for \agnobh, and triangles for the Eddington accretion simulation (\agntEdd). We only show one in every five snapshots to make the figure easier to read, and join the symbols with thin lines to account for the complete growth history. As guide for the eye, we indicate constant BH to stellar mass ratio of $10^{-3}$ ($10^{-4}$) as a dashed (dotted) line, and the empirical relation derived by \citet{Reines2015} for galaxies in the local universe as a dash-dotted line.

It appears very clearly that when SNe and AGN are included, the BH does not grow at all. In 1 Gyr, the most massive black hole of the \agnt simulation, which lies at the centre of the galaxy in Fig.~\ref{fig:AGNT-maps}, grows by less than $1000\ \Msun$. By comparison, when supernova feedback is removed, the most massive black hole in \agntnoSN grows much faster and reaches a higher mass, roughly following $M_\bullet \simeq 10^{-3}\ \Mstar$. Interestingly, in the \agnobh simulation (with only SN feedback), the black hole reaches a slightly higher mass than in the fiducial simulation. Indeed, over the course of the whole simulation, the BH grows by $\sim 3500\ \Msun$.
For the simulation where the back hole is forced to accrete at the Eddington limit when possible (\agntEdd, diamonds in Fig.~\ref{fig:BH-growth}), it grows more regularly but still much less than in the \agntnoSN case.

\begin{figure*}
  \centering
  \subfloat[][\agnt]{\includegraphics[width=0.45\textwidth]{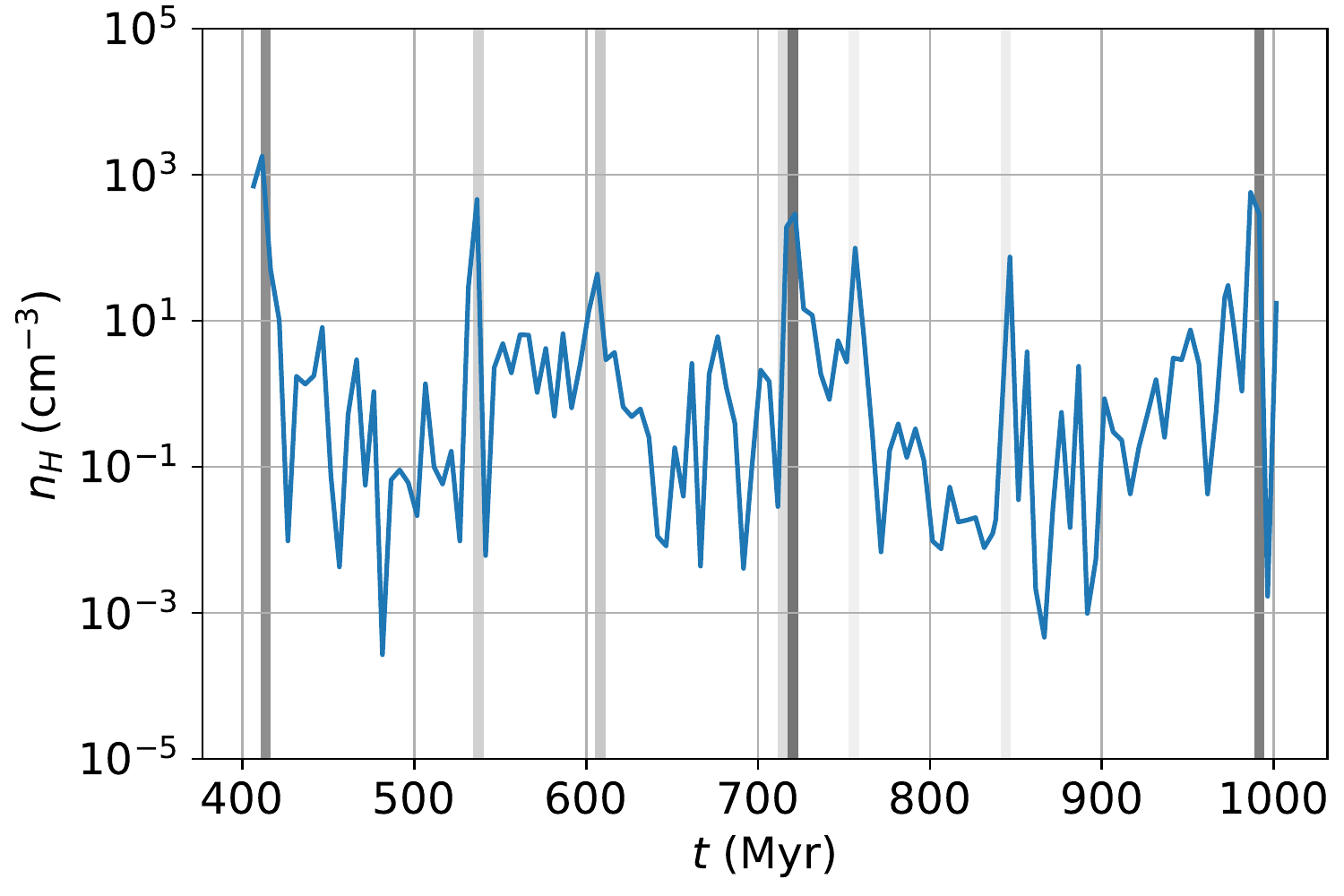}\label{fig:BH-density-AGNT}}
  \subfloat[][\agnobh]{\includegraphics[width=0.45\textwidth]{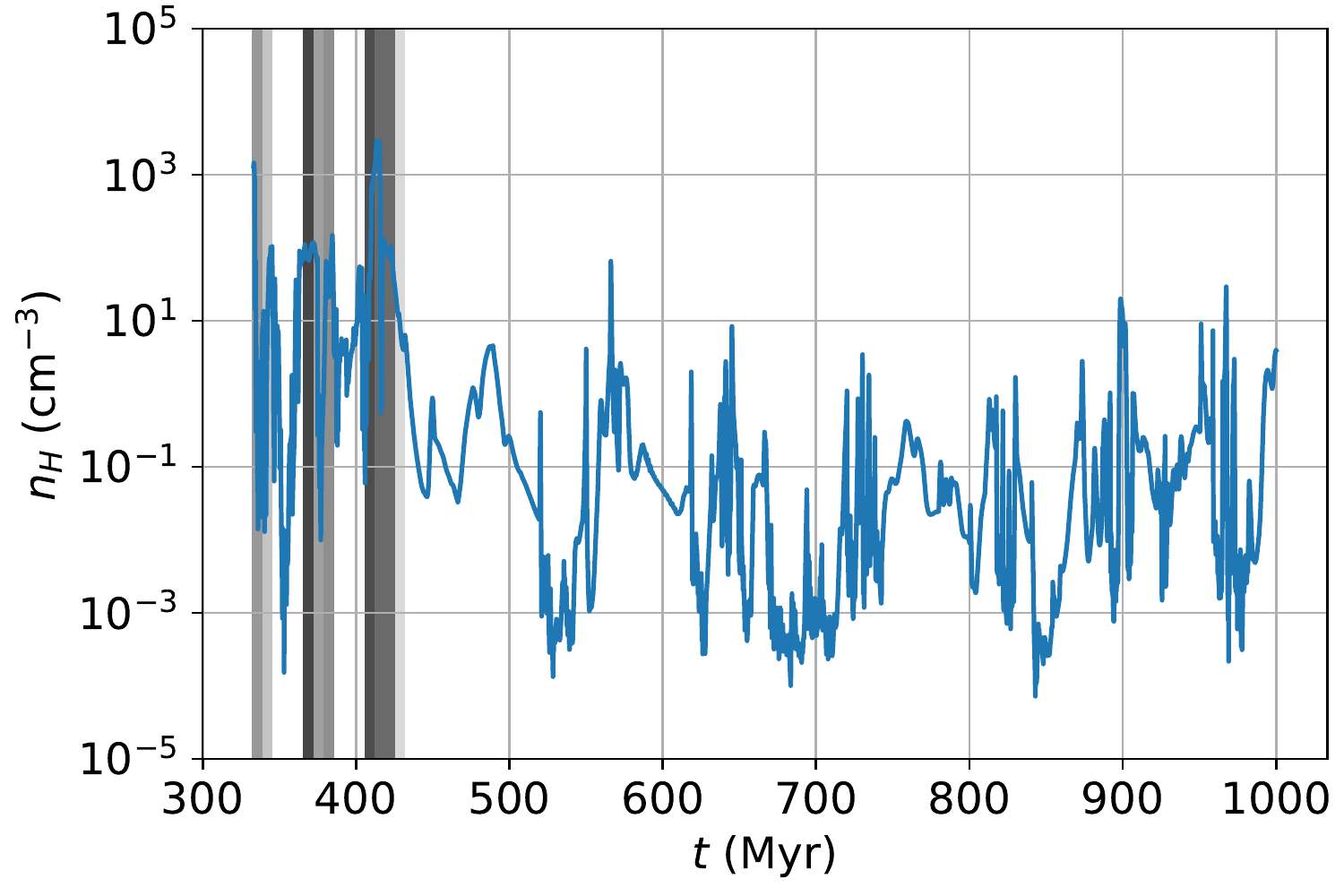}\label{fig:BH-density-AGN0_BH}}\\
  \subfloat[][\agntnoSN]{\includegraphics[width=0.45\textwidth]{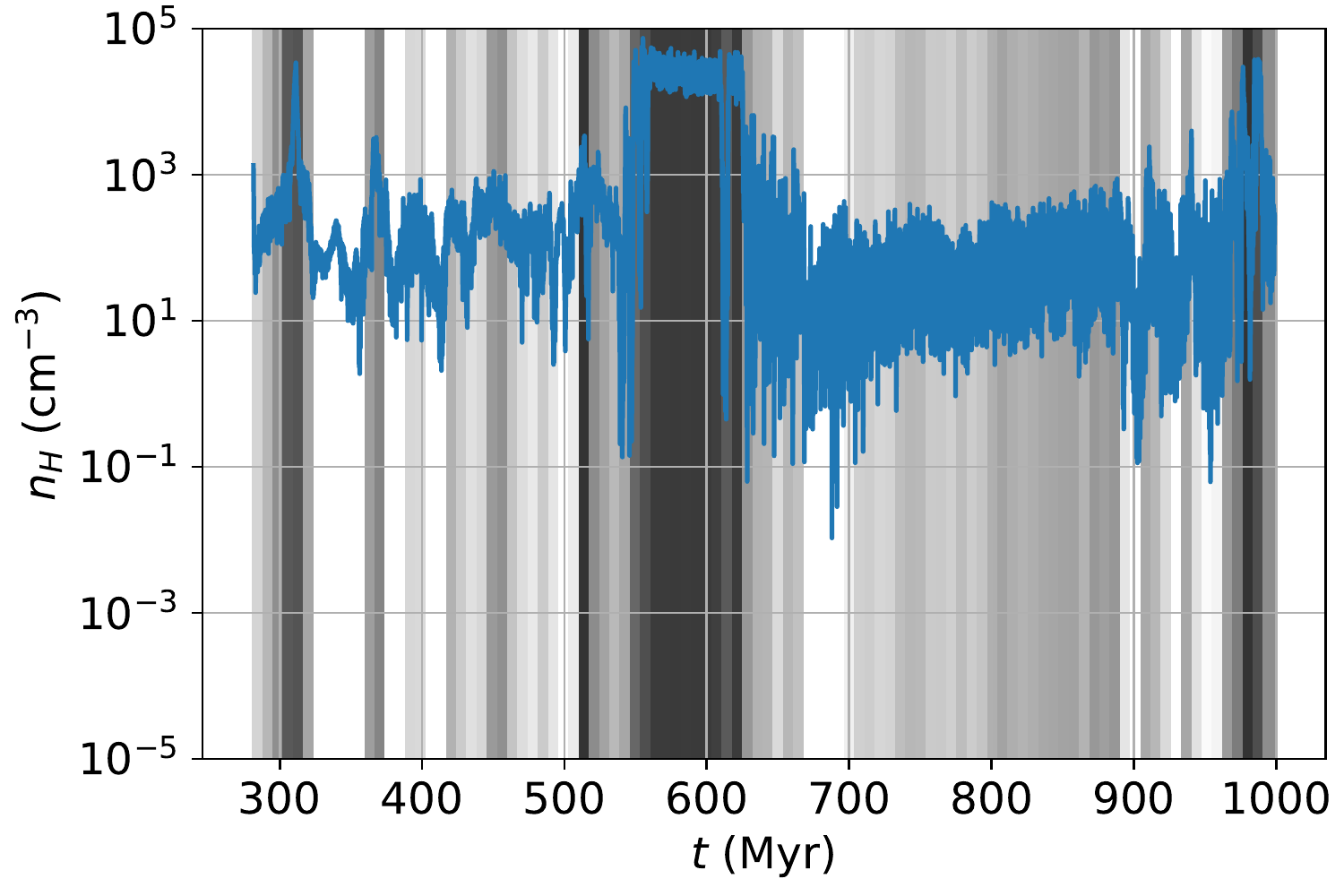}\label{fig:BH-density-AGNT_noSN}}
  \subfloat[][\agntEdd]{\includegraphics[width=0.45\textwidth]{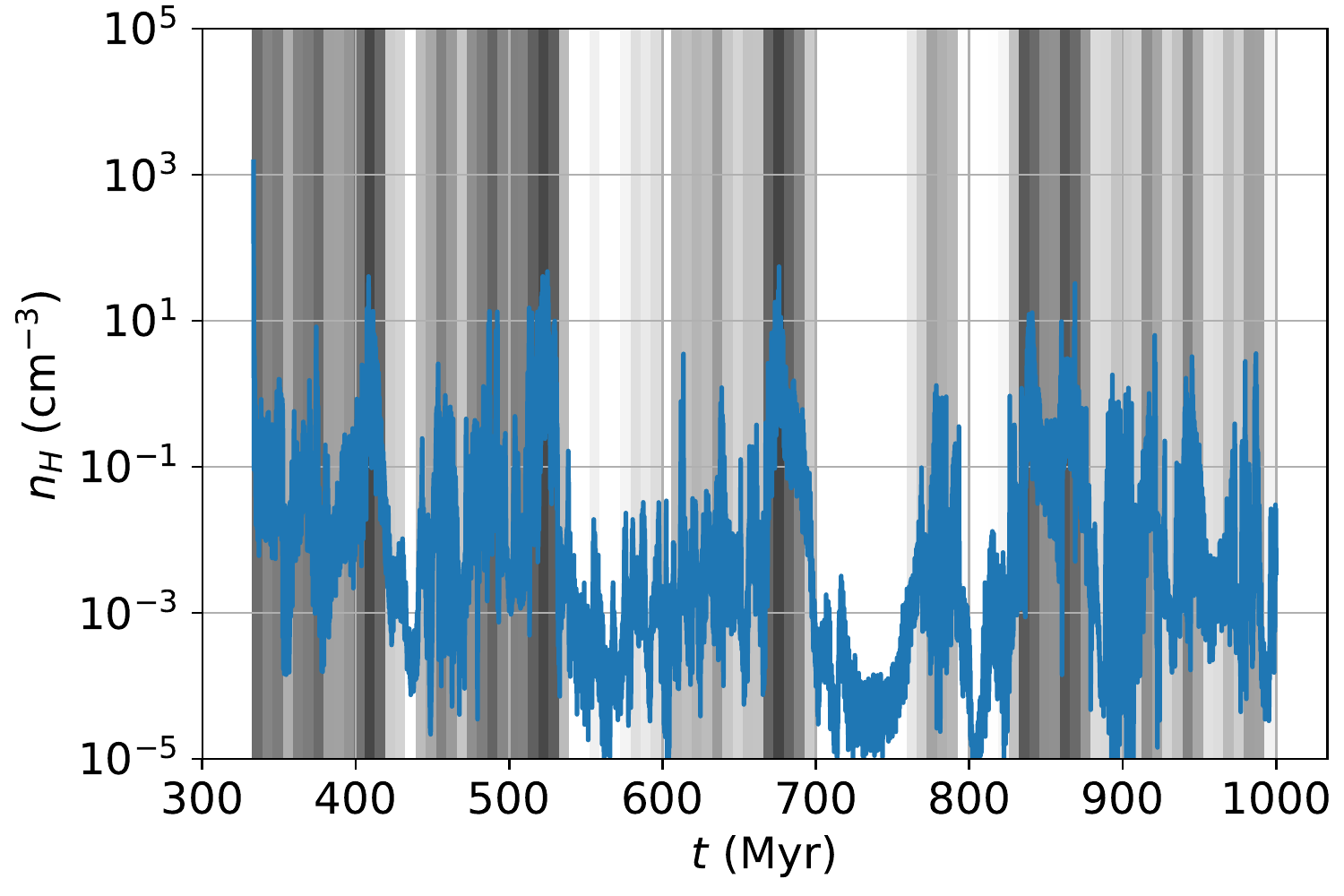}\label{fig:BH-density-AGNT_Edd}}
  \caption{Density in the vicinity of the most massive BH of each simulation. The background colour indicates the accretion rate, ranging from $10^{-2}\ \dot{M}_{\rm Edd}$ and below in white up to $\dot{M}_{\rm Edd}$ in dark grey. Note that the black holes form at slightly different times in the different simulation, as a result of a slightly different star formation history for the host galaxy.}
  \label{fig:BH-density}
\end{figure*}
This seems to indicate that the main mechanism regulating the growth of BHs in small galaxies is SN feedback. Indeed, even when the BH is allowed to accrete at the Eddington limit independently of the local gas properties, its growth is strongly limited by the lack of available gas. We develop this argument further in Fig.~\ref{fig:BH-density}, which presents the local gas density around the main BH in each simulation.
The accretion rate is shown as a grey background, with accretion rates below $10^{-2}\ \dot{M}_{\rm Edd}$ in white and a logarithmic scaling up to $\dot{M}_{\rm Edd}$ in black. The two upper plots (\agnt and \agnobh respectively on panel \ref{fig:BH-density-AGNT} and \ref{fig:BH-density-AGN0_BH}) display very similar features, with a low gas density on average, oscillating around $0.1$ to $1\ \mbox{cm}^{-3}$. The rare peaks of high densities correspond to episodes of stronger accretion, but these only last for less than $5 - 10\ \mbox{Myr}$ (the typical duration of a star formation episode). We interpret this as follows: at some point, the black hole falls into a denser gas clump, which will start to form stars. For a few Myr, the BH will accrete efficiently, until the cloud is dispersed by supernova feedback.

By comparison, without SNe (Fig.~\ref{fig:BH-density-AGNT_noSN}), the average density around the black hole is much higher, typically between $30$ and $200\ \mbox{cm}^{-3}$. In that case, the BH will accrete much more steadily, and as a result the AGN will be in ``quasar mode" for most of time. Because of this smooth energy injection, the gas surrounding the BH is continuously heated and pushed away from the very centre of the galaxy, and thus the BH does not grow at the Eddington limit all of the time. However, the BH sometimes falls into a massive clump where accretion becomes Eddington-limited, this happens for example from $t = 550\ \mbox{Myr}$ to $t = 620\ \mbox{Myr}$ in Fig.~\ref{fig:BH-density-AGNT_noSN}. The accretion will then only drop when the AGN feedback has injected enough energy to lower the density around the BH. Interestingly, the formation of this specific clump is not triggered by a {galaxy} merger (even though the central BH in that simulation does undergo a merger $\sim 30\ \mbox{Myr}$ earlier), but by the close encounter with a nearby satellite.
\begin{figure}
  \centering
  \includegraphics[width=.48\linewidth]{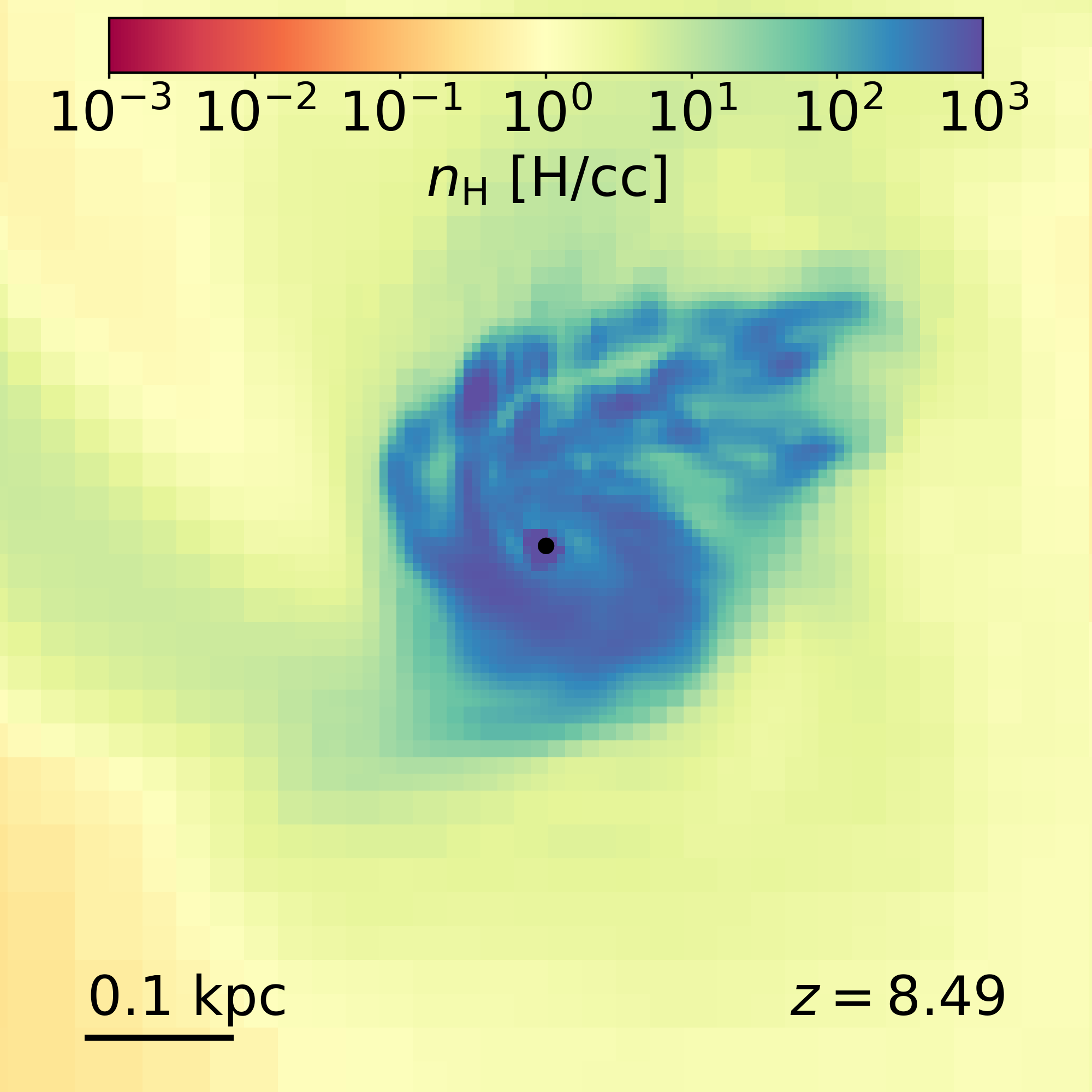}
  \includegraphics[width=.48\linewidth]{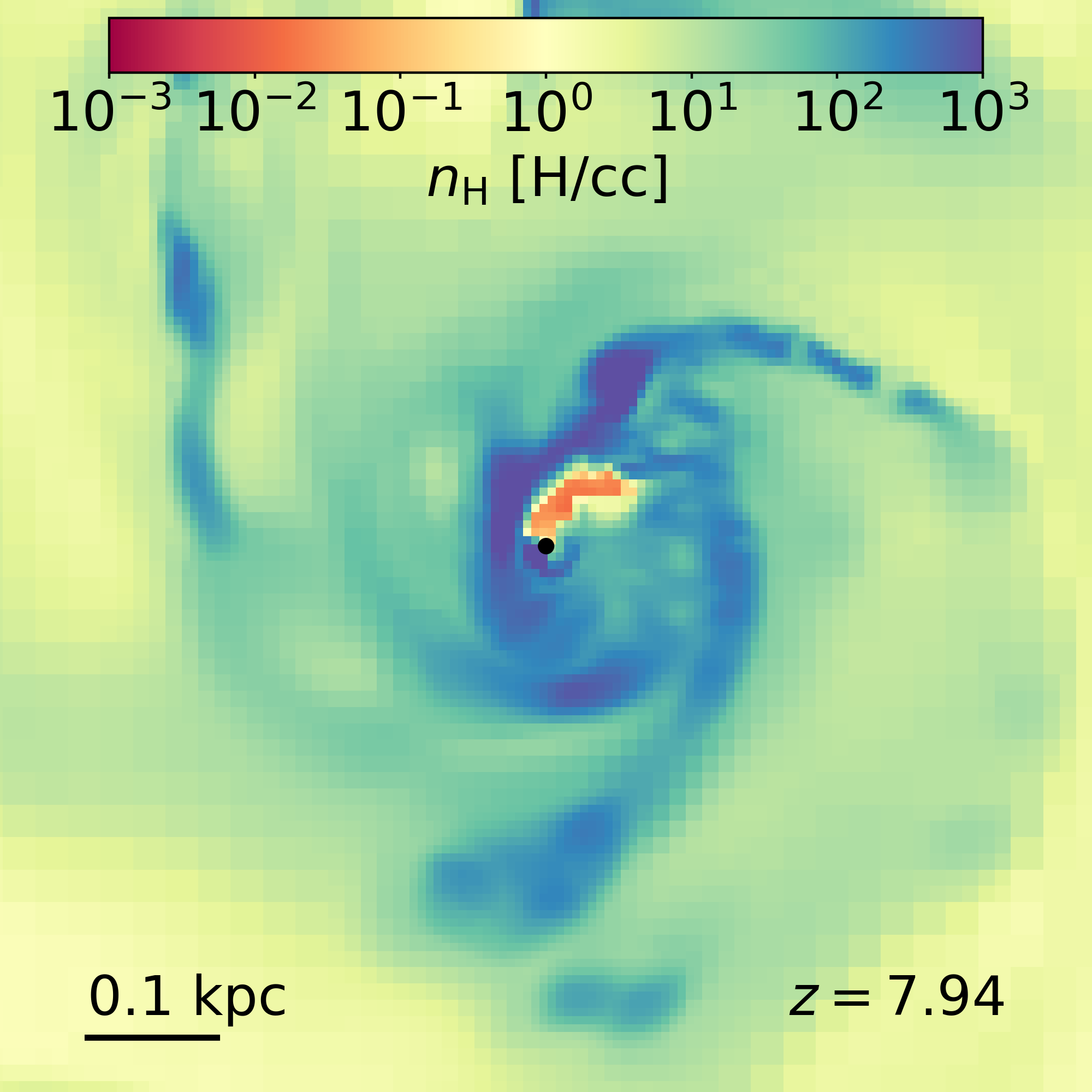}
  \caption{Projected gas distribution for two snapshots of the \agntnoSN simulation, with the BH position highlighted by a black circle. The left panel corresponds to a phase where the BH is embedded in a dense clump and accretes at the Eddington limit, while in the right panel, the Eddington accretion phase just finished.}
  \label{fig:BH-clump}
\end{figure}
We illustrate this process in Fig.~\ref{fig:BH-clump}, which shows a projected map of the gas density for two snapshots of the \agntnoSN simulation, at $t \sim 600\ \mbox{Myr}$ (left) and $t \sim 640\ \mbox{Myr}$ (right). The black circle at the centre of the figure depicts the BH position. A quick visual inspection confirms that the drop in density and accretion rate seen in Fig.~\ref{fig:BH-density-AGNT_noSN} is indeed due to feedback and not dynamics: the BH stays at the centre of the galaxy while a low-density hole is created nearby.

The situation is very different for the \agntEdd simulation (Fig.~\ref{fig:BH-density-AGNT_Edd}), where we allow for both AGN and SN feedback. Since the BH is allowed to accrete at the Eddington limit if there is enough gas available, the accretion rate is directly tied to the local gas density. At the same time, this artificially boosted accretion will result in an artificially strong AGN feedback, which will reinforce the effect of SNe to reduce the local gas density around the BH. This explains why the average density in Fig.~\ref{fig:BH-density-AGNT_Edd} is approximately one order of magnitude lower than for the fiducial \agnt simulation.

\begin{figure}
  \centering
  \includegraphics[width=\linewidth]{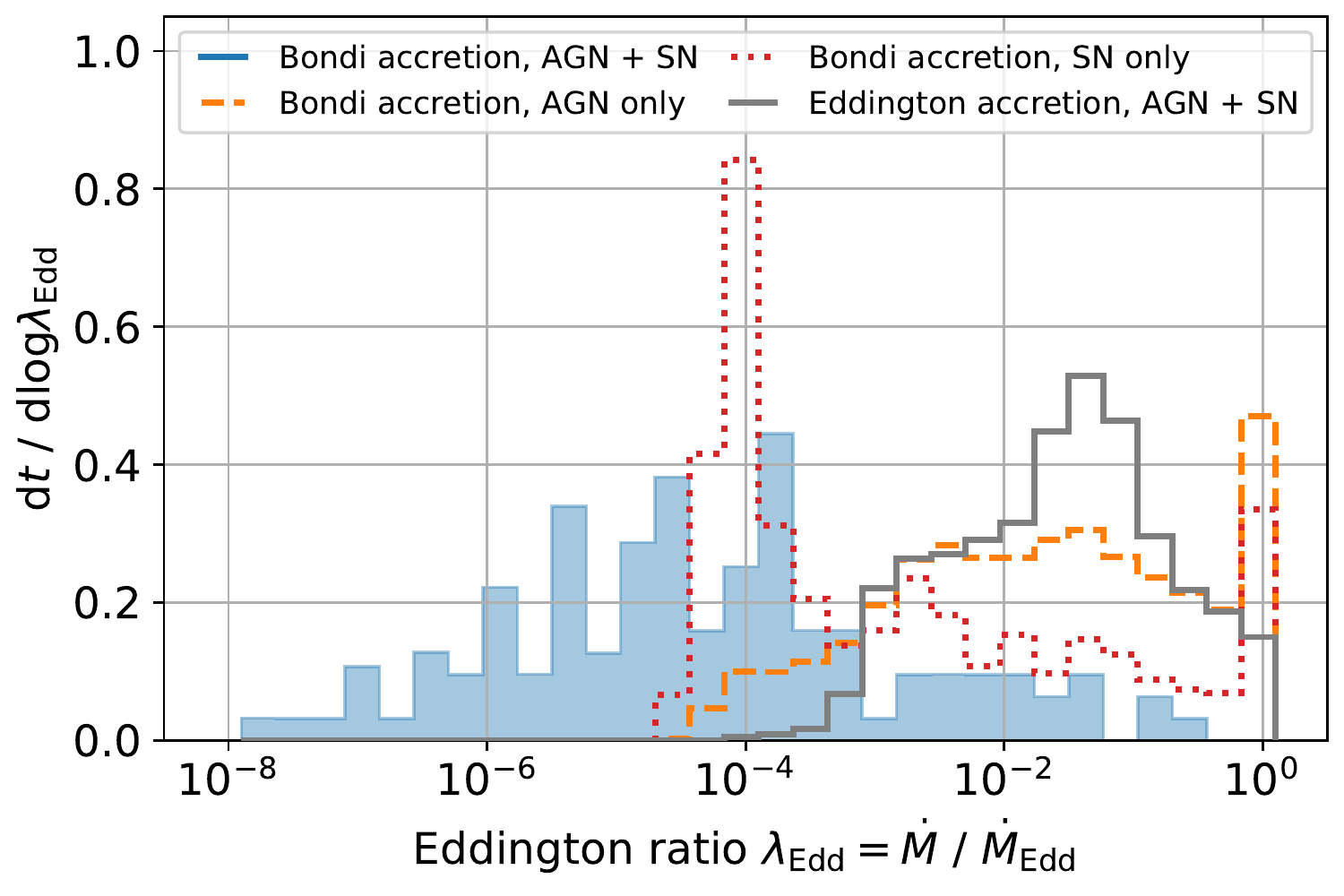}
  \caption{Probability distribution function of Eddington ratios for all the simulations: blue filled histogram for \agnt, orange dashed line for \agntnoSN, red dotted line for \agnobh and solid grey line for \agntEdd.}
  \label{fig:BH-accrate}
\end{figure}
We quantify this further in Fig.~\ref{fig:BH-accrate}, which presents the probability distribution function of Eddington ratios $\lambda_{\rm Edd} = \dot{M}/\dot{M}_{\rm Edd}$ for all the simulations. As already expected from Fig.~\ref{fig:BH-density}, the BH in the \agnt simulation (filled histogram in blue) accretes very little. All the other curves show episodes where the BH is accreting at the Eddington limit, but interestingly the \agntEdd simulation (solid grey line) is the one for which these Eddington-limited episodes represent the shortest amount of time. For the \agnobh simulation (red dotted line), the BH accretes very little for most of the time, but experiences some high accretion episodes when it falls in a star forming clump prior to its destruction via SN feedback. Finally, the BH in the simulation \agntnoSN spends a large amount of time accreting at the Eddington limit (mostly corresponding to the clump capture at $t\simeq 600\,\rm Myr$), but the $\lambda_{\rm Edd}$ distribution has a broad secondary component around $\lambda_{\rm Edd} \sim 10^{-2}$, corresponding to the phases where the BH is not embedded in a very dense clump. 

\section{Escape of ionizing radiation}
\label{sec:fesc}

\subsection{Escape of stellar radiation}
\label{sec:fesc:glx}

We now discuss the effect of AGN feedback on the escape of \hi-ionizing radiation from the galaxy where they are produced. We measure the fraction \fesc of photons that manage to escape the halo following \citet{Trebitsch2017}: we integrate the total ionizing flux escaping the halo and compare it to the ionizing photon production rate of the galaxy within the halo. We assume that all the stars are at the centre of the galaxy, and we correct the stellar ages for the light crossing time of the halo, $t_{\rm cross} = R_{\rm vir} / \tilde{c}$, where $R_{\rm vir}$ is the virial radius and $\tilde{c}$ is the reduced speed of light. The photon production rate is then measured using SED models of \citet{Bruzual2003}.

\begin{figure}
  \centering
  \includegraphics[width=\linewidth]{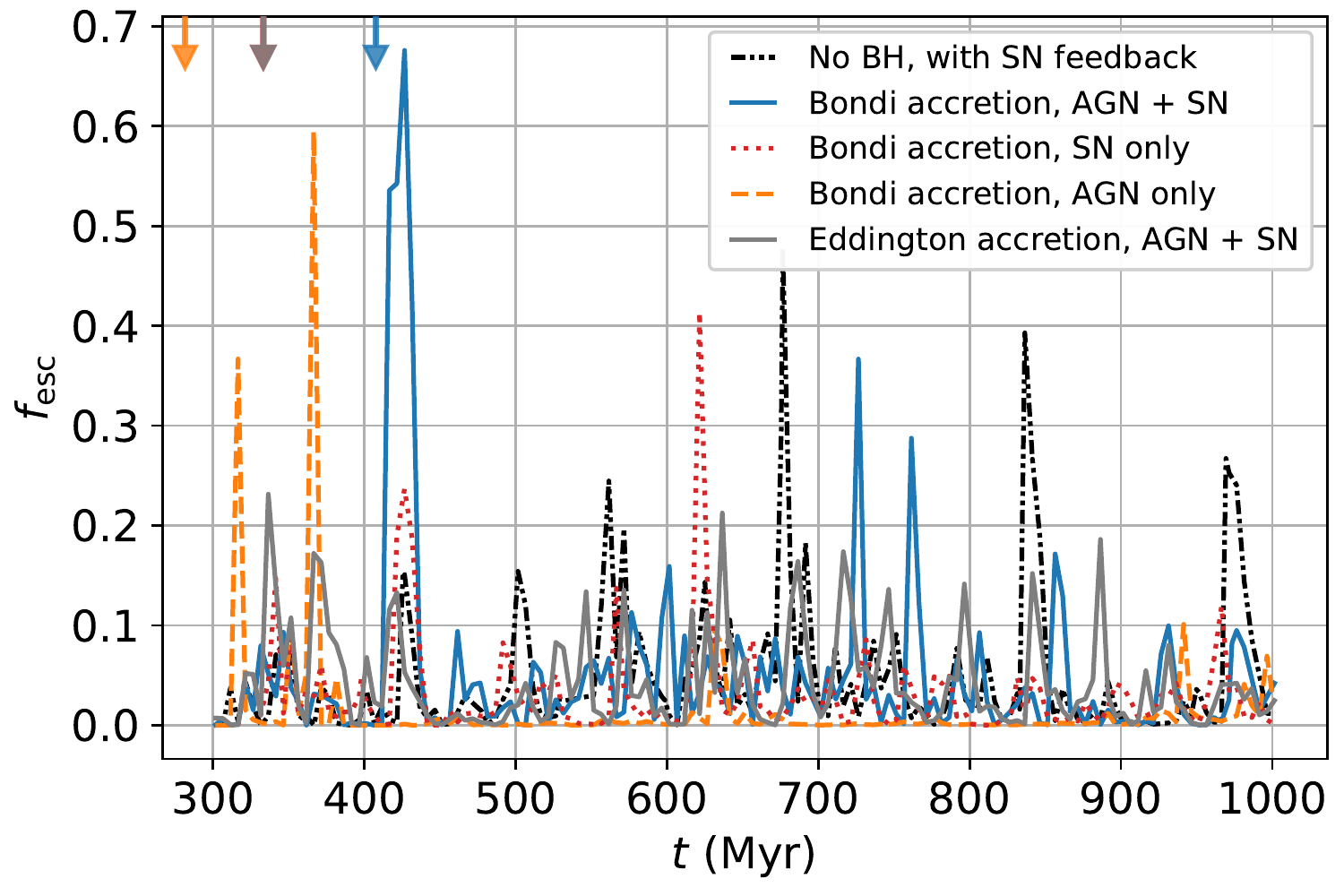}
  \caption{Evolution of the escape fraction \fesc with time for the different simulations presented in this work. When SN feedback is included, \fesc has a very bursty behaviour, almost independently of the BH accretion and feedback. Without SN feedback however, the high \fesc episodes are much less frequent and directly tied to AGN feedback.}
  \label{fig:photons-fesc}
\end{figure}

We show in Fig.~\ref{fig:photons-fesc} the escape fraction \fesc as a function of time for all the simulations presented in this work. The BH formation times are annotated as arrows at the top of the figure.
All the simulations with SN feedback (black dash-dot-dotted line for \agno, blue solid line for \agnt, red dotted line for \agnobh and grey solid line for \agntEdd) behave very similarly, with some large spikes of high \fesc episodes lasting for up to $\sim 10 - 20\ \mbox{Myr}$.
The two simulations without AGN feedback (\agno in black and \agnobh in red) follow each other even more closely, but some differences seem to appear (e.g. a peak of \fesc at $680\ \mbox{Myr}$ for \agno does not exist for \agnobh), mostly due to the intrinsic stochasticity of star formation. This tends to indicate that the mere presence of a BH accreting in the galaxy can slightly change its instantaneous star formation properties.
The fiducial \agnt simulation seems to undergo a very high escape fraction episode around the formation time of its BH, around $410\ \mbox{Myr}$, but it is actually due to a concurrent star formation event (see the jump in stellar mass around that time in Fig.~\ref{fig:AGNT-SMHM} { and appendix~\ref{sec:app-stellar-mass}}). These two events are of course related: since we impose a threshold on the local stellar density to spawn our sink particles, such a massive star formation event will necessarily result in the apparition of a new BH at that time, if not already present. The beginning of the high \fesc episode corresponds to the end of the first (and strongest) BH accretion episode (see Fig.~\ref{fig:BH-density-AGNT}), further suggesting that AGN feedback participates in clearing the ISM. However, the spike in \fesc is also present (though reduced by a factor $\sim 3$) for other simulations which include SN feedback. While this suggests that its origin is physical rather than a numerical artefact, it also means that  AGN feedback is only helping the radiation to escape at that time.
The simulation without SN feedback (\agntnoSN, the orange dashed line) exhibits a very different time evolution, which is expected based on the results of e.g. \citet{Trebitsch2017}. Yet, Fig.~\ref{fig:photons-fesc} features two spikes of very high \fesc, which indeed correspond to episodes of high accretion rate, e.g. at $t \sim 300\ \mbox{Myr}$ and $t \sim 370\ \mbox{Myr}$ (see Fig.~\ref{fig:BH-density-AGNT_noSN}). However, these episodes (at $z \sim 13-14$) correspond to a merger between the progenitor of the halo of interest at $z \sim 6$ and another halo that had already formed stars. The high \fesc episodes result from this merger, which breaks our assumption that all stars lie at the centre of the galaxy, and strongly disturbs the ISM of both galaxies and therefore facilitate the escape of ionizing radiation.
The long episode of very high accretion rate that lasts from $t \sim 550\ \mbox{Myr}$ to $t \sim 620\ \mbox{Myr}$ has apparently no effect on \fesc. It is only at the end of this episode that \fesc rises again to around $10\%$, when  AGN feedback finally manages to pierce the dense clump in which the BH is embedded.

\begin{figure}
  \centering
  \includegraphics[width=\linewidth]{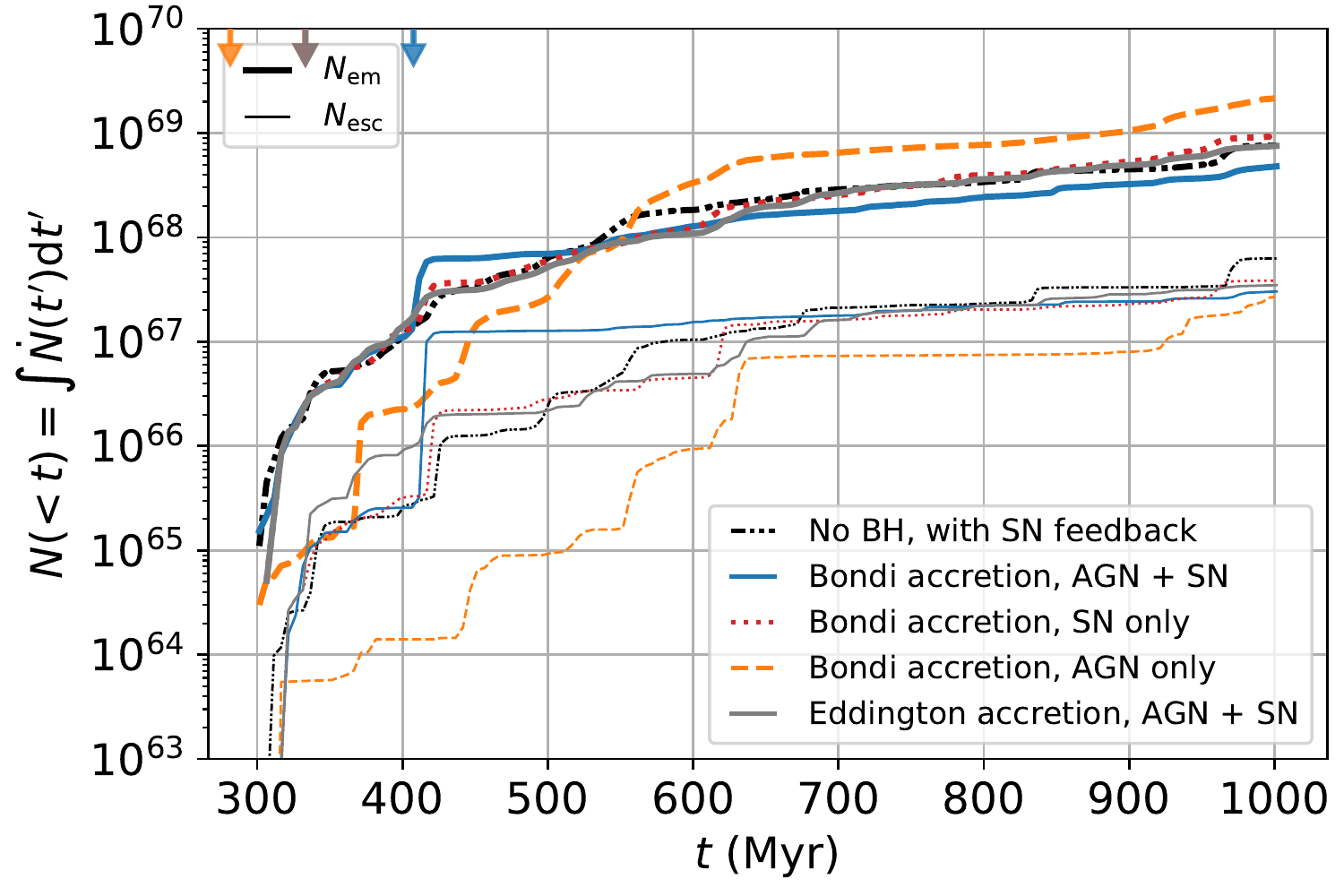} \\
  \includegraphics[width=\linewidth]{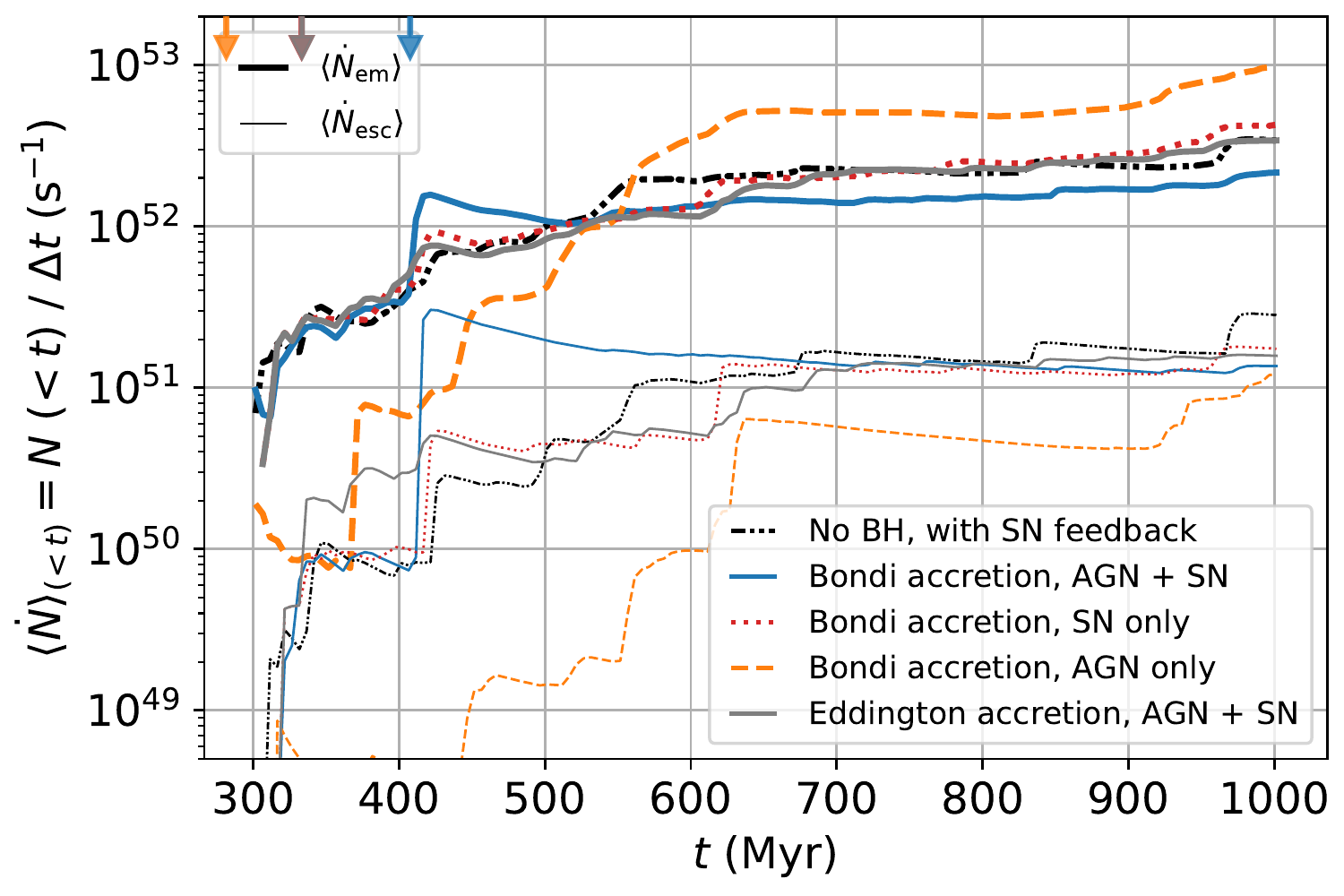}
  \caption{\emph{Top:} Cumulative number of ionizing photons that are produced (thick lines) and that escape (thin lines) the halo as a function of time. \emph{Bottom:} Averaged production rate (thick lines) and contribution rate (thin lines) of ionizing photons. This essentially reduces to the upper panel divided by the time since the formation of the galaxy. Both panels use the same legend as Fig.~\ref{fig:photons-fesc}.}
  \label{fig:photons-contrib}
\end{figure}

Because of the high variability of \fesc, it is difficult to quantitatively comment the differences between all the simulations including SN feedback, and isolate the effect of BH growth on the contribution of galaxies to reionization.
A more practical quantity for that purpose is the total number of \hi-ionizing photons produced by the galaxy and available to reionize the IGM. We compute the total number of photons produced as the integral of the photons production rate $\dot{N}_{\rm em}$ between the time of formation of a galaxy $t_{\rm form}$ and the current time $t$:
\begin{equation}
  \label{eq:Nemtot}
  N_{\rm em}\ (<t) = \int_{t_{\rm form}}^t \dot{N}_{\rm em}(t')\ {\rm d}t'.
\end{equation}
Similarly, the number of photons that escape the halo to contribute to reionization is given by
\begin{equation}
  \label{eq:Nesctot}
  N_{\rm esc}\ (<t) = \int_{t_{\rm form}}^t \dot{N}_{\rm esc}(t')\ {\rm d}t' = \int_{t_{\rm form}}^t \fesc(t')\dot{N}_{\rm em}(t' - t_{\rm cross})\ {\rm d}t',
\end{equation}
where $t_{\rm cross} = R_{\rm vir} / \tilde{c}$ is the time it takes for radiation travelling at the reduced speed of light $\tilde{c}$ to reach the virial radius of the halo.
The upper panel of Fig.~\ref{fig:photons-contrib} shows $N_{\rm em}$ and $N_{\rm esc}$ (respectively in thick and thin lines) for the various simulations presented in this work, with the same legend as in Fig.~\ref{fig:photons-fesc}. A valuable feature of Fig.~\ref{fig:photons-contrib} compared to Fig.~\ref{fig:photons-fesc} is that it shows clearly that, provided SN feedback is present, the feedback from the BH has hardly any effect on either the production or the escape of ionizing radiation, for instance $N_{\rm esc}$ seems to converge after $z\sim 9$ ($t \sim 550\ \mbox{Myr}$), when the stellar mass of the galaxy reaches $\Mstar \sim 10^{7}\ \Msun$. The simulation without SN feedback ends up producing more ionizing photons (a factor two more from $z \sim 9$), consistent with a higher stellar mass, but a lower fraction manages to escape the halo. Around $t \sim 410\ \mbox{Myr}$, the \agnt simulations exhibits a sudden increase both in $N_{\rm em}$ and $N_{\rm esc}$: this explains the high \fesc episode already discussed in Fig.~\ref{fig:photons-fesc}. The jump in $N_{\rm em}$ is caused by a massive star formation event, which in turn causes a very strong SN feedback episode.
In the \agntEdd run, even though the BH is actively growing and provides feedback continuously, the stellar radiation does not seem to escape more easily than in runs where BH growth is stalled. This suggests that SN feedback remains the dominant mechanism regulating \fesc in that galaxy.

In order to facilitate the comparison with reionization models, we show on the lower panel of Fig.~\ref{fig:photons-contrib} the averaged photon production rate (thick lines) and contribution rate (thin lines), defined as $\left\langle \dot{N}_{\rm em} \right\rangle = \frac{N_{\rm em}\ (<t)}{t-t_{\rm form}}$ and $\left\langle \dot{N}_{\rm esc} \right\rangle = \frac{N_{\rm esc}\ (<t)}{t-t_{\rm form}}$. On average, below $z\sim 9$ (when the halo is more massive than $\Mvir \gtrsim 10^9\ \Msun$ and has a stellar mass above $\Mstar \gtrsim 10^7\ \Msun$), the galaxy contributes $\left\langle \dot{N}_{\rm esc} \right\rangle \simeq 10^{51}\ \mbox{s}^{-1}$ when supernovae are included, very consistent with the findings of e.g. \citet{Kimm2014} for haloes of similar mass.

\begin{figure}
  \centering
  \includegraphics[width=\linewidth]{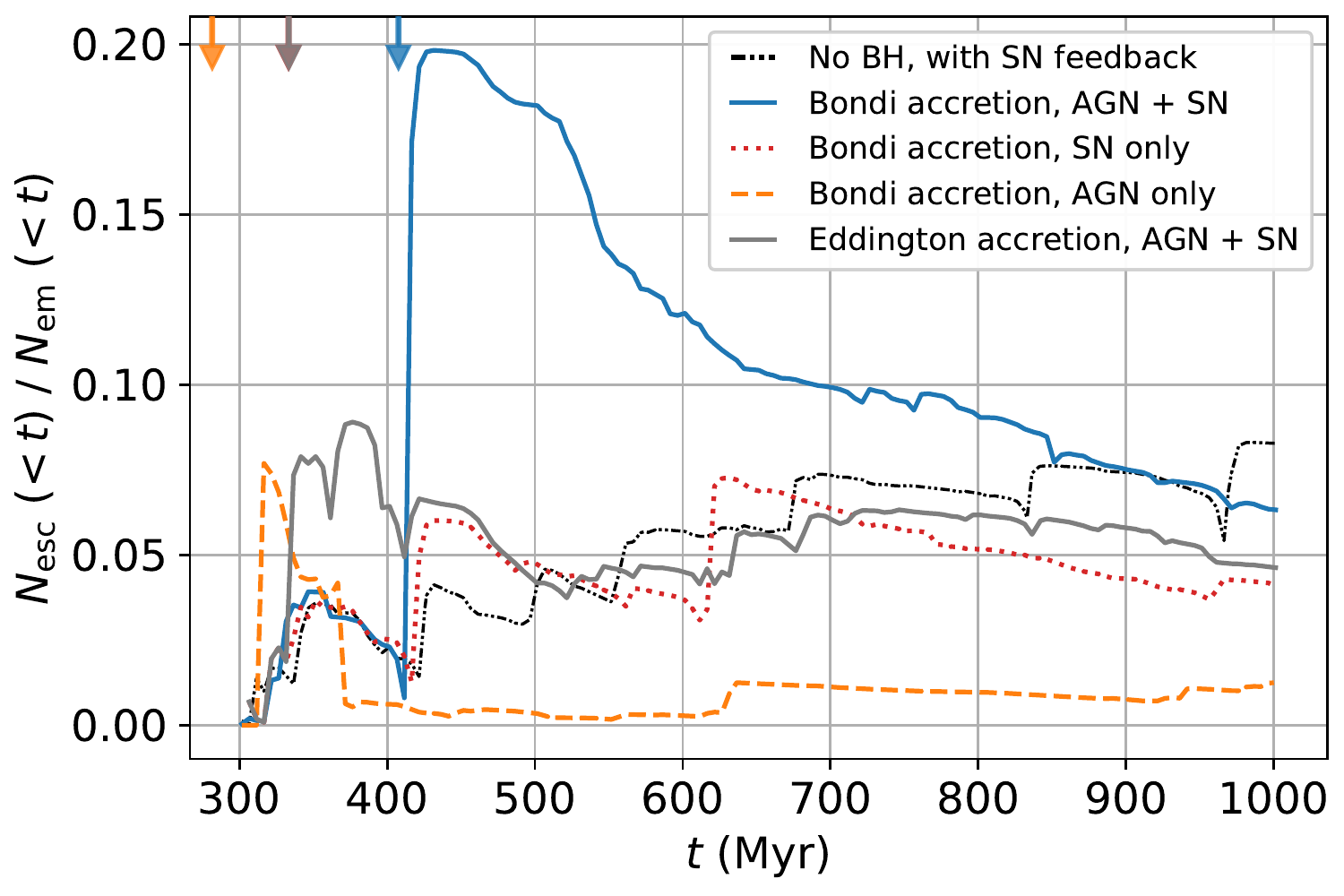}
  \caption{Escape fraction averaged over the lifetime of the galaxy. Legend is the same as Fig.~\ref{fig:photons-fesc}. Without supernovae, \fesc plateaus around $1-2\%$, except for the initial spike.}
  \label{fig:photons-fesc-cumul}
\end{figure}
While these two previous quantities are ultimately the most relevant ones to assess the contribution of an individual galaxy to the reionization of its local environment, the average escape fraction $\langle\fesc\rangle$ is a more common way to estimate the importance of a given class of galaxies to the reionization of the universe.
We present in Fig.~\ref{fig:photons-fesc-cumul} the ratio of $N_{\rm esc}$ and $N_{\rm em}$, which gives an effective escape fraction averaged over the history of the galaxy for the different simulations, using the same colour-coding as in Fig.~\ref{fig:photons-fesc}. While this has no statistical significance since it is still tied to the past history of one single galaxy, it still gives a reasonable average once the galaxy becomes massive enough such that one single burst of star formation does not change significantly the total stellar mass.

\subsection{Escape of radiation from the AGN}
\label{sec:fesc:AGN-fesc}

Even though we did not explicitly track the radiation from the AGN, it is nonetheless interesting to try to estimate its contribution to the total released ionizing radiation. We can compute this quantity in a very similar fashion $N_{\rm esc}$ discussed in the previous section by estimating an \emph{AGN escape fraction} \fescAGN and the luminosity in the \hi-ionizing band produced by the accretion onto the BH.
This luminosity $L_{\rm UV1}$ is directly proportional to the AGN bolometric luminosity
\begin{equation}
  \label{eq:LUV1}
  L_{\rm UV1} = f_{\rm UV1} L_{\rm bol} = f_{\rm UV1} \epsilon_r \dot{M}_{\bullet} c^2,
\end{equation}
where the proportionality factor $f_{\rm UV1}$ is given by the AGN spectrum integrated between $13.6\ \mbox{eV}$ and $24.59\ \mbox{eV}$, the bounds of the \hi-ionizing radiation used with \ramsesrt. We use the composite AGN spectrum of \citet{Sazonov2004} for an unobscured AGN, and we find $f_{\rm UV1} \simeq 7.9\%$ \citep[see also][]{Bieri2017}. The number of photons produced in that band is then simply given by
\begin{equation}
  \label{eq:NUV1-AGN}
  \dot{N}_{\rm em}^{\rm AGN} = \frac{L_{\rm UV1}}{\langle \epsilon\rangle},
\end{equation}
where $\langle \epsilon\rangle \simeq 18\ \mbox{eV}$ is the luminosity-weighted average energy in the band. From this, we can finally compute the total number of escaped photons per unit time:
\begin{equation}
  \label{eq:NUV1esc-AGN}
  \dot{N}_{\rm esc}^{\rm AGN} = \fescAGN \dot{N}_{\rm em}^{\rm AGN}.
\end{equation}
We then use a ray-tracing technique using the \textsc{Rascas} code (Michel-Dansac et al, 2018, in preparation) to measure the column density seen by the central BH in order to estimate the escape fraction \fescAGN: we cast $N = 100\, 000$ rays from the position of the BH and integrate the \hi column density in each direction $N_{\hi, i}$ up to the virial radius, by assuming that the BH is always at the centre of the halo. We then estimate the average \fescAGN as
\begin{equation}
  \label{eq:fescAGN}
  \fescAGN = \frac{1}{N} \sum_{i=1}^N \mathrm{e}^{-\tau_i} = \frac{1}{N} \sum_{i=1}^N \mathrm{e}^{-\sigma_{\rm cross} N_{\hi, i}},
\end{equation}
where $\sigma_{\rm cross}$ is the average cross-section within that energy band.

\begin{figure}
  \centering
  \includegraphics[width=\linewidth]{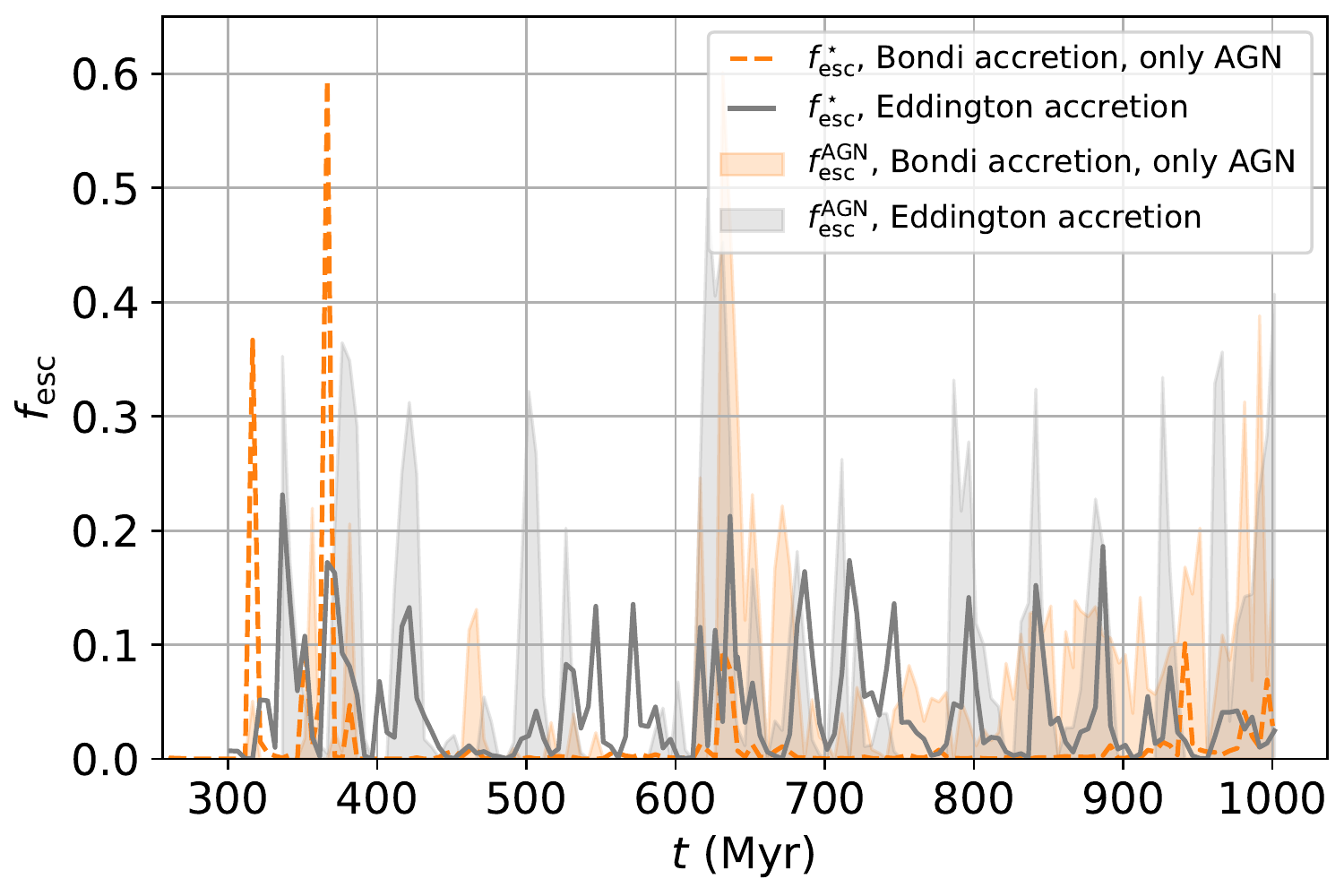} \\
  \includegraphics[width=\linewidth]{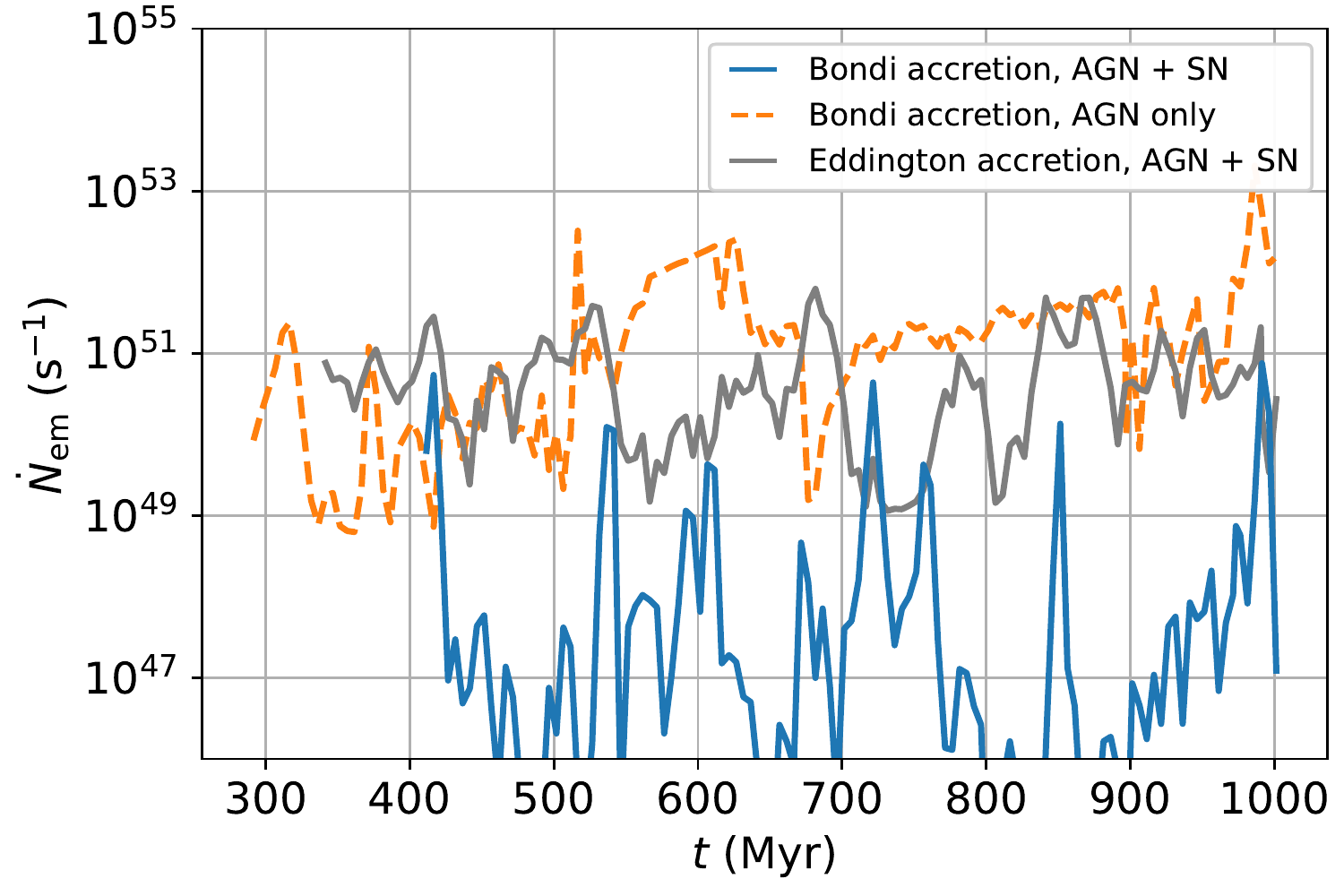} \\
  \includegraphics[width=\linewidth]{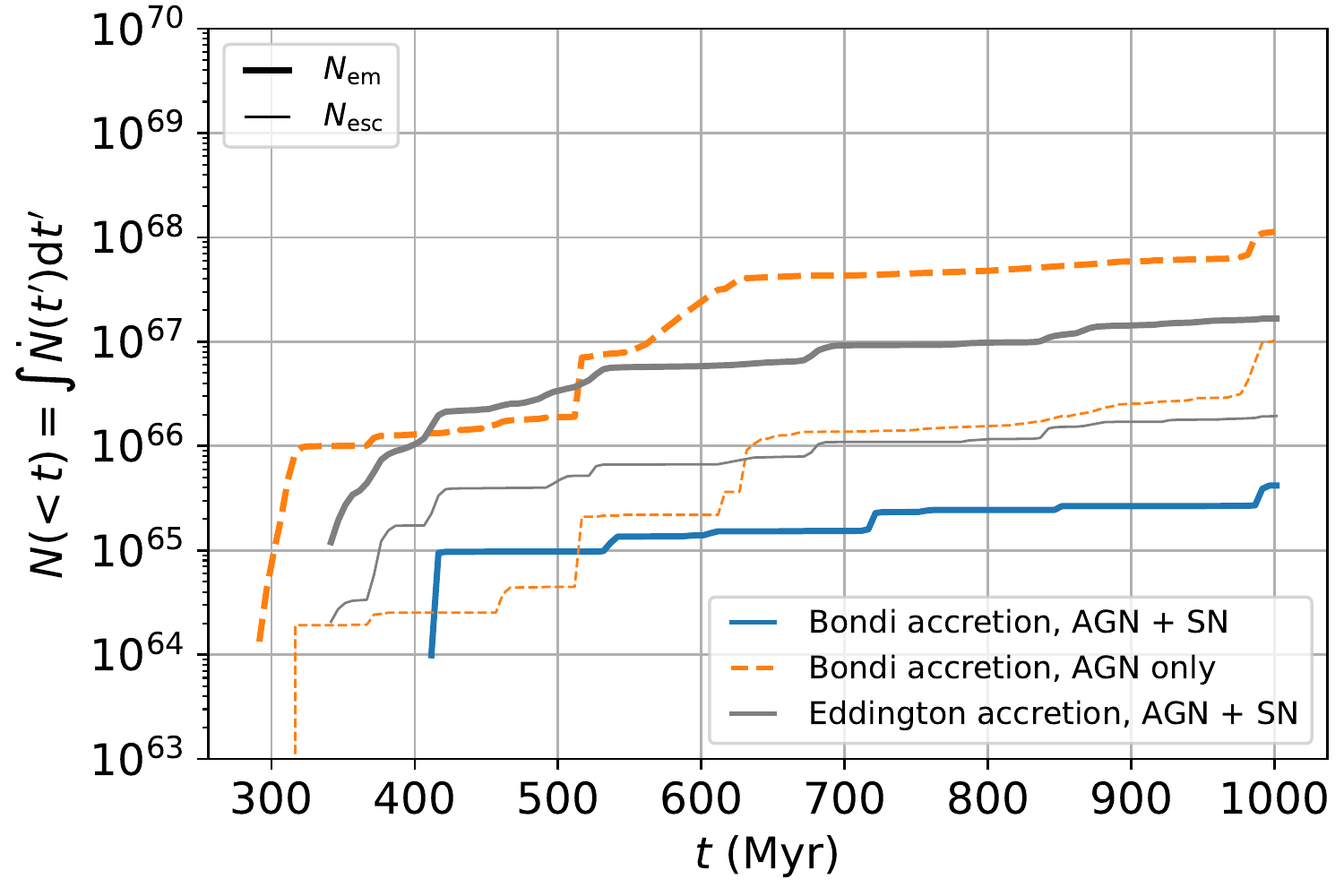}
  \caption{\emph{Top:} Instantaneous escape faction for the radiation produced by the stars (lines) and from the AGN (shaded area) for the \agntnoSN and \agntEdd simulations. For the latter, there is a good correlation between \fesc and \fescAGN, because the gas distribution in the ISM is dominated by SN feedback. \emph{Middle:} Ionizing emissivity of the AGN in the \agnt, \agntnoSN and \agntEdd simulations, which follows the accretion rate onto the BH. \emph{Bottom:} Cumulative number of ionizing photons that are produced (thick lines) and that escape the halo (thin lines) as a function of time. All panels use the same legend as Fig.~\ref{fig:photons-fesc}.}
  \label{fig:rascas}
\end{figure}

Since the BH is only growing in the \agntnoSN and \agntEdd simulations, we only compute \fescAGN for these two simulations. We present the results on the top panel of Fig.~\ref{fig:rascas}, where we compare the escape fraction for stellar radiation (noted $f_{\rm esc}^\star$, in solid and dashed lines) to the escape fraction from the AGN (\fescAGN, shaded areas). We use the same colours as Fig.~\ref{fig:photons-fesc} for the different simulations. Interestingly, the two simulations present a very different behaviour.

For the simulation with SN feedback (\agntEdd), there is a relatively good correlation between the stellar and AGN escape fraction:  most peaks of stellar \fesc correspond to a peak of similar duration of \fescAGN. This can easily be understood in light of the results of Sect.~\ref{sec:coevolution:BH-growth}. We have seen that the accretion rate is limited by the amount of available gas in the vicinity of the BH. Whenever the BH is in a clump and accretes efficiently, it will be in competition with star formation to access the local gas reservoir: after a few Myr, when the cloud starts to be disrupted by SN explosions, the column density around the BH drops and \fescAGN increases.
The tight correlation between \fesc and \fescAGN confirms that the AGN has little impact on the galaxy. The ISM is already so stirred by SN feedback that even in the extreme case of Eddington accretion, AGN feedback cannot disturb it much more \citep[see also][]{Prieto2017}. In this picture, we can explain the relative height of the \fescAGN peaks compared to the peaks in stellar \fesc: while the BH is either inside or outside of a clump, the stellar distribution in the galaxy is more complex. \citet{Trebitsch2017} showed that the stellar escape fraction is almost completely determined by the local properties around a stellar cluster, meaning that either radiation escape, or it doesn't at all. If the average \fesc over the galaxy is lower than \fescAGN, this indicates that { some stars are still embedded in dense, neutral clouds}. Further absorption of ionizing photons in the circum-galactic medium accounts for the fact that the escape fraction (\fesc or \fescAGN) do not reach $100\%$.
For the simulation without SN feedback (\agntnoSN), the situation is very different: while the stellar \fesc is almost always close to zero, \fescAGN is non-zero most of the times, especially after $600\ \mbox{Myr}$. This is again consistent with our finding: while stars are trapped in dense clumps with no feedback to create low-density channels, AGN feedback can heat and push the gas around the BH, for example at $t = 620\ \mbox{Myr}$ on the right panel of Fig.~\ref{fig:BH-clump}. This creates easy escape routes for the ionizing radiation, and results in $\fescAGN \sim 60\%$. At the same time, some radiation produced by young stars in the clump will also find low-density channels, carved by AGN feedback, explaining the value of $\fesc \sim 10\%$ at $t = 620\ \mbox{Myr}$.

We show in the central panel of Fig.~\ref{fig:rascas} the instantaneous photon production rate, $\dot{N}_{\rm em}^{\rm AGN}$, resulting from the accretion onto the BH in the \agnt, \agntnoSN and \agntEdd simulations with the same lines as before. Since $\dot{N}_{\rm em}^{\rm AGN} \propto \dot{M}_\bullet$, this scales directly with the BH accretion rate. This immediately confirms that when SN feedback is present and BH accretion is not artificially increased, the AGN contributes very little to the ionizing budget of the galaxy. As expected from the BH growth discussed in Sect.~\ref{sec:coevolution:BH-growth}, turning off SN feedback significantly increases the AGN ionizing luminosity, sometimes by more than three orders of magnitude. The high accretion episode that takes place around $600\ \mbox{Myr}$ appear very clearly, with $\dot{N}_{\rm em}^{\rm AGN}$ increasing suddenly by more than one order of magnitude to reach $\dot{N}_{\rm em}^{\rm AGN} \sim 10^{51}\ \mbox{s}^{-1}$ for the whole duration of the episode. By comparison with \fescAGN just above, we see that the peak of \fescAGN corresponds to the very end of the high accretion episode, suggesting again that the AGN struggles to pierce a hole in the dense clump in which the BH is embedded.
The \agntEdd simulation present a relatively similar behaviour, with a high ionizing luminosity all throughout the simulation. The correlation between the ionizing production rate and \fescAGN is however much less clear, further supporting the idea the turbulence in the ISM caused by SN feedback is what regulates the accretion onto the BH.

The lower panel of Fig.~\ref{fig:rascas} presents the integrated number of ionizing photons produced (thick lines) and that escape the halo (thin lines), similar to Fig.~\ref{fig:photons-contrib}. Again, the \agnt simulation is shown as a thick blue solid line, the \agntnoSN simulation as an orange dashed line and the \agntEdd simulation as a solid grey line. Without SN feedback, the raw \emph{production} of ionizing radiation by the AGN is more than three orders of magnitude below the total amount of ionizing radiation produced from stars, and still roughly two orders of magnitude below what is \emph{escaping} the halo.
When the BH grows, the situation is very different. The total ionizing radiation coming from the AGN in the \agntnoSN simulation after $1$ Gyr is only one order of magnitude lower than what is produced by stars in the same simulation, and the amount of radiation escaping is lower by only a factor of a few.
For the \agntEdd simulation, the story is similar: the AGN produces less ionizing radiation than the stars, and approximately $10\%$ of the total production actually reaches the IGM.
This suggests that even in a ``maximal case'' where our SMBH is allowed to grow steadily, its total contribution to the ionizing budget compared to its host galaxy will be relatively small (less than $10\%$ in our simulation).

\section{Discussion}
\label{sec:discussion}

\subsection{BH growth in low mass galaxies}
\label{sec:discussion:growth}
From the simulations described previously, we can confirm the picture in which SN feedback is a crucial ingredient in regulating the growth of BH in low mass galaxies. This is in line with the findings of earlier studies of \citet{Dubois2015} and \citet{Habouzit2017} who find that efficient supernova feedback gives rise to a ``supernova regulated'' phase, during which BH does not grow much and which lasts until the host galaxy is massive enough ($\Mstar \sim 10^9\ \Msun$) to feed steadily the central BH \citep[see also][who drew similar conclusions]{AnglesAlcazar2017, Bower2017, McAlpine2017, Prieto2017} and initiate a second phase, where the BH self-regulates its growth. Admittedly, we do not use any ``boost'' to account for unresolved clumping that could increase the accretion rate, as suggested e.g. \citet{Booth2009}. However, boosting the accretion rate would increase AGN feedback, and this would act to further decrease the gas density near the BH. The \agntEdd simulation is arguably the most optimistic in terms of BH growth, as the BH, even when forced to endeavour to accrete at the Eddington limit, does not reach this limit except sporadically. Even in our extreme case where BH growth is only limited by the amount of gas available (and capped at the Eddington rate), it still seems that the gas distribution in the centre of the galaxy is essentially determined by SN feedback. Clearly, the dwarf galaxies that we have been simulating are well below the BH self-regulation regime, with their stellar mass barely reaching $\Mstar \lesssim 10^8\ \Msun$ at $z \sim 6$. The situation is expected to change at later times or in more massive galaxies. \cite{Prieto2017} find that SNe create a rarefied environment from where hot gas can escape, but they cannot accelerate cold gas to more than $\sim 200\ \mbox{km\,s}^{-1}$. Once the BH has grown, however, AGN feedback affects the cold gas directly, and can accelerate it up to $\sim 600-800\ \mbox{km\,s}^{-1}$.

Assuming that galaxies such as the one presented in this work undergo a reasonably standard evolution, they would be good candidate to be progenitors of the objects observed by \citet{Reines2013}. Indeed, with a stellar mass around $\Mstar \simeq 10^9\ \Msun$, these galaxies reside in haloes of typically $\Mvir(z=0) \simeq 1-2 \times 10^{11}\ \Msun$ \citep{Behroozi2013}, for which we can expect $z = 6$ progenitors to reside in $\Mvir \simeq 3-6\times 10^{9}\ \Msun$ haloes \citep{Correa2015}. As these galaxies live a relatively quiet life, with on average one major merger between $z = 6$ and $z = 0$ \citep{Fakhouri2010}, it seems plausible to assume that BH in these galaxy did not grow very much since their formation. This is especially the case at the very low mass end, for instance for RGG 118 \citep{Baldassare2015} and its BH with $M_\bullet \simeq 5 \times 10^4\ \Msun$, for which radiative feedback from reionization could limit the amount of cool gas available in the galaxy for star formation \citep[e.g.][]{Simpson2013} and, incidentally, for BH growth. This support the idea that the observed population of BHs in dwarfs are good tracers of the BH seeding mechanism at high redshift \citep{Volonteri2008,Volonteri2010,vanWassenhove2010}.

\subsection{Weak AGN feedback in low mass galaxies}
\label{sec:discussion:feedback}
From the distribution of Eddington ratios presented in Fig.~\ref{fig:BH-accrate}, it follows that in all the runs discussed here, the BH spends most of its time accreting at very low $\lambda_{\rm Edd}$. As a result, most of the AGN feedback is in form of ``radio mode'' feedback, for which we have assumed a feedback efficiency $\epsilon_f = 100\%$. Despite this higher efficiency, we have so far found that AGN feedback has only a negligible effect on the ISM of the low mass galaxies we are studying. This is mostly due to the fact that the BH in these galaxies is just too small to have any notable effect. Indeed, for $\lambda_{\rm Edd} \sim 10^{-4}$, the AGN luminosity in ``radio mode'' is of order $\dot{E}_{\rm AGN} \simeq 3.3 \times 10^{4}\ L_\odot$. By comparison, the lower panel of Fig.~\ref{fig:photons-contrib} shows that the typical ionizing photon production rate for our galaxy is of order $\dot{N}_{\rm em} \simeq 10^{52}\ \mbox{s}^{-1}$. Assuming that $\sim 10\%$ of these photons are absorbed, the amount of energy transferred to the gas via photoionization from stellar origin is still of order $6.8 \times 10^{7}\ L_\odot$.
\begin{figure}
  \centering
  \includegraphics[width=\linewidth]{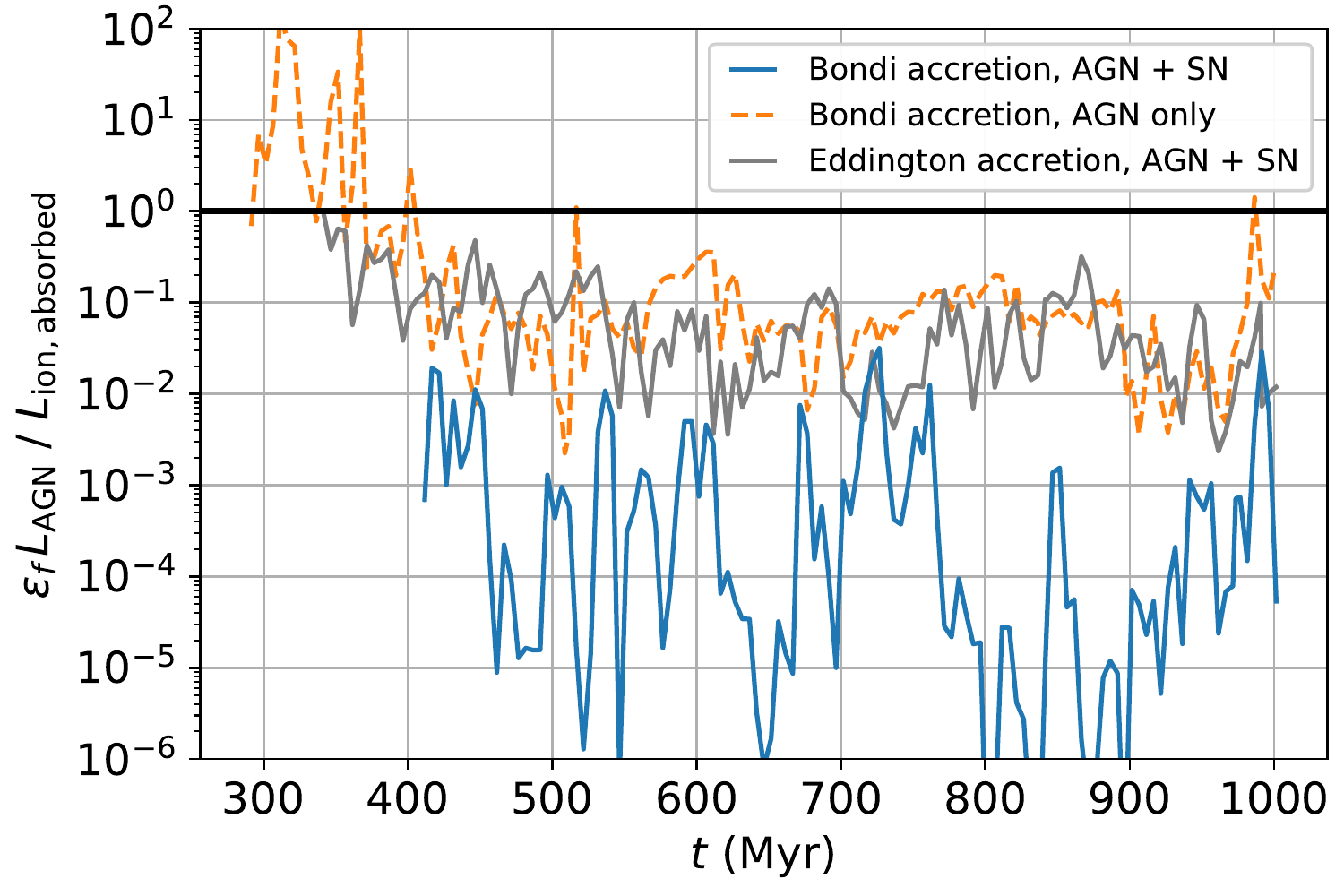}
  \caption{Ratio of the AGN feedback power to the ionizing luminosity emitted by stars and absorbed by the gas for the three simulations with AGN feedback.}
  \label{fig:discussion:feedback}
\end{figure}
More generally, we show in Fig.~\ref{fig:discussion:feedback} the ratio of AGN feedback power $\dot{E}_{\rm AGN}$ to the ionizing luminosity emitted by stars and absorbed by the gas. As stated in Sect.~\ref{sec:sims:AGN}, we assume a coupling efficiency of $15\%$ for the quasar mode and $100\%$ for the radio mode. The horizontal black line indicate equal energy injection from these two channels. Most of the time (after $t \gtrsim 400\ \mbox{Myr}$), AGN feedback transfers much less energy to the gas compared to photoionization. This is especially true for the \agnt run, where the accretion rate is very low.

\subsection{Contribution of faint AGN to the reionization}
\label{sec:discussion:eor}

In the analysis of Sect.~\ref{sec:fesc:AGN-fesc}, we assumed that the radiation emitted by the AGN followed an energy distribution based on the average non-obscured quasar spectrum of \citet{Sazonov2004}. For an accretion disc following the \citet{Shakura1973} disc model, we expect the peak of the SED to shift towards higher frequency when the Eddington ratio $\lambda_{\rm Edd}$ increases and when the BH mass decreases. When computing the ionizing luminosity of the AGN in eq.~\ref{eq:LUV1} and more specifically the $f_{\rm UV1}$, we should use a more physically motivated model for the spectrum, for instance the \textsc{oxaf} model of \citet{Thomas2016} based on \citet{Done2012}. This is however at most a sub-dominant effect: even if we assume that the ionizing fraction of the bolometric luminosity is ten times larger ($f_{\rm UV1} \simeq 80\%$), the total production of ionizing photons by the AGN would still be two orders of magnitudes below the amount of ionizing photons produced by the stars. The inclusion of secondary ionizations from the hard AGN spectrum does not change this result either, as at most 35-37\% of their energy goes into secondary reionizations \citep[with only a weak dependence on the photon energy, and the fraction dropping as the ionized hydrogen fraction increases,][]{Shull1985,Venkatesan2001}. This is because the dominant factor here is not so much the AGN spectrum, but the fact that in our most physical simulation (\agnt), the accretion rate is low at all times, yielding a low bolometric luminosity. 

The direct consequence of this is that BHs in the low-mass regime are unlikely to contribute much to the ionizing budget of the EoR. It is of course possible that we are underestimating the accretion onto the BH we are simulating here, for instance because our $\lesssim 10\ \mbox{pc}$ resolution can only resolve structures down to that scale, and we do not correct for any unresolved subgrid clumping of the gas. For instance, \citet{Negri2017} discuss how the Bondi prescription for accretion can lead to both over- and under-estimation of the real accretion rate. However, even in our \agntEdd simulation (which corresponds to an extreme model where the both the galaxy and the BH grow reasonably), the ionizing contribution from the AGN is negligible compared to the ionizing starlight. Going beyond this model and increasing even more the accretion rate would result in an over-massive BH compared to its host galaxy.

\section{Summary conclusions}
\label{sec:ccl}

Motivated by the recent suggestions that faint AGN could play an important role in the reionization of the universe, we performed a series of cosmological zooms on a dwarf galaxy ($\Mvir \simeq 5 \times 10^9\ \Msun$ at $z \simeq 6$) using a high resolution of $\sim 10\ \mbox{pc}$, fully coupled radiative hydrodynamics simulation, and following the growth of massive black holes and the associated AGN feedback.
In our reference run, we use a dual mode AGN feedback describing both the radio and quasar modes, and we account for SN feedback. We compare this reference run to simulations where we alternately turned off SN and AGN feedback and we also tried an alternative prescription where the accretion rate follows the Eddington rate whenever possible.
The main goal of our study was to determine how and how much BH can affect the ISM of low mass galaxies, and more specifically how this affects their contribution to the ionizing budget of the EoR.
We summarize our main results below:
\begin{enumerate}
\item SN feedback strongly hampers BH growth in low mass galaxies. The resulting constant stirring of the ISM removes gas from around the BH, resulting in a very low accretion rate. Even when the BH is forced to accrete at the Eddington limit, the actual accretion is limited by the amount of gas locally available, and the BH grows at sub-Eddington rates.
\item BH growth has no effect on the escape of ionizing radiation. In the run without SN feedback, the integrated \fesc is much lower than in all other runs. There is no sign of amplification of the integrated \fesc by AGN feedback, even in runs where the BH grows.
\item The instantaneous \fesc is virtually unaffected by the presence of a BH or by its growth: AGN feedback does not provide an extra source of energy injection in the ISM, and therefore does not help ionizing radiation to escape before the destruction of the star forming clouds by SN feedback.
\item When the BH grows in presence of SN feedback (\agntEdd simulation), the ionizing radiation produced by the accretion onto the BH escapes synchronously with the stellar ionizing radiation.
\item Only a fraction \fescAGN of the ionizing luminosity produced by the gas accretion onto the BH manages to escape in the IGM to contribute to reionization, with $\fescAGN < 50\%$ and on average $\left\langle \fescAGN \right\rangle \simeq 10\%$.
\end{enumerate}

\section*{Acknowledgements}

We thank J{\'e}r{\'e}my Blaizot and L{\'e}o Michel-Dansac for providing us with the \textsc{Rascas} machinery. MT and MV acknowledge funding from the European Research Council under the European Community's Seventh Framework Programme (FP7/2007-2013 Grant Agreement no. 614199, project `BLACK').
This work has made use of the Horizon Cluster hosted by Institut d'Astrophysique de Paris; we thank Stephane Rouberol for running smoothly this cluster for us.
This work was granted access to the HPC resources of CINES under the allocation A0020406955 made by GENCI.
This work has made extensive use of the \textsc{PyMSES}\footnote{\label{fn:pymses}\url{https://bitbucket.org/dchapon/pymses/}} analysis package \citep{Guillet2013}, as well as the \textsc{Matplotlib} \citep{Hunter2007}, \textsc{Numpy/Scipy} \citep{Jones2001} and \textsc{IPython} \citep{Perez2007} packages.




\bibliographystyle{mnras}
\bibliography{fesc_agnglx} 

\begin{thebibliography}{}
\makeatletter
\relax
\def\mn@urlcharsother{\let\do\@makeother \do\$\do\&\do\#\do\^\do\_\do\%\do\~}
\def\mn@doi{\begingroup\mn@urlcharsother \@ifnextchar [ {\mn@doi@}
  {\mn@doi@[]}}
\def\mn@doi@[#1]#2{\def\@tempa{#1}\ifx\@tempa\@empty \href
  {http://dx.doi.org/#2} {doi:#2}\else \href {http://dx.doi.org/#2} {#1}\fi
  \endgroup}
\def\mn@eprint#1#2{\mn@eprint@#1:#2::\@nil}
\def\mn@eprint@arXiv#1{\href {http://arxiv.org/abs/#1} {{\tt arXiv:#1}}}
\def\mn@eprint@dblp#1{\href {http://dblp.uni-trier.de/rec/bibtex/#1.xml}
  {dblp:#1}}
\def\mn@eprint@#1:#2:#3:#4\@nil{\def\@tempa {#1}\def\@tempb {#2}\def\@tempc
  {#3}\ifx \@tempc \@empty \let \@tempc \@tempb \let \@tempb \@tempa \fi \ifx
  \@tempb \@empty \def\@tempb {arXiv}\fi \@ifundefined
  {mn@eprint@\@tempb}{\@tempb:\@tempc}{\expandafter \expandafter \csname
  mn@eprint@\@tempb\endcsname \expandafter{\@tempc}}}

\bibitem[\protect\citeauthoryear{{Angl{\'e}s-Alc{\'a}zar},
  {Faucher-Gigu{\`e}re}, {Quataert}, {Hopkins}, {Feldmann}, {Torrey}, {Wetzel}
  \& {Kere{\v s}}}{{Angl{\'e}s-Alc{\'a}zar} et~al.}{2017}]{AnglesAlcazar2017}
{Angl{\'e}s-Alc{\'a}zar} D.,  {Faucher-Gigu{\`e}re} C.-A.,  {Quataert} E.,
  {Hopkins} P.~F.,  {Feldmann} R.,  {Torrey} P.,  {Wetzel} A.,   {Kere{\v s}}
  D.,  2017, preprint, \href
  {http://adsabs.harvard.edu/abs/2017arXiv170703832A} {} (\mn@eprint {arXiv}
  {1707.03832})

\bibitem[\protect\citeauthoryear{{Atek} et~al.,}{{Atek}
  et~al.}{2015}]{Atek2015}
{Atek} H.,  et~al., 2015, \mn@doi [\apj] {10.1088/0004-637X/814/1/69}, \href
  {http://adsabs.harvard.edu/abs/2015ApJ...814...69A} {814, 69}

\bibitem[\protect\citeauthoryear{{Aubert}, {Pichon}  \& {Colombi}}{{Aubert}
  et~al.}{2004}]{Aubert2004}
{Aubert} D.,  {Pichon} C.,   {Colombi} S.,  2004, \mn@doi [\mnras]
  {10.1111/j.1365-2966.2004.07883.x}, \href
  {http://adsabs.harvard.edu/abs/2004MNRAS.352..376A} {352, 376}

\bibitem[\protect\citeauthoryear{{Baldassare}, {Reines}, {Gallo}  \&
  {Greene}}{{Baldassare} et~al.}{2015}]{Baldassare2015}
{Baldassare} V.~F.,  {Reines} A.~E.,  {Gallo} E.,   {Greene} J.~E.,  2015,
  \mn@doi [\apjl] {10.1088/2041-8205/809/1/L14}, \href
  {http://adsabs.harvard.edu/abs/2015ApJ...809L..14B} {809, L14}

\bibitem[\protect\citeauthoryear{{Becker} et~al.,}{{Becker}
  et~al.}{2001}]{Becker2001}
{Becker} R.~H.,  et~al., 2001, \mn@doi [\aj] {10.1086/324231}, \href
  {http://adsabs.harvard.edu/abs/2001AJ....122.2850B} {122, 2850}

\bibitem[\protect\citeauthoryear{{Becker}, {Bolton}, {Madau}, {Pettini},
  {Ryan-Weber}  \& {Venemans}}{{Becker} et~al.}{2015}]{Becker2015}
{Becker} G.~D.,  {Bolton} J.~S.,  {Madau} P.,  {Pettini} M.,  {Ryan-Weber}
  E.~V.,   {Venemans} B.~P.,  2015, \mn@doi [\mnras] {10.1093/mnras/stu2646},
  \href {http://adsabs.harvard.edu/abs/2015MNRAS.447.3402B} {447, 3402}

\bibitem[\protect\citeauthoryear{{Behroozi}, {Wechsler}  \&
  {Conroy}}{{Behroozi} et~al.}{2013}]{Behroozi2013}
{Behroozi} P.~S.,  {Wechsler} R.~H.,   {Conroy} C.,  2013, \mn@doi [\apj]
  {10.1088/0004-637X/770/1/57}, \href
  {http://adsabs.harvard.edu/abs/2013ApJ...770...57B} {770, 57}

\bibitem[\protect\citeauthoryear{{Bergvall}, {Leitet}, {Zackrisson}  \&
  {Marquart}}{{Bergvall} et~al.}{2013}]{Bergvall2013}
{Bergvall} N.,  {Leitet} E.,  {Zackrisson} E.,   {Marquart} T.,  2013, \mn@doi
  [\aap] {10.1051/0004-6361/201118433}, \href
  {http://adsabs.harvard.edu/abs/2013A%26A...554A..38B} {554, A38}

\bibitem[\protect\citeauthoryear{{Bieri}, {Dubois}, {Rosdahl}, {Wagner}, {Silk}
   \& {Mamon}}{{Bieri} et~al.}{2017}]{Bieri2017}
{Bieri} R.,  {Dubois} Y.,  {Rosdahl} J.,  {Wagner} A.,  {Silk} J.,   {Mamon}
  G.~A.,  2017, \mn@doi [\mnras] {10.1093/mnras/stw2380}, \href
  {http://adsabs.harvard.edu/abs/2017MNRAS.464.1854B} {464, 1854}

\bibitem[\protect\citeauthoryear{{Bondi}}{{Bondi}}{1952}]{Bondi1952}
{Bondi} H.,  1952, \mn@doi [\mnras] {10.1093/mnras/112.2.195}, \href
  {http://adsabs.harvard.edu/abs/1952MNRAS.112..195B} {112, 195}

\bibitem[\protect\citeauthoryear{{Booth} \& {Schaye}}{{Booth} \&
  {Schaye}}{2009}]{Booth2009}
{Booth} C.~M.,  {Schaye} J.,  2009, \mn@doi [\mnras]
  {10.1111/j.1365-2966.2009.15043.x}, \href
  {http://adsabs.harvard.edu/abs/2009MNRAS.398...53B} {398, 53}

\bibitem[\protect\citeauthoryear{{Bower}, {Schaye}, {Frenk}, {Theuns},
  {Schaller}, {Crain}  \& {McAlpine}}{{Bower} et~al.}{2017}]{Bower2017}
{Bower} R.~G.,  {Schaye} J.,  {Frenk} C.~S.,  {Theuns} T.,  {Schaller} M.,
  {Crain} R.~A.,   {McAlpine} S.,  2017, \mn@doi [\mnras]
  {10.1093/mnras/stw2735}, \href
  {http://adsabs.harvard.edu/abs/2017MNRAS.465...32B} {465, 32}

\bibitem[\protect\citeauthoryear{{Bridge} et~al.,}{{Bridge}
  et~al.}{2010}]{Bridge2010}
{Bridge} C.~R.,  et~al., 2010, \mn@doi [\apj] {10.1088/0004-637X/720/1/465},
  \href {http://adsabs.harvard.edu/abs/2010ApJ...720..465B} {720, 465}

\bibitem[\protect\citeauthoryear{{Bruzual} \& {Charlot}}{{Bruzual} \&
  {Charlot}}{2003}]{Bruzual2003}
{Bruzual} G.,  {Charlot} S.,  2003, \mn@doi [\mnras]
  {10.1046/j.1365-8711.2003.06897.x}, \href
  {http://adsabs.harvard.edu/abs/2003MNRAS.344.1000B} {344, 1000}

\bibitem[\protect\citeauthoryear{{Chabrier}}{{Chabrier}}{2003}]{Chabrier2003}
{Chabrier} G.,  2003, \mn@doi [\pasp] {10.1086/376392}, \href
  {http://adsabs.harvard.edu/abs/2003PASP..115..763C} {115, 763}

\bibitem[\protect\citeauthoryear{{Chapon}, {Mayer}  \& {Teyssier}}{{Chapon}
  et~al.}{2013}]{Chapon2013}
{Chapon} D.,  {Mayer} L.,   {Teyssier} R.,  2013, \mn@doi [\mnras]
  {10.1093/mnras/sts568}, \href
  {http://adsabs.harvard.edu/abs/2013MNRAS.429.3114C} {429, 3114}

\bibitem[\protect\citeauthoryear{{Chen} et~al.,}{{Chen}
  et~al.}{2017}]{Chen2017}
{Chen} C.-T.~J.,  et~al., 2017, \mn@doi [\apj] {10.3847/1538-4357/aa5d5b},
  \href {http://adsabs.harvard.edu/abs/2017ApJ...837...48C} {837, 48}

\bibitem[\protect\citeauthoryear{{Correa}, {Wyithe}, {Schaye}  \&
  {Duffy}}{{Correa} et~al.}{2015}]{Correa2015}
{Correa} C.~A.,  {Wyithe} J.~S.~B.,  {Schaye} J.,   {Duffy} A.~R.,  2015,
  \mn@doi [\mnras] {10.1093/mnras/stv689}, \href
  {http://adsabs.harvard.edu/abs/2015MNRAS.450.1514C} {450, 1514}

\bibitem[\protect\citeauthoryear{{Cristiani}, {Serrano}, {Fontanot}, {Vanzella}
   \& {Monaco}}{{Cristiani} et~al.}{2016}]{Cristiani2016}
{Cristiani} S.,  {Serrano} L.~M.,  {Fontanot} F.,  {Vanzella} E.,   {Monaco}
  P.,  2016, \mn@doi [\mnras] {10.1093/mnras/stw1810}, \href
  {http://adsabs.harvard.edu/abs/2016MNRAS.462.2478C} {462, 2478}

\bibitem[\protect\citeauthoryear{{Deharveng}, {Buat}, {Le Brun}, {Milliard},
  {Kunth}, {Shull}  \& {Gry}}{{Deharveng} et~al.}{2001}]{Deharveng2001}
{Deharveng} J.-M.,  {Buat} V.,  {Le Brun} V.,  {Milliard} B.,  {Kunth} D.,
  {Shull} J.~M.,   {Gry} C.,  2001, \mn@doi [\aap]
  {10.1051/0004-6361:20010920}, \href
  {http://adsabs.harvard.edu/abs/2001A%26A...375..805D} {375, 805}

\bibitem[\protect\citeauthoryear{{Dom{\'{\i}}nguez}, {Siana}, {Brooks},
  {Christensen}, {Bruzual}, {Stark}  \& {Alavi}}{{Dom{\'{\i}}nguez}
  et~al.}{2015}]{Dominguez2015}
{Dom{\'{\i}}nguez} A.,  {Siana} B.,  {Brooks} A.~M.,  {Christensen} C.~R.,
  {Bruzual} G.,  {Stark} D.~P.,   {Alavi} A.,  2015, \mn@doi [\mnras]
  {10.1093/mnras/stv1001}, \href
  {http://adsabs.harvard.edu/abs/2015MNRAS.451..839D} {451, 839}

\bibitem[\protect\citeauthoryear{{Done}, {Davis}, {Jin}, {Blaes}  \&
  {Ward}}{{Done} et~al.}{2012}]{Done2012}
{Done} C.,  {Davis} S.~W.,  {Jin} C.,  {Blaes} O.,   {Ward} M.,  2012, \mn@doi
  [\mnras] {10.1111/j.1365-2966.2011.19779.x}, \href
  {http://adsabs.harvard.edu/abs/2012MNRAS.420.1848D} {420, 1848}

\bibitem[\protect\citeauthoryear{{Dubois}, {Devriendt}, {Slyz}  \&
  {Teyssier}}{{Dubois} et~al.}{2012}]{Dubois2012}
{Dubois} Y.,  {Devriendt} J.,  {Slyz} A.,   {Teyssier} R.,  2012, \mn@doi
  [\mnras] {10.1111/j.1365-2966.2011.20236.x}, \href
  {http://adsabs.harvard.edu/abs/2012MNRAS.420.2662D} {420, 2662}

\bibitem[\protect\citeauthoryear{{Dubois}, {Pichon}, {Devriendt}, {Silk},
  {Haehnelt}, {Kimm}  \& {Slyz}}{{Dubois} et~al.}{2013}]{Dubois2013}
{Dubois} Y.,  {Pichon} C.,  {Devriendt} J.,  {Silk} J.,  {Haehnelt} M.,  {Kimm}
  T.,   {Slyz} A.,  2013, \mn@doi [\mnras] {10.1093/mnras/sts224}, \href
  {http://adsabs.harvard.edu/abs/2013MNRAS.428.2885D} {428, 2885}

\bibitem[\protect\citeauthoryear{{Dubois}, {Volonteri}, {Silk}, {Devriendt},
  {Slyz}  \& {Teyssier}}{{Dubois} et~al.}{2015}]{Dubois2015}
{Dubois} Y.,  {Volonteri} M.,  {Silk} J.,  {Devriendt} J.,  {Slyz} A.,
  {Teyssier} R.,  2015, \mn@doi [\mnras] {10.1093/mnras/stv1416}, \href
  {http://adsabs.harvard.edu/abs/2015MNRAS.452.1502D} {452, 1502}

\bibitem[\protect\citeauthoryear{{Dubroca} \& {Feugeas}}{{Dubroca} \&
  {Feugeas}}{1999}]{Dubroca1999}
{Dubroca} B.,  {Feugeas} J.,  1999, \mn@doi [Academie des Sciences Paris
  Comptes Rendus Serie Sciences Mathematiques] {10.1016/S0764-4442(00)87499-6},
  \href {http://adsabs.harvard.edu/abs/1999CRASM.329..915D} {329, 915}

\bibitem[\protect\citeauthoryear{{Fakhouri}, {Ma}  \&
  {Boylan-Kolchin}}{{Fakhouri} et~al.}{2010}]{Fakhouri2010}
{Fakhouri} O.,  {Ma} C.-P.,   {Boylan-Kolchin} M.,  2010, \mn@doi [\mnras]
  {10.1111/j.1365-2966.2010.16859.x}, \href
  {http://adsabs.harvard.edu/abs/2010MNRAS.406.2267F} {406, 2267}

\bibitem[\protect\citeauthoryear{{Fan} et~al.,}{{Fan} et~al.}{2001}]{Fan2001}
{Fan} X.,  et~al., 2001, \mn@doi [\aj] {10.1086/324111}, \href
  {http://adsabs.harvard.edu/abs/2001AJ....122.2833F} {122, 2833}

\bibitem[\protect\citeauthoryear{{Federrath} \& {Klessen}}{{Federrath} \&
  {Klessen}}{2012}]{Federrath2012}
{Federrath} C.,  {Klessen} R.~S.,  2012, \mn@doi [\apj]
  {10.1088/0004-637X/761/2/156}, \href
  {http://adsabs.harvard.edu/abs/2012ApJ...761..156F} {761, 156}

\bibitem[\protect\citeauthoryear{{Ferland} et~al.,}{{Ferland}
  et~al.}{2017}]{Ferland2017}
{Ferland} G.~J.,  et~al., 2017, \rmxaa, \href
  {http://adsabs.harvard.edu/abs/2017RMxAA..53..385F} {53, 385}

\bibitem[\protect\citeauthoryear{{Filippenko} \& {Sargent}}{{Filippenko} \&
  {Sargent}}{1989}]{Filippenko1989}
{Filippenko} A.~V.,  {Sargent} W.~L.~W.,  1989, \mn@doi [\apjl]
  {10.1086/185472}, \href {http://adsabs.harvard.edu/abs/1989ApJ...342L..11F}
  {342, L11}

\bibitem[\protect\citeauthoryear{{Fontanot}, {Cristiani}  \&
  {Vanzella}}{{Fontanot} et~al.}{2012}]{Fontanot2012}
{Fontanot} F.,  {Cristiani} S.,   {Vanzella} E.,  2012, \mn@doi [\mnras]
  {10.1111/j.1365-2966.2012.21594.x}, \href
  {http://adsabs.harvard.edu/abs/2012MNRAS.425.1413F} {425, 1413}

\bibitem[\protect\citeauthoryear{{Geen}, {Rosdahl}, {Blaizot}, {Devriendt}  \&
  {Slyz}}{{Geen} et~al.}{2015}]{Geen2015}
{Geen} S.,  {Rosdahl} J.,  {Blaizot} J.,  {Devriendt} J.,   {Slyz} A.,  2015,
  \mn@doi [\mnras] {10.1093/mnras/stv251}, \href
  {http://adsabs.harvard.edu/abs/2015MNRAS.448.3248G} {448, 3248}

\bibitem[\protect\citeauthoryear{{Giallongo} et~al.,}{{Giallongo}
  et~al.}{2015}]{Giallongo2015}
{Giallongo} E.,  et~al., 2015, \mn@doi [\aap] {10.1051/0004-6361/201425334},
  \href {http://adsabs.harvard.edu/abs/2015A%26A...578A..83G} {578, A83}

\bibitem[\protect\citeauthoryear{{Gnedin}}{{Gnedin}}{2016}]{Gnedin2016}
{Gnedin} N.~Y.,  2016, \mn@doi [\apjl] {10.3847/2041-8205/825/2/L17}, \href
  {http://adsabs.harvard.edu/abs/2016ApJ...825L..17G} {825, L17}

\bibitem[\protect\citeauthoryear{{Grazian} et~al.,}{{Grazian}
  et~al.}{2017}]{Grazian2017}
{Grazian} A.,  et~al., 2017, \mn@doi [\aap] {10.1051/0004-6361/201730447},
  \href {http://adsabs.harvard.edu/abs/2017A%26A...602A..18G} {602, A18}

\bibitem[\protect\citeauthoryear{{Grissom}, {Ballantyne}  \& {Wise}}{{Grissom}
  et~al.}{2014}]{Grissom2014}
{Grissom} R.~L.,  {Ballantyne} D.~R.,   {Wise} J.~H.,  2014, \mn@doi [\aap]
  {10.1051/0004-6361/201322637}, \href
  {http://adsabs.harvard.edu/abs/2014A%26A...561A..90G} {561, A90}

\bibitem[\protect\citeauthoryear{{Guillet}, {Chapon}  \& {Labadens}}{{Guillet}
  et~al.}{2013}]{Guillet2013}
{Guillet} T.,  {Chapon} D.,   {Labadens} M.,  2013, {PyMSES: Python modules for
  RAMSES}, Astrophysics Source Code Library (\mn@eprint {ascl} {1310.002})

\bibitem[\protect\citeauthoryear{{Gunn} \& {Peterson}}{{Gunn} \&
  {Peterson}}{1965}]{Gunn1965}
{Gunn} J.~E.,  {Peterson} B.~A.,  1965, \mn@doi [\apj] {10.1086/148444}, \href
  {http://adsabs.harvard.edu/abs/1965ApJ...142.1633G} {142, 1633}

\bibitem[\protect\citeauthoryear{{Haardt} \& {Madau}}{{Haardt} \&
  {Madau}}{1996}]{Haardt1996}
{Haardt} F.,  {Madau} P.,  1996, \mn@doi [\apj] {10.1086/177035}, \href
  {http://adsabs.harvard.edu/abs/1996ApJ...461...20H} {461, 20}

\bibitem[\protect\citeauthoryear{{Haardt} \& {Madau}}{{Haardt} \&
  {Madau}}{2012}]{Haardt2012}
{Haardt} F.,  {Madau} P.,  2012, \mn@doi [\apj] {10.1088/0004-637X/746/2/125},
  \href {http://adsabs.harvard.edu/abs/2012ApJ...746..125H} {746, 125}

\bibitem[\protect\citeauthoryear{{Habouzit}, {Volonteri}  \&
  {Dubois}}{{Habouzit} et~al.}{2017}]{Habouzit2017}
{Habouzit} M.,  {Volonteri} M.,   {Dubois} Y.,  2017, \mn@doi [\mnras]
  {10.1093/mnras/stx666}, \href
  {http://adsabs.harvard.edu/abs/2017MNRAS.468.3935H} {468, 3935}

\bibitem[\protect\citeauthoryear{{Hahn} \& {Abel}}{{Hahn} \&
  {Abel}}{2011}]{Hahn2011}
{Hahn} O.,  {Abel} T.,  2011, \mn@doi [\mnras]
  {10.1111/j.1365-2966.2011.18820.x}, \href
  {http://adsabs.harvard.edu/abs/2011MNRAS.415.2101H} {415, 2101}

\bibitem[\protect\citeauthoryear{{Hassan}, {Dav{\'e}}, {Mitra}, {Finlator},
  {Ciardi}  \& {Santos}}{{Hassan} et~al.}{2017}]{Hassan2017}
{Hassan} S.,  {Dav{\'e}} R.,  {Mitra} S.,  {Finlator} K.,  {Ciardi} B.,
  {Santos} M.~G.,  2017, preprint, \href
  {http://adsabs.harvard.edu/abs/2017arXiv170505398H} {} (\mn@eprint {arXiv}
  {1705.05398})

\bibitem[\protect\citeauthoryear{{Hunter}}{{Hunter}}{2007}]{Hunter2007}
{Hunter} J.~D.,  2007, \mn@doi [Computing in Science and Engineering]
  {10.1109/MCSE.2007.55}, \href
  {http://adsabs.harvard.edu/abs/2007CSE.....9...90H} {9, 90}

\bibitem[\protect\citeauthoryear{{Ishigaki}, {Kawamata}, {Ouchi}, {Oguri}  \&
  {Shimasaku}}{{Ishigaki} et~al.}{2017}]{Ishigaki2017}
{Ishigaki} M.,  {Kawamata} R.,  {Ouchi} M.,  {Oguri} M.,   {Shimasaku} K.,
  2017, preprint, \href {http://adsabs.harvard.edu/abs/2017arXiv170204867I} {}
  (\mn@eprint {arXiv} {1702.04867})

\bibitem[\protect\citeauthoryear{{Izotov}, {Schaerer}, {Thuan}, {Worseck},
  {Guseva}, {Orlitov{\'a}}  \& {Verhamme}}{{Izotov} et~al.}{2016}]{Izotov2016}
{Izotov} Y.~I.,  {Schaerer} D.,  {Thuan} T.~X.,  {Worseck} G.,  {Guseva} N.~G.,
   {Orlitov{\'a}} I.,   {Verhamme} A.,  2016, \mn@doi [\mnras]
  {10.1093/mnras/stw1205}, \href
  {http://adsabs.harvard.edu/abs/2016MNRAS.461.3683I} {461, 3683}

\bibitem[\protect\citeauthoryear{{Japelj} et~al.,}{{Japelj}
  et~al.}{2017}]{Japelj2017}
{Japelj} J.,  et~al., 2017, \mn@doi [\mnras] {10.1093/mnras/stx477}, \href
  {http://adsabs.harvard.edu/abs/2017MNRAS.468..389J} {468, 389}

\bibitem[\protect\citeauthoryear{Jones, Oliphant, Peterson  et~al.}{Jones
  et~al.}{2001}]{Jones2001}
Jones E.,  Oliphant T.,  Peterson P.,   et~al., 2001, {SciPy}: Open source
  scientific tools for {Python}, \url {http://www.scipy.org/}

\bibitem[\protect\citeauthoryear{{Kaaret}, {Brorby}, {Casella}  \&
  {Prestwich}}{{Kaaret} et~al.}{2017}]{Kaaret2017}
{Kaaret} P.,  {Brorby} M.,  {Casella} L.,   {Prestwich} A.~H.,  2017, \mn@doi
  [\mnras] {10.1093/mnras/stx1945}, \href
  {http://adsabs.harvard.edu/abs/2017MNRAS.471.4234K} {471, 4234}

\bibitem[\protect\citeauthoryear{{Kauffmann}, {Pillai}  \&
  {Goldsmith}}{{Kauffmann} et~al.}{2013}]{Kauffmann2013}
{Kauffmann} J.,  {Pillai} T.,   {Goldsmith} P.~F.,  2013, \mn@doi [\apj]
  {10.1088/0004-637X/779/2/185}, \href
  {http://adsabs.harvard.edu/abs/2013ApJ...779..185K} {779, 185}

\bibitem[\protect\citeauthoryear{{Kennicutt} \& {Evans}}{{Kennicutt} \&
  {Evans}}{2012}]{Kennicutt2012}
{Kennicutt} R.~C.,  {Evans} N.~J.,  2012, \mn@doi [\araa]
  {10.1146/annurev-astro-081811-125610}, \href
  {http://adsabs.harvard.edu/abs/2012ARA%26A..50..531K} {50, 531}

\bibitem[\protect\citeauthoryear{{Kimm} \& {Cen}}{{Kimm} \&
  {Cen}}{2014}]{Kimm2014}
{Kimm} T.,  {Cen} R.,  2014, \mn@doi [\apj] {10.1088/0004-637X/788/2/121},
  \href {http://adsabs.harvard.edu/abs/2014ApJ...788..121K} {788, 121}

\bibitem[\protect\citeauthoryear{{Kimm}, {Cen}, {Devriendt}, {Dubois}  \&
  {Slyz}}{{Kimm} et~al.}{2015}]{Kimm2015}
{Kimm} T.,  {Cen} R.,  {Devriendt} J.,  {Dubois} Y.,   {Slyz} A.,  2015,
  \mn@doi [\mnras] {10.1093/mnras/stv1211}, \href
  {http://adsabs.harvard.edu/abs/2015MNRAS.451.2900K} {451, 2900}

\bibitem[\protect\citeauthoryear{{Kimm}, {Katz}, {Haehnelt}, {Rosdahl},
  {Devriendt}  \& {Slyz}}{{Kimm} et~al.}{2017}]{Kimm2017}
{Kimm} T.,  {Katz} H.,  {Haehnelt} M.,  {Rosdahl} J.,  {Devriendt} J.,   {Slyz}
  A.,  2017, \mn@doi [\mnras] {10.1093/mnras/stx052}, \href
  {http://adsabs.harvard.edu/abs/2017MNRAS.466.4826K} {466, 4826}

\bibitem[\protect\citeauthoryear{{Kroupa} \& {Weidner}}{{Kroupa} \&
  {Weidner}}{2003}]{Kroupa2003}
{Kroupa} P.,  {Weidner} C.,  2003, \mn@doi [\apj] {10.1086/379105}, \href
  {http://adsabs.harvard.edu/abs/2003ApJ...598.1076K} {598, 1076}

\bibitem[\protect\citeauthoryear{{Kuhlen} \& {Faucher-Gigu{\`e}re}}{{Kuhlen} \&
  {Faucher-Gigu{\`e}re}}{2012}]{Kuhlen2012}
{Kuhlen} M.,  {Faucher-Gigu{\`e}re} C.-A.,  2012, \mn@doi [\mnras]
  {10.1111/j.1365-2966.2012.20924.x}, \href
  {http://adsabs.harvard.edu/abs/2012MNRAS.423..862K} {423, 862}

\bibitem[\protect\citeauthoryear{{Kunth}, {Sargent}  \& {Bothun}}{{Kunth}
  et~al.}{1987}]{Kunth1987}
{Kunth} D.,  {Sargent} W.~L.~W.,   {Bothun} G.~D.,  1987, \mn@doi [\aj]
  {10.1086/114287}, \href {http://adsabs.harvard.edu/abs/1987AJ.....93...29K}
  {93, 29}

\bibitem[\protect\citeauthoryear{{Leitherer}, {Ferguson}, {Heckman}  \&
  {Lowenthal}}{{Leitherer} et~al.}{1995}]{Leitherer1995}
{Leitherer} C.,  {Ferguson} H.~C.,  {Heckman} T.~M.,   {Lowenthal} J.~D.,
  1995, \mn@doi [\apjl] {10.1086/309760}, \href
  {http://adsabs.harvard.edu/abs/1995ApJ...454L..19L} {454, L19}

\bibitem[\protect\citeauthoryear{{Levermore}}{{Levermore}}{1984}]{Levermore1984}
{Levermore} C.~D.,  1984, \mn@doi [\jqsrt] {10.1016/0022-4073(84)90112-2},
  \href {http://adsabs.harvard.edu/abs/1984JQSRT..31..149L} {31, 149}

\bibitem[\protect\citeauthoryear{{Livermore}, {Finkelstein}  \&
  {Lotz}}{{Livermore} et~al.}{2017}]{Livermore2017}
{Livermore} R.~C.,  {Finkelstein} S.~L.,   {Lotz} J.~M.,  2017, \mn@doi [\apj]
  {10.3847/1538-4357/835/2/113}, \href
  {http://adsabs.harvard.edu/abs/2017ApJ...835..113L} {835, 113}

\bibitem[\protect\citeauthoryear{{Ma}, {Kasen}, {Hopkins},
  {Faucher-Gigu{\`e}re}, {Quataert}, {Kere{\v s}}  \& {Murray}}{{Ma}
  et~al.}{2015}]{Ma2015}
{Ma} X.,  {Kasen} D.,  {Hopkins} P.~F.,  {Faucher-Gigu{\`e}re} C.-A.,
  {Quataert} E.,  {Kere{\v s}} D.,   {Murray} N.,  2015, \mn@doi [\mnras]
  {10.1093/mnras/stv1679}, \href
  {http://adsabs.harvard.edu/abs/2015MNRAS.453..960M} {453, 960}

\bibitem[\protect\citeauthoryear{{Madau} \& {Haardt}}{{Madau} \&
  {Haardt}}{2015}]{Madau2015}
{Madau} P.,  {Haardt} F.,  2015, \mn@doi [\apjl] {10.1088/2041-8205/813/1/L8},
  \href {http://adsabs.harvard.edu/abs/2015ApJ...813L...8M} {813, L8}

\bibitem[\protect\citeauthoryear{{Madau}, {Rees}, {Volonteri}, {Haardt}  \&
  {Oh}}{{Madau} et~al.}{2004}]{Madau2004}
{Madau} P.,  {Rees} M.~J.,  {Volonteri} M.,  {Haardt} F.,   {Oh} S.~P.,  2004,
  \mn@doi [\apj] {10.1086/381935}, \href
  {http://adsabs.harvard.edu/abs/2004ApJ...604..484M} {604, 484}

\bibitem[\protect\citeauthoryear{{Mao} \& {Kim}}{{Mao} \&
  {Kim}}{2016}]{Mao2016}
{Mao} J.,  {Kim} M.,  2016, \mn@doi [\apj] {10.3847/0004-637X/828/2/96}, \href
  {http://adsabs.harvard.edu/abs/2016ApJ...828...96M} {828, 96}

\bibitem[\protect\citeauthoryear{{McAlpine}, {Bower}, {Harrison}, {Crain},
  {Schaller}, {Schaye}  \& {Theuns}}{{McAlpine} et~al.}{2017}]{McAlpine2017}
{McAlpine} S.,  {Bower} R.~G.,  {Harrison} C.~M.,  {Crain} R.~A.,  {Schaller}
  M.,  {Schaye} J.,   {Theuns} T.,  2017, \mn@doi [\mnras]
  {10.1093/mnras/stx658}, \href
  {http://adsabs.harvard.edu/abs/2017MNRAS.468.3395M} {468, 3395}

\bibitem[\protect\citeauthoryear{{Mezcua}, {Civano}, {Fabbiano}, {Miyaji}  \&
  {Marchesi}}{{Mezcua} et~al.}{2016}]{Mezcua2016}
{Mezcua} M.,  {Civano} F.,  {Fabbiano} G.,  {Miyaji} T.,   {Marchesi} S.,
  2016, \mn@doi [\apj] {10.3847/0004-637X/817/1/20}, \href
  {http://adsabs.harvard.edu/abs/2016ApJ...817...20M} {817, 20}

\bibitem[\protect\citeauthoryear{{Micheva}, {Iwata}  \& {Inoue}}{{Micheva}
  et~al.}{2017}]{Micheva2017}
{Micheva} G.,  {Iwata} I.,   {Inoue} A.~K.,  2017, \mn@doi [\mnras]
  {10.1093/mnras/stw1329}, \href
  {http://adsabs.harvard.edu/abs/2017MNRAS.465..302M} {465, 302}

\bibitem[\protect\citeauthoryear{{Miller}, {Ellis}, {Newman}  \&
  {Benson}}{{Miller} et~al.}{2014}]{Miller2014}
{Miller} S.~H.,  {Ellis} R.~S.,  {Newman} A.~B.,   {Benson} A.,  2014, \mn@doi
  [\apj] {10.1088/0004-637X/782/2/115}, \href
  {http://adsabs.harvard.edu/abs/2014ApJ...782..115M} {782, 115}

\bibitem[\protect\citeauthoryear{{Moran}, {Shahinyan}, {Sugarman}, {V{\'e}lez}
  \& {Eracleous}}{{Moran} et~al.}{2014}]{Moran2014}
{Moran} E.~C.,  {Shahinyan} K.,  {Sugarman} H.~R.,  {V{\'e}lez} D.~O.,
  {Eracleous} M.,  2014, \mn@doi [\aj] {10.1088/0004-6256/148/6/136}, \href
  {http://adsabs.harvard.edu/abs/2014AJ....148..136M} {148, 136}

\bibitem[\protect\citeauthoryear{{Moster}, {Naab}  \& {White}}{{Moster}
  et~al.}{2017}]{Moster2017}
{Moster} B.~P.,  {Naab} T.,   {White} S.~D.~M.,  2017, preprint, \href
  {http://adsabs.harvard.edu/abs/2017arXiv170505373M} {} (\mn@eprint {arXiv}
  {1705.05373})

\bibitem[\protect\citeauthoryear{{Negri} \& {Volonteri}}{{Negri} \&
  {Volonteri}}{2017}]{Negri2017}
{Negri} A.,  {Volonteri} M.,  2017, \mn@doi [\mnras] {10.1093/mnras/stx362},
  \href {http://adsabs.harvard.edu/abs/2017MNRAS.467.3475N} {467, 3475}

\bibitem[\protect\citeauthoryear{{O'Shea}, {Wise}, {Xu}  \& {Norman}}{{O'Shea}
  et~al.}{2015}]{OShea2015}
{O'Shea} B.~W.,  {Wise} J.~H.,  {Xu} H.,   {Norman} M.~L.,  2015, \mn@doi
  [\apjl] {10.1088/2041-8205/807/1/L12}, \href
  {http://adsabs.harvard.edu/abs/2015ApJ...807L..12O} {807, L12}

\bibitem[\protect\citeauthoryear{{Ocvirk} et~al.,}{{Ocvirk}
  et~al.}{2016}]{Ocvirk2016}
{Ocvirk} P.,  et~al., 2016, \mn@doi [\mnras] {10.1093/mnras/stw2036}, \href
  {http://adsabs.harvard.edu/abs/2016MNRAS.463.1462O} {463, 1462}

\bibitem[\protect\citeauthoryear{{Onoue} et~al.,}{{Onoue}
  et~al.}{2017}]{Onoue2017}
{Onoue} M.,  et~al., 2017, \mn@doi [\apjl] {10.3847/2041-8213/aa8cc6}, \href
  {http://adsabs.harvard.edu/abs/2017ApJ...847L..15O} {847, L15}

\bibitem[\protect\citeauthoryear{{Ostriker}}{{Ostriker}}{1999}]{Ostriker1999}
{Ostriker} E.~C.,  1999, \mn@doi [\apj] {10.1086/306858}, \href
  {http://adsabs.harvard.edu/abs/1999ApJ...513..252O} {513, 252}

\bibitem[\protect\citeauthoryear{{Paardekooper}, {Khochfar}  \& {Dalla
  Vecchia}}{{Paardekooper} et~al.}{2015}]{Paardekooper2015}
{Paardekooper} J.-P.,  {Khochfar} S.,   {Dalla Vecchia} C.,  2015, \mn@doi
  [\mnras] {10.1093/mnras/stv1114}, \href
  {http://adsabs.harvard.edu/abs/2015MNRAS.451.2544P} {451, 2544}

\bibitem[\protect\citeauthoryear{{Padoan} \& {Nordlund}}{{Padoan} \&
  {Nordlund}}{2011}]{Padoan2011}
{Padoan} P.,  {Nordlund} {\AA}.,  2011, \mn@doi [\apj]
  {10.1088/0004-637X/730/1/40}, \href
  {http://adsabs.harvard.edu/abs/2011ApJ...730...40P} {730, 40}

\bibitem[\protect\citeauthoryear{{Padoan}, {Haugb{\o}lle}, {Nordlund}  \&
  {Frimann}}{{Padoan} et~al.}{2017}]{Padoan2017}
{Padoan} P.,  {Haugb{\o}lle} T.,  {Nordlund} {\AA}.,   {Frimann} S.,  2017,
  \mn@doi [\apj] {10.3847/1538-4357/aa6afa}, \href
  {http://adsabs.harvard.edu/abs/2017ApJ...840...48P} {840, 48}

\bibitem[\protect\citeauthoryear{{Parsa}, {Dunlop}  \& {McLure}}{{Parsa}
  et~al.}{2017}]{Parsa2017}
{Parsa} S.,  {Dunlop} J.~S.,   {McLure} R.~J.,  2017, preprint, \href
  {http://adsabs.harvard.edu/abs/2017arXiv170407750P} {} (\mn@eprint {arXiv}
  {1704.07750})

\bibitem[\protect\citeauthoryear{{Perez} \& {Granger}}{{Perez} \&
  {Granger}}{2007}]{Perez2007}
{Perez} F.,  {Granger} B.~E.,  2007, \mn@doi [Computing in Science Engineering]
  {10.1109/MCSE.2007.53}, 9

\bibitem[\protect\citeauthoryear{{Planck Collaboration} et~al.,}{{Planck
  Collaboration} et~al.}{2016a}]{Planck2015}
{Planck Collaboration} et~al., 2016a, \mn@doi [\aap]
  {10.1051/0004-6361/201525830}, \href
  {http://adsabs.harvard.edu/abs/2016A%26A...594A..13P} {594, A13}

\bibitem[\protect\citeauthoryear{{Planck Collaboration} et~al.,}{{Planck
  Collaboration} et~al.}{2016b}]{PlanckCollaboration2016}
{Planck Collaboration} et~al., 2016b, \mn@doi [\aap]
  {10.1051/0004-6361/201628890}, \href
  {http://adsabs.harvard.edu/abs/2016A%26A...596A.107P} {596, A107}

\bibitem[\protect\citeauthoryear{{Prieto}, {Escala}, {Volonteri}  \&
  {Dubois}}{{Prieto} et~al.}{2017}]{Prieto2017}
{Prieto} J.,  {Escala} A.,  {Volonteri} M.,   {Dubois} Y.,  2017, \mn@doi
  [\apj] {10.3847/1538-4357/aa5be5}, \href
  {http://adsabs.harvard.edu/abs/2017ApJ...836..216P} {836, 216}

\bibitem[\protect\citeauthoryear{{Qin} et~al.,}{{Qin} et~al.}{2017}]{Qin2017}
{Qin} Y.,  et~al., 2017, \mn@doi [\mnras] {10.1093/mnras/stx1909}, \href
  {http://adsabs.harvard.edu/abs/2017MNRAS.472.2009Q} {472, 2009}

\bibitem[\protect\citeauthoryear{{Rasera} \& {Teyssier}}{{Rasera} \&
  {Teyssier}}{2006}]{Rasera2006}
{Rasera} Y.,  {Teyssier} R.,  2006, \mn@doi [\aap]
  {10.1051/0004-6361:20053116}, \href
  {http://adsabs.harvard.edu/abs/2006A%26A...445....1R} {445, 1}

\bibitem[\protect\citeauthoryear{{Reines} \& {Comastri}}{{Reines} \&
  {Comastri}}{2016}]{Reines2016}
{Reines} A.~E.,  {Comastri} A.,  2016, \mn@doi [\pasa] {10.1017/pasa.2016.46},
  \href {http://adsabs.harvard.edu/abs/2016PASA...33...54R} {33, e054}

\bibitem[\protect\citeauthoryear{{Reines} \& {Volonteri}}{{Reines} \&
  {Volonteri}}{2015}]{Reines2015}
{Reines} A.~E.,  {Volonteri} M.,  2015, \mn@doi [\apj]
  {10.1088/0004-637X/813/2/82}, \href
  {http://adsabs.harvard.edu/abs/2015ApJ...813...82R} {813, 82}

\bibitem[\protect\citeauthoryear{{Reines}, {Greene}  \& {Geha}}{{Reines}
  et~al.}{2013}]{Reines2013}
{Reines} A.~E.,  {Greene} J.~E.,   {Geha} M.,  2013, \mn@doi [\apj]
  {10.1088/0004-637X/775/2/116}, \href
  {http://adsabs.harvard.edu/abs/2013ApJ...775..116R} {775, 116}

\bibitem[\protect\citeauthoryear{{Ricci}, {Marchesi}, {Shankar}, {La Franca}
  \& {Civano}}{{Ricci} et~al.}{2017}]{Ricci2017}
{Ricci} F.,  {Marchesi} S.,  {Shankar} F.,  {La Franca} F.,   {Civano} F.,
  2017, \mn@doi [\mnras] {10.1093/mnras/stw2909}, \href
  {http://adsabs.harvard.edu/abs/2017MNRAS.465.1915R} {465, 1915}

\bibitem[\protect\citeauthoryear{{Robertson}, {Ellis}, {Dunlop}, {McLure}  \&
  {Stark}}{{Robertson} et~al.}{2010}]{Robertson2010}
{Robertson} B.~E.,  {Ellis} R.~S.,  {Dunlop} J.~S.,  {McLure} R.~J.,   {Stark}
  D.~P.,  2010, \mn@doi [\nat] {10.1038/nature09527}, \href
  {http://adsabs.harvard.edu/abs/2010Natur.468...49R} {468, 49}

\bibitem[\protect\citeauthoryear{{Robertson} et~al.,}{{Robertson}
  et~al.}{2013}]{Robertson2013}
{Robertson} B.~E.,  et~al., 2013, \mn@doi [\apj] {10.1088/0004-637X/768/1/71},
  \href {http://adsabs.harvard.edu/abs/2013ApJ...768...71R} {768, 71}

\bibitem[\protect\citeauthoryear{{Rosdahl} \& {Teyssier}}{{Rosdahl} \&
  {Teyssier}}{2015}]{Rosdahl2015}
{Rosdahl} J.,  {Teyssier} R.,  2015, \mn@doi [\mnras] {10.1093/mnras/stv567},
  \href {http://adsabs.harvard.edu/abs/2015MNRAS.449.4380R} {449, 4380}

\bibitem[\protect\citeauthoryear{{Rosdahl}, {Blaizot}, {Aubert}, {Stranex}  \&
  {Teyssier}}{{Rosdahl} et~al.}{2013}]{Rosdahl2013}
{Rosdahl} J.,  {Blaizot} J.,  {Aubert} D.,  {Stranex} T.,   {Teyssier} R.,
  2013, \mn@doi [\mnras] {10.1093/mnras/stt1722}, \href
  {http://adsabs.harvard.edu/abs/2013MNRAS.436.2188R} {436, 2188}

\bibitem[\protect\citeauthoryear{{Rosdahl}, {Schaye}, {Teyssier}  \&
  {Agertz}}{{Rosdahl} et~al.}{2015}]{Rosdahl2015a}
{Rosdahl} J.,  {Schaye} J.,  {Teyssier} R.,   {Agertz} O.,  2015, \mn@doi
  [\mnras] {10.1093/mnras/stv937}, \href
  {http://adsabs.harvard.edu/abs/2015MNRAS.451...34R} {451, 34}

\bibitem[\protect\citeauthoryear{{Rosdahl}, {Schaye}, {Dubois}, {Kimm}  \&
  {Teyssier}}{{Rosdahl} et~al.}{2017}]{Rosdahl2017}
{Rosdahl} J.,  {Schaye} J.,  {Dubois} Y.,  {Kimm} T.,   {Teyssier} R.,  2017,
  \mn@doi [\mnras] {10.1093/mnras/stw3034}, \href
  {http://adsabs.harvard.edu/abs/2017MNRAS.466...11R} {466, 11}

\bibitem[\protect\citeauthoryear{{Rosen} \& {Bregman}}{{Rosen} \&
  {Bregman}}{1995}]{Rosen1995}
{Rosen} A.,  {Bregman} J.~N.,  1995, \mn@doi [\apj] {10.1086/175303}, \href
  {http://adsabs.harvard.edu/abs/1995ApJ...440..634R} {440, 634}

\bibitem[\protect\citeauthoryear{{Salmon} et~al.,}{{Salmon}
  et~al.}{2015}]{Salmon2015}
{Salmon} B.,  et~al., 2015, \mn@doi [\apj] {10.1088/0004-637X/799/2/183}, \href
  {http://adsabs.harvard.edu/abs/2015ApJ...799..183S} {799, 183}

\bibitem[\protect\citeauthoryear{{Salpeter}}{{Salpeter}}{1955}]{Salpeter1955}
{Salpeter} E.~E.,  1955, \mn@doi [\apj] {10.1086/145971}, \href
  {http://adsabs.harvard.edu/abs/1955ApJ...121..161S} {121, 161}

\bibitem[\protect\citeauthoryear{{Santini} et~al.,}{{Santini}
  et~al.}{2017}]{Santini2017}
{Santini} P.,  et~al., 2017, \mn@doi [\apj] {10.3847/1538-4357/aa8874}, \href
  {http://adsabs.harvard.edu/abs/2017ApJ...847...76S} {847, 76}

\bibitem[\protect\citeauthoryear{{Sazonov}, {Ostriker}  \& {Sunyaev}}{{Sazonov}
  et~al.}{2004}]{Sazonov2004}
{Sazonov} S.~Y.,  {Ostriker} J.~P.,   {Sunyaev} R.~A.,  2004, \mn@doi [\mnras]
  {10.1111/j.1365-2966.2004.07184.x}, \href
  {http://adsabs.harvard.edu/abs/2004MNRAS.347..144S} {347, 144}

\bibitem[\protect\citeauthoryear{{Shakura} \& {Sunyaev}}{{Shakura} \&
  {Sunyaev}}{1973}]{Shakura1973}
{Shakura} N.~I.,  {Sunyaev} R.~A.,  1973, \aap, \href
  {http://adsabs.harvard.edu/abs/1973A%26A....24..337S} {24, 337}

\bibitem[\protect\citeauthoryear{{Shull} \& {van Steenberg}}{{Shull} \& {van
  Steenberg}}{1985}]{Shull1985}
{Shull} J.~M.,  {van Steenberg} M.~E.,  1985, \mn@doi [\apj] {10.1086/163605},
  \href {http://adsabs.harvard.edu/abs/1985ApJ...298..268S} {298, 268}

\bibitem[\protect\citeauthoryear{{Siana} et~al.,}{{Siana}
  et~al.}{2015}]{Siana2015}
{Siana} B.,  et~al., 2015, \mn@doi [\apj] {10.1088/0004-637X/804/1/17}, \href
  {http://adsabs.harvard.edu/abs/2015ApJ...804...17S} {804, 17}

\bibitem[\protect\citeauthoryear{{Simpson}, {Bryan}, {Johnston}, {Smith}, {Mac
  Low}, {Sharma}  \& {Tumlinson}}{{Simpson} et~al.}{2013}]{Simpson2013}
{Simpson} C.~M.,  {Bryan} G.~L.,  {Johnston} K.~V.,  {Smith} B.~D.,  {Mac Low}
  M.-M.,  {Sharma} S.,   {Tumlinson} J.,  2013, \mn@doi [\mnras]
  {10.1093/mnras/stt474}, \href
  {http://adsabs.harvard.edu/abs/2013MNRAS.432.1989S} {432, 1989}

\bibitem[\protect\citeauthoryear{{Song} et~al.,}{{Song}
  et~al.}{2016}]{Song2016}
{Song} M.,  et~al., 2016, \mn@doi [\apj] {10.3847/0004-637X/825/1/5}, \href
  {http://adsabs.harvard.edu/abs/2016ApJ...825....5S} {825, 5}

\bibitem[\protect\citeauthoryear{{Speagle}, {Steinhardt}, {Capak}  \&
  {Silverman}}{{Speagle} et~al.}{2014}]{Speagle2014}
{Speagle} J.~S.,  {Steinhardt} C.~L.,  {Capak} P.~L.,   {Silverman} J.~D.,
  2014, \mn@doi [\apjs] {10.1088/0067-0049/214/2/15}, \href
  {http://adsabs.harvard.edu/abs/2014ApJS..214...15S} {214, 15}

\bibitem[\protect\citeauthoryear{{Teyssier}}{{Teyssier}}{2002}]{Teyssier2002}
{Teyssier} R.,  2002, \mn@doi [\aap] {10.1051/0004-6361:20011817}, \href
  {http://adsabs.harvard.edu/abs/2002A%26A...385..337T} {385, 337}

\bibitem[\protect\citeauthoryear{{Thomas}, {Groves}, {Sutherland}, {Dopita},
  {Kewley}  \& {Jin}}{{Thomas} et~al.}{2016}]{Thomas2016}
{Thomas} A.~D.,  {Groves} B.~A.,  {Sutherland} R.~S.,  {Dopita} M.~A.,
  {Kewley} L.~J.,   {Jin} C.,  2016, \mn@doi [\apj]
  {10.3847/1538-4357/833/2/266}, \href
  {http://adsabs.harvard.edu/abs/2016ApJ...833..266T} {833, 266}

\bibitem[\protect\citeauthoryear{{Toro}, {Spruce}  \& {Speares}}{{Toro}
  et~al.}{1994}]{Toro1994}
{Toro} E.~F.,  {Spruce} M.,   {Speares} W.,  1994, \mn@doi [Shock Waves]
  {10.1007/BF01414629}, \href
  {http://adsabs.harvard.edu/abs/1994ShWav...4...25T} {4, 25}

\bibitem[\protect\citeauthoryear{{Trebitsch}, {Blaizot}, {Rosdahl}, {Devriendt}
   \& {Slyz}}{{Trebitsch} et~al.}{2017}]{Trebitsch2017}
{Trebitsch} M.,  {Blaizot} J.,  {Rosdahl} J.,  {Devriendt} J.,   {Slyz} A.,
  2017, \mn@doi [\mnras] {10.1093/mnras/stx1060}, \href
  {http://adsabs.harvard.edu/abs/2017MNRAS.470..224T} {470, 224}

\bibitem[\protect\citeauthoryear{{Tweed}, {Devriendt}, {Blaizot}, {Colombi}  \&
  {Slyz}}{{Tweed} et~al.}{2009}]{Tweed2009}
{Tweed} D.,  {Devriendt} J.,  {Blaizot} J.,  {Colombi} S.,   {Slyz} A.,  2009,
  \mn@doi [\aap] {10.1051/0004-6361/200911787}, \href
  {http://adsabs.harvard.edu/abs/2009A%26A...506..647T} {506, 647}

\bibitem[\protect\citeauthoryear{{Vanzella} et~al.,}{{Vanzella}
  et~al.}{2016}]{Vanzella2016}
{Vanzella} E.,  et~al., 2016, \mn@doi [\apj] {10.3847/0004-637X/825/1/41},
  \href {http://adsabs.harvard.edu/abs/2016ApJ...825...41V} {825, 41}

\bibitem[\protect\citeauthoryear{{Vasei} et~al.,}{{Vasei}
  et~al.}{2016}]{Vasei2016}
{Vasei} K.,  et~al., 2016, \mn@doi [\apj] {10.3847/0004-637X/831/1/38}, \href
  {http://adsabs.harvard.edu/abs/2016ApJ...831...38V} {831, 38}

\bibitem[\protect\citeauthoryear{{Venkatesan}, {Giroux}  \&
  {Shull}}{{Venkatesan} et~al.}{2001}]{Venkatesan2001}
{Venkatesan} A.,  {Giroux} M.~L.,   {Shull} J.~M.,  2001, \mn@doi [\apj]
  {10.1086/323691}, \href {http://adsabs.harvard.edu/abs/2001ApJ...563....1V}
  {563, 1}

\bibitem[\protect\citeauthoryear{{Vito} et~al.,}{{Vito}
  et~al.}{2016}]{Vito2016}
{Vito} F.,  et~al., 2016, \mn@doi [\mnras] {10.1093/mnras/stw1998}, \href
  {http://adsabs.harvard.edu/abs/2016MNRAS.463..348V} {463, 348}

\bibitem[\protect\citeauthoryear{{Volonteri}}{{Volonteri}}{2010}]{Volonteri2010}
{Volonteri} M.,  2010, \mn@doi [\aapr] {10.1007/s00159-010-0029-x}, \href
  {http://adsabs.harvard.edu/abs/2010A%26ARv..18..279V} {18, 279}

\bibitem[\protect\citeauthoryear{{Volonteri} \& {Gnedin}}{{Volonteri} \&
  {Gnedin}}{2009}]{Volonteri2009}
{Volonteri} M.,  {Gnedin} N.~Y.,  2009, \mn@doi [\apj]
  {10.1088/0004-637X/703/2/2113}, \href
  {http://adsabs.harvard.edu/abs/2009ApJ...703.2113V} {703, 2113}

\bibitem[\protect\citeauthoryear{{Volonteri}, {Lodato}  \&
  {Natarajan}}{{Volonteri} et~al.}{2008}]{Volonteri2008}
{Volonteri} M.,  {Lodato} G.,   {Natarajan} P.,  2008, \mn@doi [\mnras]
  {10.1111/j.1365-2966.2007.12589.x}, \href
  {http://adsabs.harvard.edu/abs/2008MNRAS.383.1079V} {383, 1079}

\bibitem[\protect\citeauthoryear{{Volonteri}, {Reines}, {Atek}, {Stark}  \&
  {Trebitsch}}{{Volonteri} et~al.}{2017}]{Volonteri2017}
{Volonteri} M.,  {Reines} A.,  {Atek} H.,  {Stark} D.~P.,   {Trebitsch} M.,
  2017, preprint, \href {http://adsabs.harvard.edu/abs/2017arXiv170400753V} {}
  (\mn@eprint {arXiv} {1704.00753})

\bibitem[\protect\citeauthoryear{{Weigel}, {Schawinski}, {Treister}, {Urry},
  {Koss}  \& {Trakhtenbrot}}{{Weigel} et~al.}{2015}]{Weigel2015}
{Weigel} A.~K.,  {Schawinski} K.,  {Treister} E.,  {Urry} C.~M.,  {Koss} M.,
  {Trakhtenbrot} B.,  2015, \mn@doi [\mnras] {10.1093/mnras/stv184}, \href
  {http://adsabs.harvard.edu/abs/2015MNRAS.448.3167W} {448, 3167}

\bibitem[\protect\citeauthoryear{{Willott} et~al.,}{{Willott}
  et~al.}{2010}]{Willott2010}
{Willott} C.~J.,  et~al., 2010, \mn@doi [\aj] {10.1088/0004-6256/139/3/906},
  \href {http://adsabs.harvard.edu/abs/2010AJ....139..906W} {139, 906}

\bibitem[\protect\citeauthoryear{{Wise}, {Demchenko}, {Halicek}, {Norman},
  {Turk}, {Abel}  \& {Smith}}{{Wise} et~al.}{2014}]{Wise2014}
{Wise} J.~H.,  {Demchenko} V.~G.,  {Halicek} M.~T.,  {Norman} M.~L.,  {Turk}
  M.~J.,  {Abel} T.,   {Smith} B.~D.,  2014, \mn@doi [\mnras]
  {10.1093/mnras/stu979}, \href
  {http://adsabs.harvard.edu/abs/2014MNRAS.442.2560W} {442, 2560}

\bibitem[\protect\citeauthoryear{{Yoshiura}, {Hasegawa}, {Ichiki}, {Tashiro},
  {Shimabukuro}  \& {Takahashi}}{{Yoshiura} et~al.}{2017}]{Yoshiura2017}
{Yoshiura} S.,  {Hasegawa} K.,  {Ichiki} K.,  {Tashiro} H.,  {Shimabukuro} H.,
   {Takahashi} K.,  2017, \mn@doi [\mnras] {10.1093/mnras/stx1754}, \href
  {http://adsabs.harvard.edu/abs/2017MNRAS.471.3713Y} {471, 3713}

\bibitem[\protect\citeauthoryear{{van Wassenhove}, {Volonteri}, {Walker}  \&
  {Gair}}{{van Wassenhove} et~al.}{2010}]{vanWassenhove2010}
{van Wassenhove} S.,  {Volonteri} M.,  {Walker} M.~G.,   {Gair} J.~R.,  2010,
  \mn@doi [\mnras] {10.1111/j.1365-2966.2010.17189.x}, \href
  {http://adsabs.harvard.edu/abs/2010MNRAS.408.1139V} {408, 1139}

\makeatother
\end{thebibliography}



\appendix

\section{Intrinsic versus observed galaxy properties}
\label{sec:app-obs}

In Sect.~\ref{sec:coevolution:galaxy}, we discussed the assembly history of our simulated galaxy based on properties measured using its inferred UV magnitude. We now wish to discuss the impact of this approach, and how it compares to quantities directly measured in the simulation.

\subsection{Stellar to halo mass relation}
\label{sec:app-obs:SMHM}

\begin{figure}
  \centering
  \includegraphics[width=\linewidth]{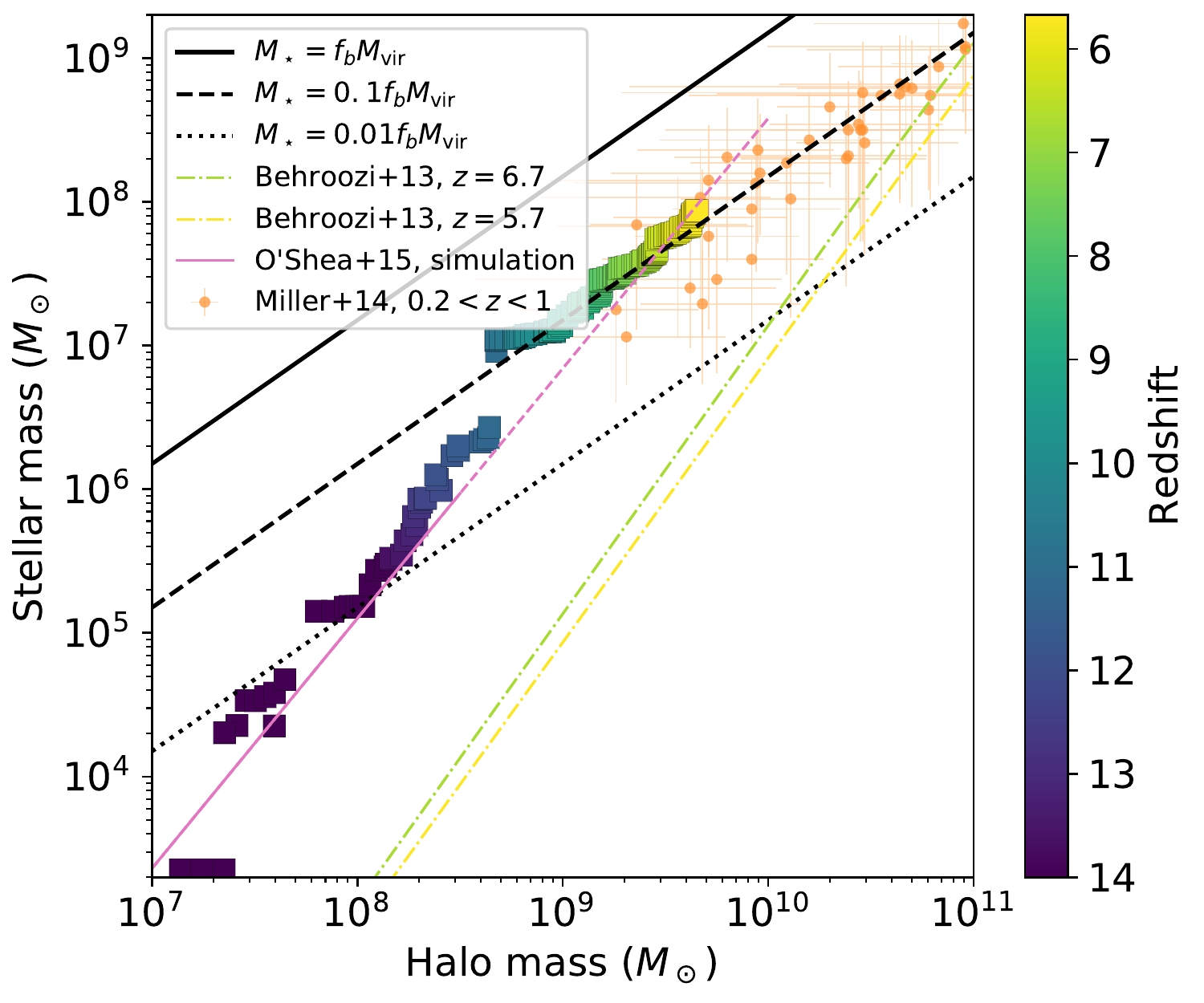}
  \caption{Stellar to halo mass relation for our simulated galaxy, using the stellar mass directly measured in the simulation. The solid, dashed and dotted line correspond respectively to 100\%, 10\% and 1\% of the baryons converted into stars. For the last 500 Myr of the simulation, the galaxy converts 10\% of its baryons in stars.}
  \label{fig:AGNT-SMHM-direct}
\end{figure}
We show in Fig.~\ref{fig:AGNT-SMHM-direct} the stellar to halo mass relation, similar to Fig.~\ref{fig:AGNT-SMHM}, but this time using the stellar mass directly measured in the simulation.
At very early times, the galaxy converts less than 10\% of its baryons into stars, and this fraction saturates after 500 Myr ($z \sim 9.5$). 
When the true stellar mass is used, our simulation lies well above the \citet{Behroozi2013} model, but gives results very similar to other simulations including similar subgrid models at equivalent spatial resolution, such as the \emph{Renaissance} simulation \citep{OShea2015} shown here in pink, with the fit to their stellar to halo mass relation shown as a solid line and an extrapolation to higher masses as a dashed line.

\begin{figure}
  \centering
  \includegraphics[width=.9\linewidth]{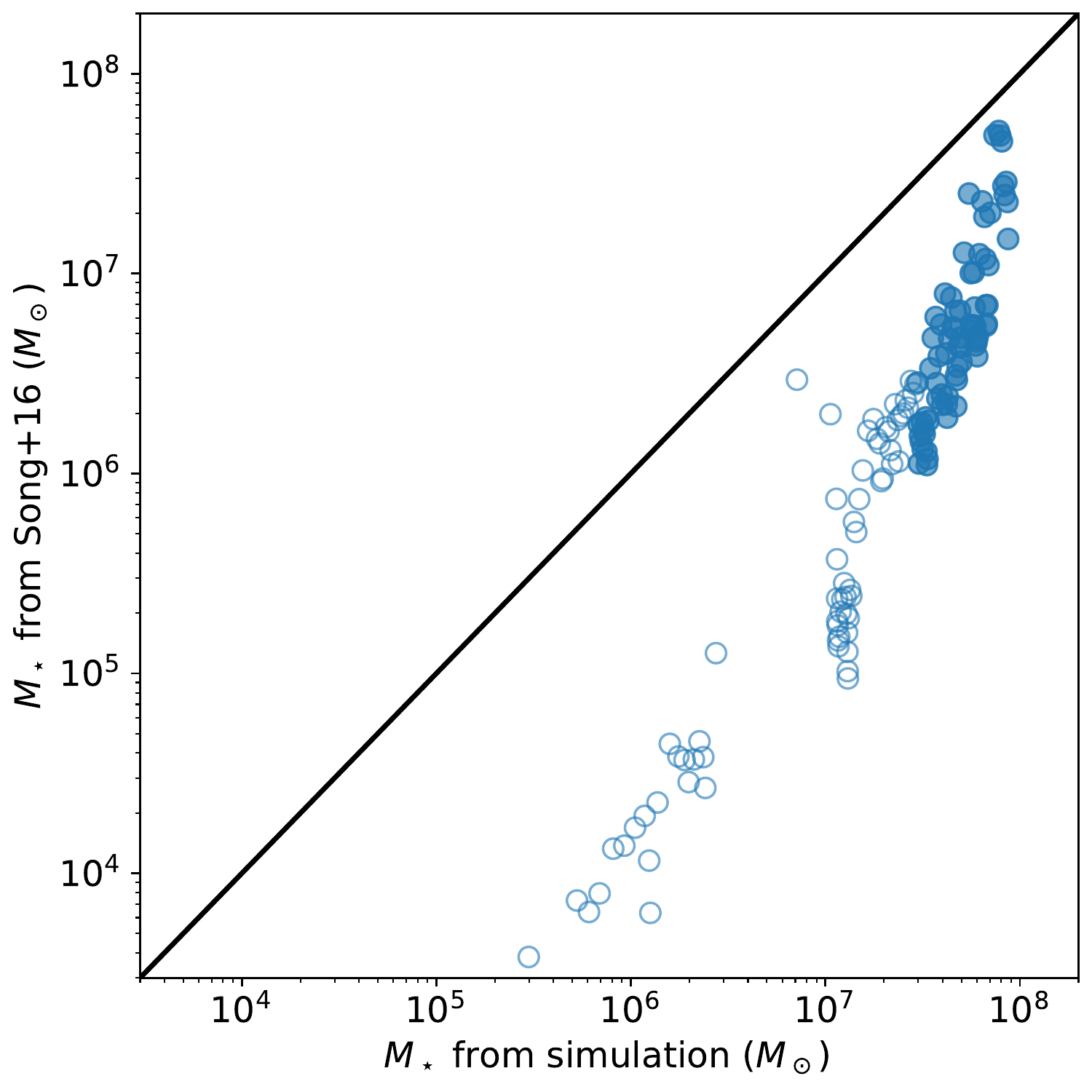}
  \caption{Stellar mass recovered from the UV magnitude using the \citet{Song2016} empirical relation versus the measured stellar mass in the simulation. The $\log\Mstar - \muv$ is extrapolated above $z>8$, as indicated by open symbols.}
  \label{fig:AGNT-mstarSimUV}
\end{figure}

To further investigate the origin of the apparent discrepancy between the stellar mass directly measured in the simulation and that derived from the UV luminosity of the galaxy, we compare in Fig.~\ref{fig:AGNT-mstarSimUV} both stellar mass estimates. The open symbols denote snapshots at $z > 8$, where the $\log\Mstar - \muv$ relation from \citet{Song2016} is extrapolated rather than interpolated.
The observational estimate of \Mstar underestimates by sometimes one order of magnitude the actual stellar mass of the galaxy, and taking into account flux losses and dust would reduce even more the UV-based stellar mass. The large scatter in the UV-based \Mstar at fixed actual mass results from the burstiness of the star formation history that translate in a UV magnitude varying more rapidly that the stellar mass \citep[see for instance][]{Dominguez2015}.

\subsection{Star formation main sequence}
\label{sec:app-obs:MS}

\begin{figure}
  \centering
  \includegraphics[width=\linewidth]{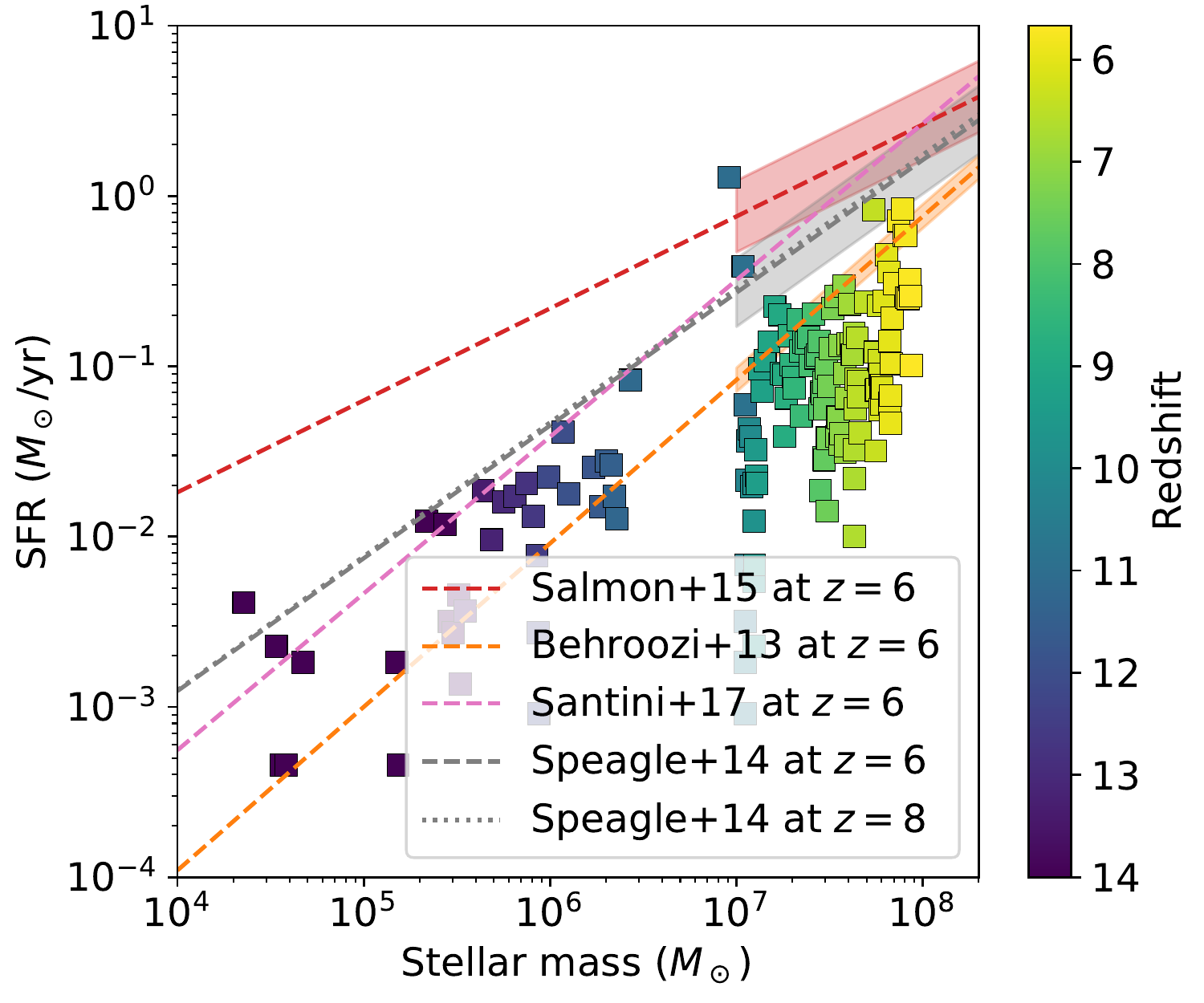}
  \caption{Star formation main sequence, similar to Fig.~\ref{fig:AGNT-MS}, but with the stellar masses and star formation rates directly measured in the simulation. The large scatter in the simulation is due to the intrinsic burstiness of star formation.}
  \label{fig:AGNT-MS-direct}
\end{figure}

In Fig.~\ref{fig:AGNT-MS-direct}, we present the star formation MS using stellar masses and instantaneous SFR directly from the simulation. As expected, the scatter is much larger than with UV-derived properties. As star formation proceeds by bursts in low mass galaxies, the SFR can range from very low to very high at fixed stellar mass, depending on whether the galaxy is caught during or after a star formation episode. This scatter is reduced when measuring the SFR from the UV luminosity as in Fig.~\ref{fig:AGNT-MS}, since the far UV at $\lambda \sim 1500\ \mbox{\AA}$ traces timescales typically longer than the interval between two bursts of star formation.

\begin{figure}
  \centering
  \includegraphics[width=\linewidth]{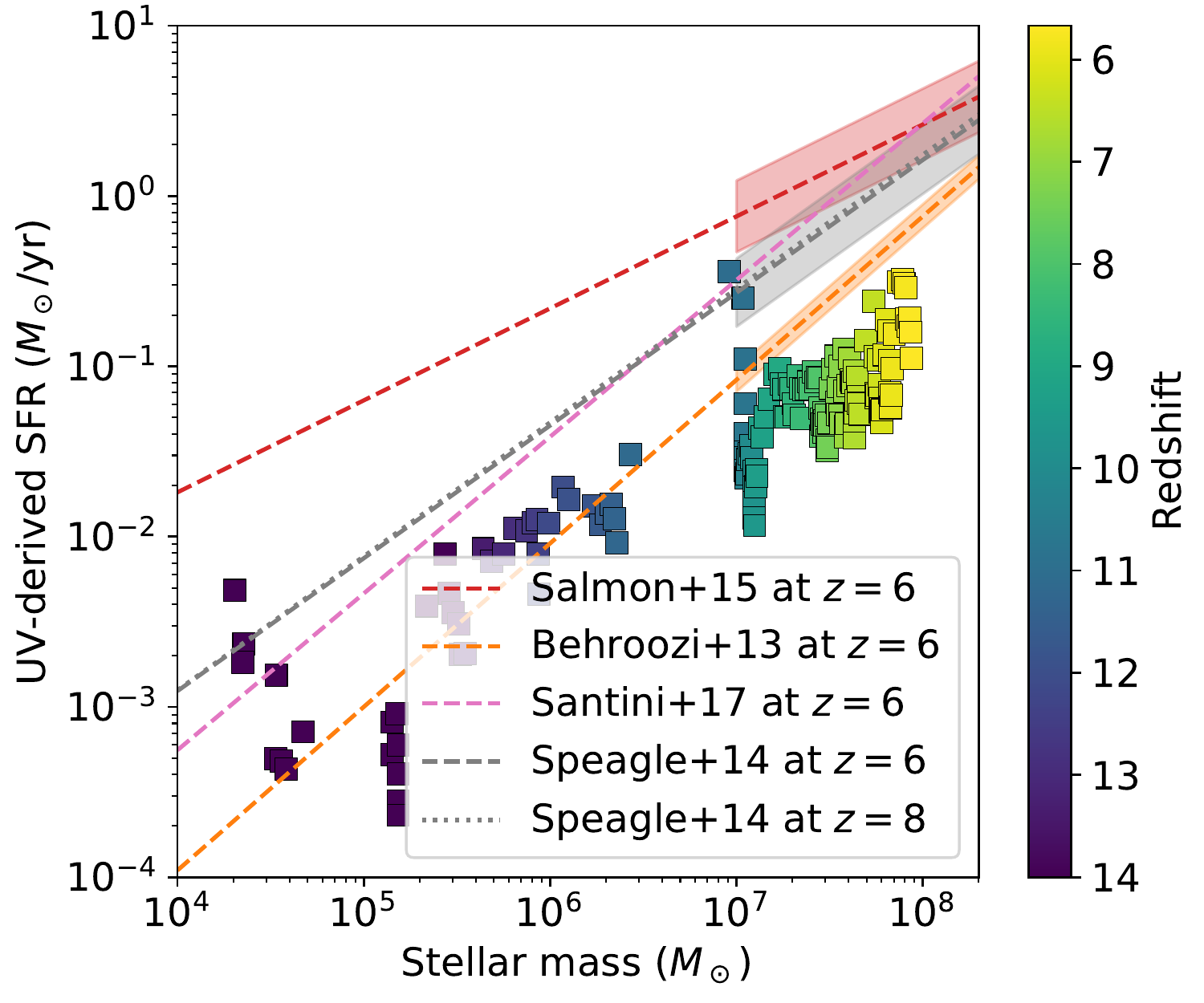}
  \caption{Star formation main sequence, similar to Fig.~\ref{fig:AGNT-MS-direct}, but with only the stellar masses directly measured in the simulation.}
  \label{fig:AGNT-MS-Mdirect-sfrUV}
\end{figure}

\begin{figure}
  \centering
  \includegraphics[width=\linewidth]{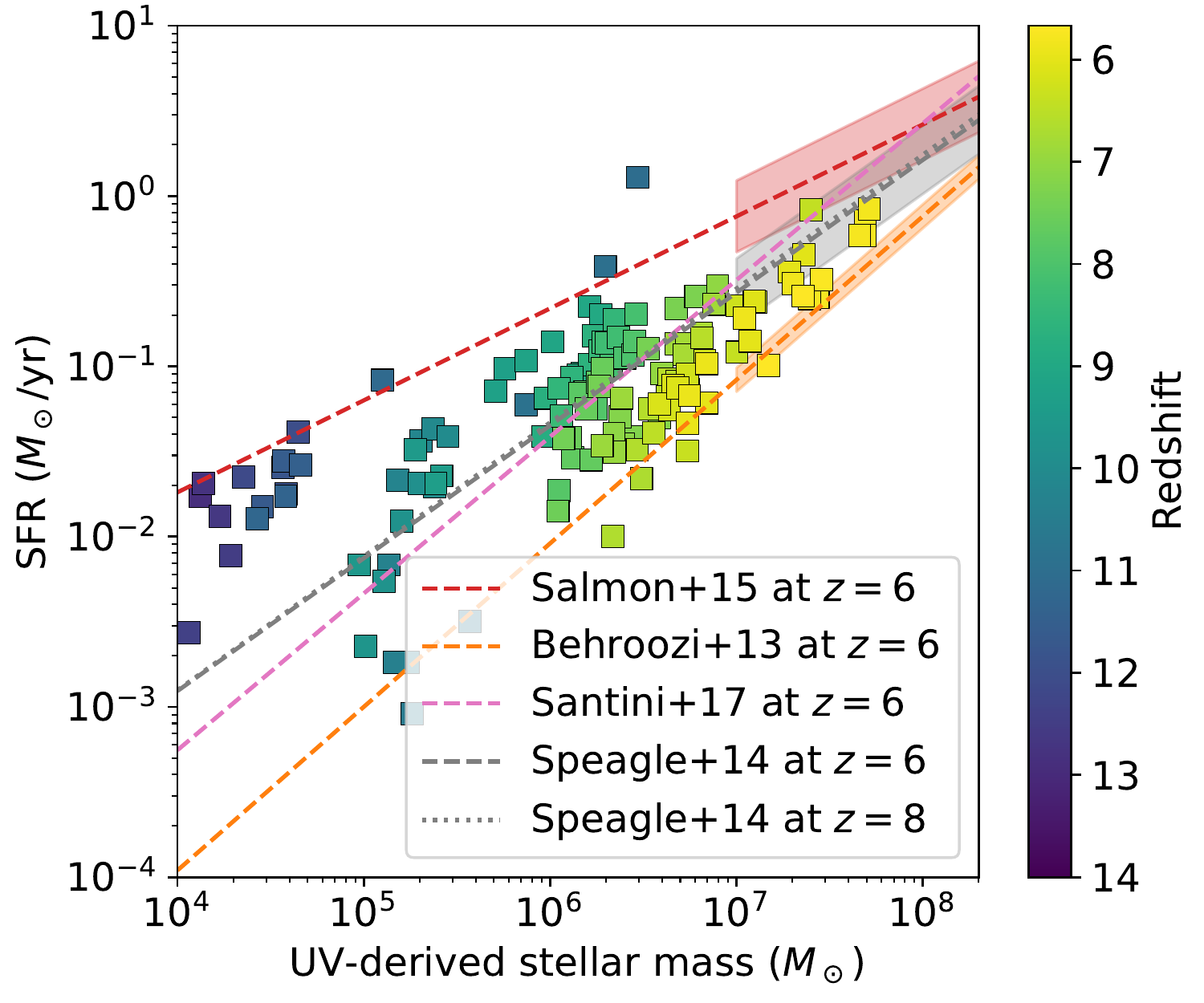}
  \caption{Star formation main sequence, similar to Fig.~\ref{fig:AGNT-MS-direct}, but with only the SFR directly measured in the simulation.}
  \label{fig:AGNT-MS-Msong-sfrdirect}
\end{figure}

We confirm this in Fig.~\ref{fig:AGNT-MS-Mdirect-sfrUV}, which shows again the MS for the \agnt simulation, but using this time the direct estimate of the stellar mass and the SFR estimate based on \citet{Kennicutt2012}. The coloured lines corresponds to the same observations as in Fig.~\ref{fig:AGNT-MS}.
At fixed stellar mass, the scatter is strongly reduced compared to Fig.~\ref{fig:AGNT-MS-direct}. By comparison, Fig.~\ref{fig:AGNT-MS-Msong-sfrdirect} shows the same MS, but with the stellar mass derived using the relation from \citet{Song2016} and the SFR directly measured in the simulation. This can be thought of as a SFR measured from emission lines such as H$\alpha$, which trace very recent star formation.
There, the scatter in stellar mass is large and originates in the burstiness of the star formation rate. This explains why the simulation points match the range of observations relatively well. 

This all suggests that great care should be taken when comparing simulations to high redshift observations, probably even more than presented here, since we did not account for dust or nebular emission. Conversely, this points to the expectation that there should not generically be a  perfect match between simulated quantities such as the stellar mass or the SFR directly measured in simulations and those derived from high redshift observations, especially in the low mass regime.

\section{Stellar mass assembly}
\label{sec:app-stellar-mass}

\begin{figure}
  \centering
  \includegraphics[width=\linewidth]{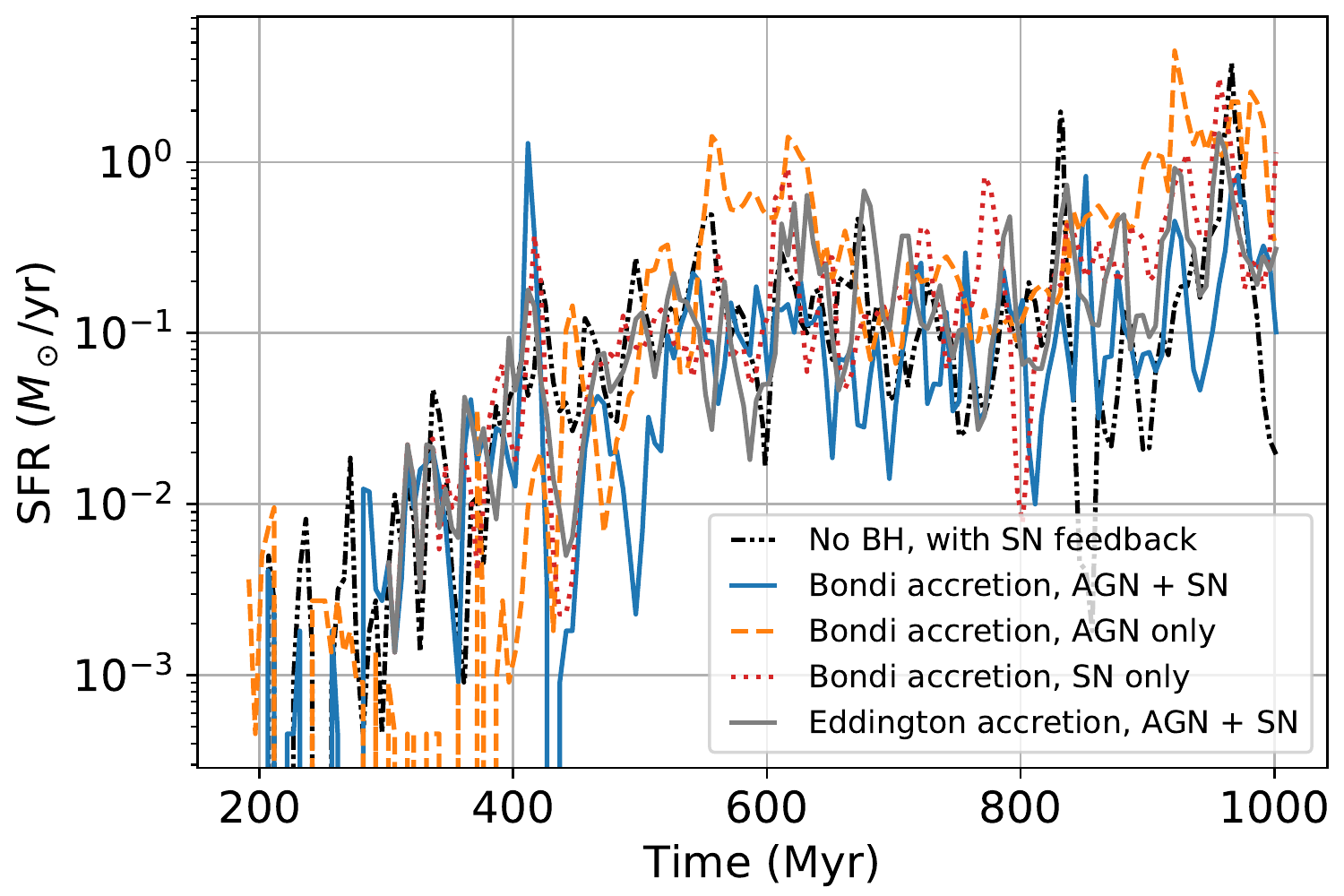}
  \caption{Evolution of the star formation rate for all the simulations presented in this work. The legend is the same as in Fig.~\ref{fig:photons-fesc}.}
  \label{fig:app:sfr}
\end{figure}

\begin{figure}
  \centering
  \includegraphics[width=\linewidth]{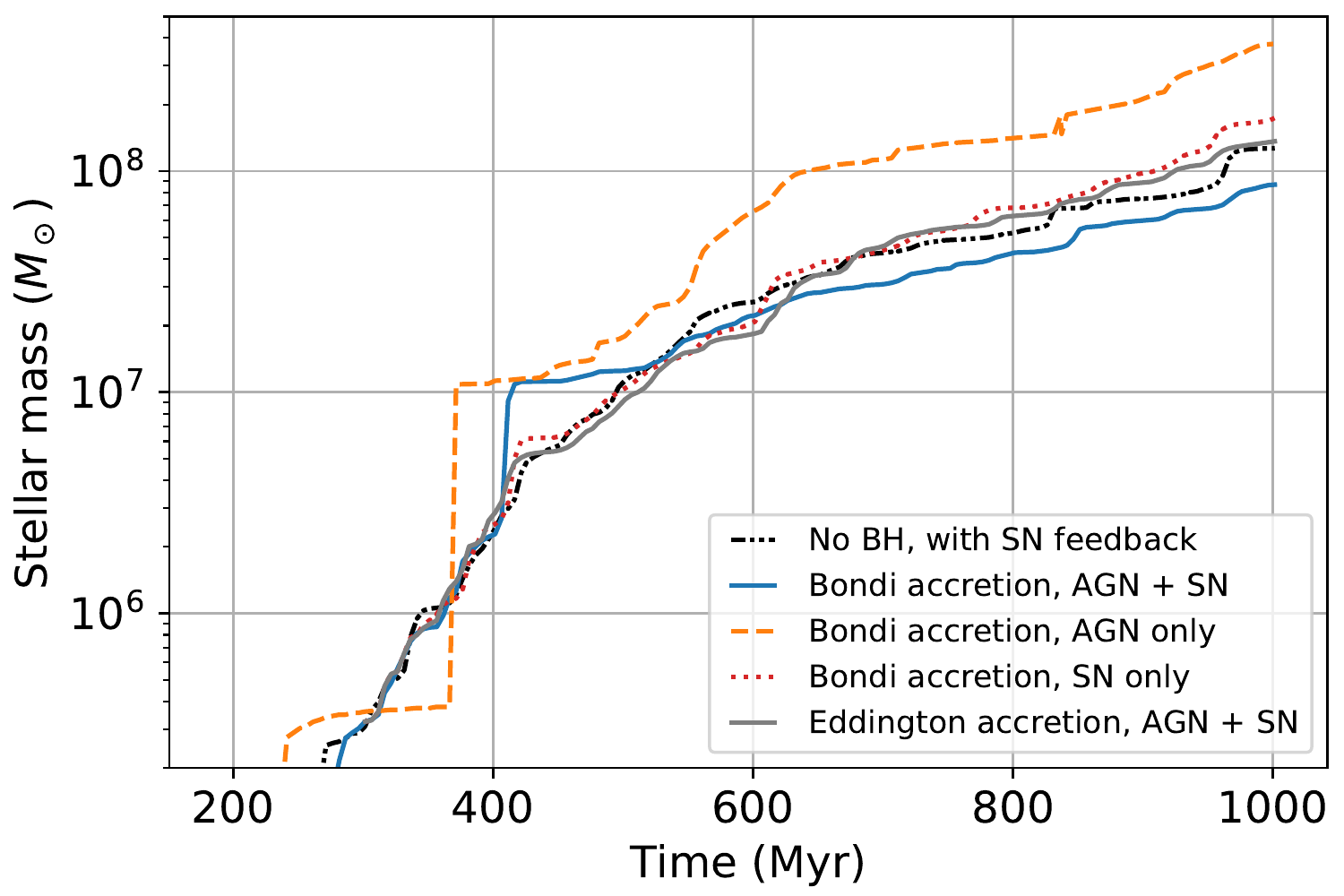}
  \caption{Stellar mass as a function of time for all the simulations presented in this work. The legend is the same as in Fig.~\ref{fig:photons-fesc}.}
  \label{fig:app:mstar}
\end{figure}

In this appendix, we present the star formation history of the main galaxy in all the runs discussed in this work. We compare the evolution of the instantaneous SFR in Fig.~\ref{fig:app:sfr} and the evolution of the stellar masses in Fig.~\ref{fig:app:mstar}. The quantities presented here are those directly measured in the simulation, rather than what one would derive using the UV luminosity (see Sect.~\ref{sec:coevolution:galaxy}).
Despite a large variability of the instantaneous SFR, it seems that the presence of an AGN has very little effect on the final stellar mass. Indeed, even though the galaxy in the run without AGN feedback (\agnobh) ends up with a slightly higher stellar mass than the reference run (\agnt), it also has roughly the same stellar mass as the run where the AGN is actually active (\agntEdd). As expected for low mass galaxies, SN feedback is the main regulator of star formation: the simulation without SN (\agntnoSN) forms many more stars.
The sudden jump in stellar mass for the \agntnoSN simulation around $t = 400\ \mbox{Myr}$ is due to the mis-identification of the galaxy progenitor's host halo.

\bsp	
\label{lastpage}
\end{document}